\documentclass[aps,prb,10pt,twocolumn,floatfix,nofootinbib,superscriptaddress,longbibliography,nobibnotes]{revtex4-2}

\usepackage{amsmath}
\usepackage{amssymb}
\usepackage{mathtools}
\usepackage{scalerel,stackengine,wasysym}

\usepackage[T1]{fontenc}
\usepackage{amsfonts}
\usepackage{newtxtext}
\usepackage[varvw]{newtxmath}
\usepackage{dsfont}                 
\usepackage{bbold}                  
\usepackage[normalem]{ulem}         

\usepackage{enumerate}
\usepackage[shortlabels]{enumitem}

\usepackage[dvipsnames,table]{xcolor}
\usepackage{graphicx}
\graphicspath{{figs/}{supp_figs/}}  

\usepackage{physics}                
\usepackage{csquotes}               

\usepackage[caption=false]{subfig}
\usepackage[colorlinks=true,citecolor=cyan,linkcolor=magenta,filecolor=magenta,bookmarksnumbered]{hyperref}
\usepackage[capitalize]{cleveref}

\usepackage{orcidlink}

\newcommand{\e}{\mathrm{e}}                 
\renewcommand{\i}{\mathrm{i}}               

\newcommand{\approxq}{{\sim}}               

\newcommand\equalhat{\mathrel{\stackon[1.5pt]{=}{\stretchto{%
    \scalerel*[\widthof{=}]{\wedge}{\rule{1ex}{3ex}}}{0.5ex}}}}

\newcommand{\cconj}[1]{{#1}^{*}}            
\newcommand{\adjo}[1]{{#1}^{\dagger}}       

\renewcommand{\vec}[1]{{\vb{#1}}}           

\DeclarePairedDelimiter{\gindexfences}{|}{|}
\newcommand{\gindex}[2]{\gindexfences{#1:#2}}
\newcommand{\gpres}[2]{\left.\left\langle #1\vphantom{#2} \right| #2 \right\rangle} 
\newcommand{\nsubgeq}{\trianglelefteq}      
\newcommand{\nsubg}{\vartriangleleft}       

\newcommand{\nn}[1]{\langle #1 \rangle}     

\newcommand{\Ham}{\mathcal{H}}              
\newcommand{\BlochHam}{H}                   
\newcommand{\TRS}{\mathcal{T}}              

\newcommand{\HTG}{\Gamma}                   
\newcommand{\HTGsc}[1]{\HTG^{(#1)}}         

\newcommand{\genus}{\mathfrak{g}}           
\newcommand{\tgquot}[2]{\mathrm{T}#1.#2}

\newcommand{\torus}[1]{\mathsf{T}^{#1}}     

\newcommand{\gap}{\textsc{gap}}

\newcounter{lettersection}
\newcommand{\sectitle}[1]{%
    \refstepcounter{lettersection}%
    \phantomsection%
    \pdfbookmark[1]{#1}{Letter-Sec-\thelettersection}%
    \emph{{#1}.---}%
\ignorespaces}

\newcounter{appendix}
\makeatletter
\@addtoreset{equation}{appendix}
\makeatother
\renewcommand{\theappendix}{\Alph{appendix}}

\crefname{appendix}{App.}{Apps.}
\Crefname{appendix}{Appendix}{Appendices}
\newcommand{\appsection}[1]{%
    \refstepcounter{appendix}%
    \renewcommand{\theequation}{\theappendix\arabic{equation}}
    \phantomsection%
    \pdfbookmark[1]{App. \theappendix~#1}{End-Matter-App-\theappendix}%
    \textit{Appendix \theappendix: #1.---}%
}

\newcommand{\labeledsection}[2]{%
  \section{#1} 
  \addtocontents{toc}{\vspace{-0.6em}} 
  \addtocontents{toc}{\protect\contentsline{section}{\noindent #2}{}{}}
}

\makeatletter
\def\l@subsection#1#2{}
\def\l@subsubsection#1#2{}
\makeatother

\begin{document}


\title{Hyperbolic Spin Liquids}

\author{Patrick M. Lenggenhager\,\orcidlink{0000-0001-6746-1387}}\email{plengg@pks.mpg.de}
\affiliation{Max Planck Institute for the Physics of Complex Systems, Nöthnitzer Str. 38, 01187 Dresden, Germany}

\author{Santanu Dey\,\orcidlink{0000-0003-2125-6163}}
\affiliation{Department of Physics, University of Alberta, Edmonton, Alberta T6G 2E1, Canada}
\affiliation{Laboratoire de Physique de l'\'{E}cole normale sup\'{e}rieure, ENS, Universit\'{e} PSL, CNRS, Sorbonne Universit\'{e}, Universit\'{e} Paris Cit\'{e}, F-75005 Paris, France}

\author{Tom\'a\v{s} Bzdu\v{s}ek\,\orcidlink{0000-0001-6904-5264}}
\affiliation{Department of Physics, University of Z\"urich, Winterthurerstrasse 190, 8057 Z\"urich, Switzerland}

\author{Joseph Maciejko\,\orcidlink{0000-0002-6946-1492}}\email{maciejko@ualberta.ca}
\affiliation{Department of Physics, University of Alberta, Edmonton, Alberta T6G 2E1, Canada}
\affiliation{Theoretical Physics Institute \& Quantum Horizons Alberta, University of Alberta, Edmonton, Alberta T6G 2E1, Canada}

\date{\today}

\begin{abstract}%
Hyperbolic lattices present a unique opportunity to venture beyond the conventional paradigm of crystalline many-body physics and explore correlated phenomena in negatively curved space.
As a theoretical benchmark for such investigations, we extend Kitaev's spin-1/2 honeycomb model to hyperbolic lattices and exploit their non-Euclidean space-group symmetries to solve the model exactly.
We elucidate the ground-state phase diagram on the $\{8,3\}$ lattice and find a gapped $\mathbb{Z}_2$ spin liquid with Abelian anyons, a gapped chiral spin liquid with non-Abelian anyons and chiral edge states, and a Majorana metal whose finite low-energy density of states is dominated by non-Abelian Bloch states.
\end{abstract}

\maketitle


\sectitle{Introduction}
Among the factors that influence the collective behavior of quantum materials, lattice geometry plays a crucial role, from determining the electronic band structure for weak correlations to geometrically frustrating conventional orders for strong correlations~\cite{SachdevQPM}.
Hyperbolic $\{p,q\}$ lattices~\cite{Kollar:2019,Lenggenhager:2022,Chen:2023,Zhang:2022,Zhang:2023,Huang:2024,Chen:2024}---synthetic materials that emulate regular tilings of two-dimensional (2D) hyperbolic space by $p$-sided polygons with coordination $q$, with $(p\!-\!2)(q\!-\!2)\:{>}\:4$~\cite{Balazs:1986}---present a unique opportunity to explore many-body physics in unusual, non-Euclidean lattice geometries.
While a wealth of phenomena have been investigated on hyperbolic lattices at the single-particle level~\cite{Boettcher:2020,Maciejko:2021,Maciejko:2022,Lenggenhager:2023,cheng2022,kienzle2022,nagy2022,attar2022,shankar2023,Boettcher:2022,Chen:2023b,kollar2020,bzdusek2022,Mosseri:2022,Mosseri:2023,Lux:2022,Lux:2023,Yu:2020,Liu:2022,Urwyler:2022,Liu:2023,Tao:2023,pei2023,yuan2024,Tummuru:2024,Canon:2024,lv2022,sun2023b,ikeda2021,stegmaier2022,Curtis:2023,Chen:2023c,li2023}, much less is known about the interplay of negative curvature and many-body correlations.

Hyperbolic analogs of prototypical interacting Hamiltonians such as the quantum Ising, XY, and Heisenberg models~\cite{daniska2016,daniska2018,gotz2024} and the Bose~\cite{Zhu:2021} and Fermi~\cite{Gluscevich:2023,gluscevich2023b,gotz2024} Hubbard models have been studied recently using mean-field theory, spin-wave theory, and quantum Monte Carlo (QMC).
However, the ability of such methods to reliably capture the bulk properties of hyperbolic lattices must be critically assessed. For example, finite $\{10,3\}$ lattices display a low-energy density of states (DOS) that appears semimetallic~\cite{Gluscevich:2023,gotz2024}, but the thermodynamic-limit DOS is known to be finite~\cite{Mosseri:2023}, with important consequences for many-body physics.
Thus, even numerically exact methods such as QMC may suffer from unusually severe finite-size effects in the hyperbolic context.
This motivates a search for exactly solvable models, to not only discover interesting emergent phenomena but also benchmark approximate many-body theories of hyperbolic lattices.

\begin{figure}[t]
    \subfloat{\label{main:fig:coloring}}
    \subfloat{\label{main:fig:lieb}}
    \centering
    \includegraphics{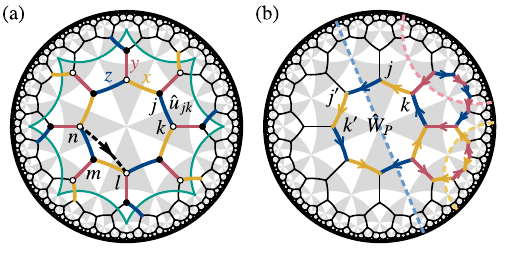}
    \caption{
        (a) Hyperbolic Kitaev model on the $\{8,3\}$ lattice with two sublattices (white/black dots). 
        Symmetric three-edge coloring (yellow, red, and blue, representing $x$, $y$, and $z$) shown inside the primitive cell (green octagon with opposite edges identified).  
        Adjacent sites $j,k$ form a bond $\nn{j,k}_z$; sites $n,m,l$ form an oriented triplet $[lmn]_{xyz}^+$.
        In the Majorana representation, these generate the bond operator $\hat{u}_{jk}$ and a next-nearest-neighbor term (dashed black arrow), respectively.
        The symmetry of the model is depicted by the gray/white triangles.
        (b) Application of Lieb's lemma to determine the ground-state flux sector for three representative plaquettes.
        Three independent mirror lines (dashed geodesics) cut bonds of a different color.
        \emph{Separately} for each plaquette, reflection positivity with respect to one of them implies ground-state bond eigenvalues $u_{jk}\:{=}\:{+}1$ as indicated by arrows from $k$ to $j$ (up to gauge transformations).
        This constrains the gauge-invariant plaquette operators $\hat{W}_P$ consistently throughout the lattice: here $W_P\:{=}\:{-}1$ for all $P$.
    }
    \label{main:fig:model}
\end{figure}

Here, we introduce for the first time an exactly solvable model of strongly correlated spins on hyperbolic lattices (Fig.~\ref{main:fig:model}).
Our model generalizes Kitaev's honeycomb lattice model~\cite{Kitaev:2006} to $\{p,3\}$ lattices and can be solved exactly for any even $p\:{\geq}\:8$.
Although the Kitaev model can be generalized to arbitrary three-coordinated graphs, exact solvability does not immediately follow.
First, a three-edge coloring of the graph must exist and be explicitly constructed, which is in general an NP-complete problem~\cite{stockmeyer1973}.
Second, even with conserved plaquette fluxes~\cite{Kitaev:2006}, the flux optimization problem is generically hard because of the exponential growth of flux configurations with system size.
While Lieb's lemma~\cite{Lieb:1994,Lieb:1993,Macris:1996,Jaffe:2013,Chesi:2013} can simplify the problem if reflection symmetries are present, unlike Euclidean lattices, noncrystalline structures typically possess at most finitely many such symmetries, thus exponentially many flux configurations must still be sampled numerically~\cite{Cassella:2023,Kim:2024}.
Here, we resolve both issues by exploiting the space-group symmetries of hyperbolic lattices~\cite{Maciejko:2021,Boettcher:2022,Chen:2023b}.
First, infinitely many non-Euclidean reflection symmetries allow us to simultaneously solve the three-edge coloring problem \emph{and} determine the ground-state flux configuration analytically.
Second, the (noncommutative) translation symmetry enables us to efficiently approximate the thermodynamic limit via hyperbolic band theory (HBT)~\cite{Maciejko:2021,Maciejko:2022,Lenggenhager:2023}.
We study the model at zero temperature on the $\{8,3\}$ lattice and find two gapped topological phases: a $\mathbb{Z}_2$ spin liquid with Abelian anyons, and a chiral spin liquid with non-Abelian anyons and chiral Majorana edge modes.
Around the isotropic point in the phase diagram, we also find a Majorana metal which---unlike Kitaev's Dirac spin liquid~\cite{Kitaev:2006}---has a finite low-energy DOS dominated by non-Abelian Bloch states~\cite{Maciejko:2022}.

\sectitle{Hyperbolic Kitaev model}
We consider hyperbolic $\{p,3\}$ lattices with a three-edge coloring, i.e., an assignment of one of three colors (yellow, red, blue, labeled as $\alpha\:{=}\:x,y,z$, respectively) to each edge such that coincident edges have different colors (\cref{main:fig:coloring}).
With an $s\:{=}\:1/2$ spin on each site, we define the ferromagnetic ($J_\alpha\:{>}\:0$) hyperbolic Kitaev model (HKM)~as:
\begin{equation}
    \hat{\Ham} = -\sum_{\nn{j,k}_\alpha}J_\alpha\hat{\sigma}_j^\alpha\hat{\sigma}_k^\alpha - K\sum_{[lmn]_{\alpha\beta\gamma}^+}\varepsilon_{\alpha\beta\gamma}\hat{\sigma}_l^\alpha\hat{\sigma}_m^\beta\hat{\sigma}_n^\gamma.
    \label{main:eq:Hamiltonian}
\end{equation}
The $J_\alpha$ term is an anisotropic exchange interaction between adjacent sites $j,k$ sharing an $\alpha$-edge $\nn{j,k}_\alpha$. The term involving the totally antisymmetric tensor $\varepsilon_{\alpha\beta\gamma}$ is an interaction among a counterclockwise-oriented triplet of sites $n,m,l$ (denoted $[lmn]_{\alpha\beta\gamma}^+$) that are connected by bonds $\nn{l,m}_\alpha$ and $\nn{m,n}_\gamma$, respectively, with $\beta\:{\neq}\:\alpha,\gamma$ the color of the third bond adjacent to site $m$.
This term can arise as the leading-order nontrivial effect of a perturbation $-\sum_{j,\alpha}h_\alpha\hat{\sigma}_j^\alpha$ by an external magnetic field $\vec{h}$~\cite{Kitaev:2006}, or through Floquet engineering~\cite{Sun:2023}.

Not all graphs are three-edge colorable, but any three-coordinated \emph{bipartite} simple graph is according to K\H{o}nig's theorem~\cite{Diestel:2017}.
Although this applies to any infinite $\{p,3\}$ lattice with even $p$, such a coloring is not unique.
In \cref{App:coloring}, we describe an algorithm for constructing a three-edge coloring for any hyperbolic $\{p,3\}$ lattice with $p$ even (see \cref{main:fig:coloring} for $p\:{=}\:8$) such that \cref{main:eq:Hamiltonian} is symmetric with respect to any (non-Euclidean) bond-cutting reflection, of which there are three types (\cref{main:fig:lieb}).
The coloring is also compatible with translation symmetry and appropriately chosen periodic boundary conditions (PBC), and can be seen as a hyperbolic generalization of the Kekul\'e pattern on the honeycomb lattice~\cite{Kamfor:2010}.

\sectitle{Majorana representation}
We now solve the HKM exactly. At each site $j$, we introduce the Majorana fermions $\hat{b}_j^\alpha$, $\alpha\:{\in}\:\{x,y,z\}$ and $\hat{c}_j$ such that $\hat{\sigma}_j^\alpha\:{=}\:\i\hat{b}_j^\alpha\hat{c}_j$~\cite{Kitaev:2006}.
Defining the bond operator $\hat{u}_{jk}\:{=}\:\i\hat{b}_j^\alpha\hat{b}_k^\alpha$ on edge $\nn{j,k}_\alpha$, the Hamiltonian becomes~\cite{Kitaev:2006,SM}\nocite{Robinson:1996,HallBook,GAP4,Humphreys,Wen:1989,Pedrocchi:2011,Haldane:1988,Gaspard:1973,Turchi:1982,Knetter:2000,Grunbaum:1987,Sandvik2010,Wright:2013,Wen:2003,Ebisu:2022,Yan:2019,Yan:2022,Yan:2023,Mitchell:2018}
\begin{equation}
    \hat{\Ham} = \sum_{\nn{j,k}_\alpha}J_\alpha\hat{u}_{jk}\i\hat{c}_j\hat{c}_k + K\sum_{[lmn]_{\alpha\beta\gamma}^+}\hat{u}_{lm}\hat{u}_{mn}\i\hat{c}_l\hat{c}_n.
    \label{main:eq:Majorana-Hamiltonian}
\end{equation}
While $\hat{\Ham}$ in \cref{main:eq:Majorana-Hamiltonian} acts on the extended Hilbert space, $\hat{\Ham}$ in \cref{main:eq:Hamiltonian} only acts on the physical Hilbert space of the spin system, defined as the common $+1$ eigenspace of the $\mathbb{Z}_2$ gauge transformations $\hat{D}_j\:{=}\:\hat{b}_j^x\hat{b}_j^y\hat{b}_j^z\hat{c}_j$.

Because the $\hat{u}_{jk}$ commute with $\hat{\Ham}$ and each other, we replace them by their eigenvalues $u_{jk}\:{=}\:{\pm}1$ and study the resulting quadratic Majorana Hamiltonian.
Since the bond operators are not gauge invariant, we consider the Wilson loops $\hat{W}(\ell)\:{=}\:\prod_{\nn{j,k}_\alpha{\in}\ell}\hat{\sigma}_j^\alpha\hat{\sigma}_k^\alpha$ along closed paths $\ell$.
In the Majorana representation, they take the form $\hat{W}(\ell)\:{=}\:\prod_{\nn{j,k}_\alpha{\in}\ell}\left(-\i\hat{u}_{jk}\right)$.
On an infinite hyperbolic lattice, all $u_{jk}$ are (up to gauge transformations) fully determined by the Wilson loops $\hat{W}_P$ around the individual plaquettes $P$, measuring the corresponding flux.
On compactified PBC clusters with genus $\genus{}$, plaquette fluxes can only be changed in pairs, and there also exist Wilson loops along $2\genus{}$ noncontractible paths~\cite{bzdusek2022}.

\sectitle{Exact solution of the flux problem}
For $K\:{=}\:0$, the ground-state configuration of plaquette fluxes can be determined analytically from symmetry.
First, Lieb's lemma on reflection positivity~\cite{Lieb:1994,Lieb:1993,Macris:1996,Jaffe:2013,Chesi:2013} implies that, in the ground state, the gauge variables $\hat{u}_{jk}$ lying on either side of a mirror line are related by reflection, up to gauge transformations.
Since our model is reflection symmetric with respect to \emph{any} bond-cutting mirror line for \emph{any} choice of parameters $J_\alpha$ (\cref{main:fig:lieb}), we can consider each plaquette separately.

Given a plaquette, we select one of the reflection symmetries and denote by $j'$ the image of site $j$ under that reflection.
We can always choose a gauge where $u_{j'j}\:{=}\:{+}1$ for the bonds crossing the mirror line (dashed geodesics in \cref{main:fig:lieb}).
Then, Lieb's lemma implies that the remaining reflection-related bonds satisfy $u_{jk}\:{=}\:u_{k'j'}$.
Indeed, under reflection symmetry the term $u_{jk}\i \hat{c}_j\hat{c}_k$ is mapped to $u_{jk}(-\i)\hat{c}_{j'}\hat{c}_{k'}\:{=}\:u_{jk}\i\hat{c}_{k'}\hat{c}_{j'}$ (reflection is represented antiunitarily for Majorana fermions).
Thus, for a $\{p,3\}$ lattice with $p$ even,
\begin{equation}
    W_P = (-\i)^p\times(-1)\times (+1)^{p/2-1} = -(-1)^{p/2},
    \label{main:eq:GS-flux-config}
\end{equation}
where $(-\i)^p$ follows from the definition of $W_P$, $(-1)$ from the opposite orientation (relative to the oriented Wilson loop) of the two bonds cut by the mirror line, and $(+1)^{p/2-1}$ from the remaining $(p/2\!-\!1)$ reflection-related pairs of bonds each having equal orientation.
Unlike in the case of the coloring studied traditionally~\cite{Kitaev:2006}, \cref{main:eq:GS-flux-config} applies for any choice of couplings $J_\alpha$.

The honeycomb ($\{6,3\}$) lattice has $p/2\:{=}\:3$, such that $W_P\:{=}\:{+}1$, while in our example, $p/2\:{=}\:4$, thus the ground state has homogeneous $\pi$-flux ($W_P\:{=}\:{-}1$).
By further exploring all $2^{6-1}\:{=}\:32$ possible translation-invariant flux configurations on the infinite $\{8,3\}$ lattice,
we find that, in agreement with \cref{main:eq:GS-flux-config}, the homogeneous $\pi$-flux configuration results in the lowest many-fermion ground-state energy, see \cref{App:GS-flux-sector}.
For concreteness, we subsequently focus on the $\{8,3\}$ lattice.

\begin{figure}[t]
    \subfloat{\label{main:fig:phase-diagram_K=0:triangle}}
    \subfloat{\label{main:fig:DOS_K=0}}
    \centering
    \includegraphics{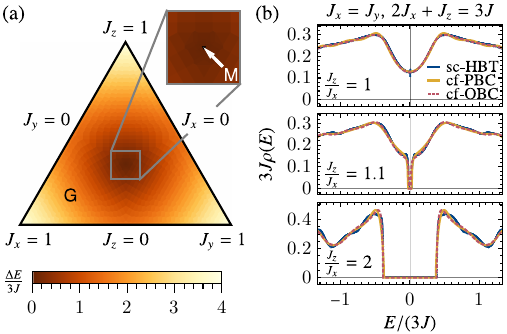}
    \caption{
        (a) Phase diagram of the spectral gap $\Delta E$ for $K\:{=}\:0$ in the plane $J_x{+}J_y{+}J_z\:{=}\:3J$. Inset: region near the isotropic point ($J_x\,{=}\,J_y\,{=}\,J_z\,{=}\,J$) where the gap vanishes (black), separating a Majorana metal ($\mathsf{M}$) from the gapped ($\mathsf{G}$) spin-liquid phase.
        (b) Low-energy fermionic DOS (top) at the isotropic point, (middle) slightly away from it, and (bottom) deep in the anisotropic region, calculated using the supercell method (sc-HBT; $2048$ sites) and the continued-fraction method applied to clusters with periodic (cf-PBC) and to flakes with open boundary conditions (cf-OBC) with $\approxq 10^8$ sites.
    }
    \label{main:fig:phase-diagram_K=0}
\end{figure}

\sectitle{Fermionic spectrum}
Having determined the ground-state flux sector, we next study the spectrum of fermionic excitations as a function of the couplings $J_\alpha$.
The relevant quadratic Majorana Hamiltonian $\hat{\Ham}\!\:{=}\:\!\frac{\i}{4}\sum_{j,k}A_{jk}\hat{c}_j\hat{c}_k$ possesses hyperbolic translation symmetry, thus we diagonalize it using HBT~\cite{Maciejko:2021,Maciejko:2022,Lenggenhager:2023}.
To capture the non-Abelian Bloch states~\cite{Maciejko:2022} characteristic of hyperbolic reciprocal space, we generalize the supercell method~\cite{Lenggenhager:2023} to quadratic Majorana Hamiltonians, see \cref{App:HBT-Majorana}.
We use a coherent sequence~\cite{Lux:2022,Lux:2023,Lenggenhager:2023} of five supercells containing up to $2048$ sites, obtained from \textsc{HyperCells}~\cite{Lenggenhager:2023,HyperCells,Lenggenhager:PhDThesis,Conder:2007}, and perform random sampling of momenta using \textsc{HyperBloch}~\cite{HyperBloch}.
From the fermionic spectrum, we deduce the DOS $\rho(E)$ and corresponding spectral gap $\Delta E$ as described in \cref{App:supercell-extrapolation}.
To complement the supercell method based on HBT, we additionally compute $\rho(E)$ at selected points in the phase diagram using the real-space continued-fraction method~\cite{Haydock:1972,Haydock:1975,Mosseri:2023} on PBC clusters and finite flakes with open boundary conditions (OBC) containing $\approxq 10^8$ sites~\cite{SM}.

\begin{figure}[t]
    \subfloat{\label{main:fig:gap_vs_Jz_for_K}}
    \subfloat{\label{main:fig:gap_vs_Jz}}
    \subfloat{\label{main:fig:gap_vs_K}}
    \centering
    \includegraphics{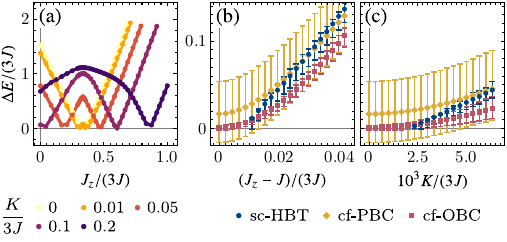}
    \caption{
        Spectral gap $\Delta E$ as a function of $J_z$ and $K$.
        (a) Vertical cut through the phase diagram in \cref{main:fig:phase-diagram_K=0:triangle} for different values of $K$ (see legend below panel), with $J_x\,{=}\,J_y\,{=}\,(3J\:{-}\:J_z)/2$.
        (b) Close-up of (a) for $K=0$ showing the gap opening as a function of $J_z$ obtained using the three methods (see legend below panels and caption of \cref{main:fig:phase-diagram_K=0}).
        (c) Gap opening with $K$ at the isotropic point $J_x\,{=}\,J_y\,{=}\,J_z\,{=}\,J$.
    }
    \label{main:fig:spectral-gap}
\end{figure}

\sectitle{Majorana metal}
We first consider the case $K\:{=}\:0$.
The $\Delta E$ phase diagram in \cref{main:fig:phase-diagram_K=0:triangle} shows a gapless phase ($\mathsf{M}$) around the isotropic point $J_x\,{=}\,J_y\,{=}\,J_z$ (see inset) and a gapped phase away from it ($\mathsf{G}$).
Representative DOS computed from different methods are in excellent agreement (\cref{main:fig:DOS_K=0}).
Our data suggests the gapless phase $\mathsf{M}$ is confined to the isotropic point, or at most a small region around it (\cref{main:fig:gap_vs_Jz}).
The phase is characterized by a finite DOS at $E\:{=}\:0$ in sharp contrast to the linearly vanishing DOS $\rho(E)\:{\propto}\:|E|$ associated with the Dirac spectrum on the honeycomb lattice~\cite{Kitaev:2006}.
Thus, unlike Kitaev's Dirac spin liquid, the $\{8,3\}$ HKM realizes a \emph{Majorana metal}.
Crucially, Abelian HBT alone incorrectly predicts a vanishing DOS $\rho(E)\:{\propto}\:|E|^3$ at low energies arising from conical singularities in the 4D Brillouin zone of Abelian Bloch states.
However, the latter only capture particular slices through the full reciprocal space which is dominated by non-Abelian Bloch states~\cite{shankar2023}.
Thus, the finite DOS here is a direct consequence of non-Abelian Bloch physics, which is absent for Euclidean lattices.
Similar phenomenology, where non-Abelian Bloch states drastically alter the low-energy DOS, has been observed in Ref.~\onlinecite{Tummuru:2024}.

\sectitle{$\mathbb{Z}_2$ spin liquid}
To better understand the nature of the gapped ($\mathsf{G}$) phase away from the isotropic point, we study the HKM in the limit of extreme coupling anisotropy, $J_x,J_y\:{\ll}\:J_z$, where the fermion gap $\Delta E/(3J)\:{\approx}\:4$ (\cref{main:fig:phase-diagram_K=0:triangle}).
When $J_x\,{=}\,J_y\,{=}\,0$, the model reduces to decoupled Ising dimers on $z$-bonds, each of which minimizes its energy by adopting one of two ferromagnetic configurations ($\uparrow\uparrow$ or $\downarrow\downarrow$), resulting in a macroscopic ground-state degeneracy.
This degeneracy is lifted at small but nonzero $J_x,J_y$, and the nature and spectrum of the resulting low-energy excitations can be determined from an effective Hamiltonian obtained by degenerate perturbation theory~\cite{Kitaev:2006,Vidal:2008,Schmidt:2008}.
We first find that the HKM on the $\{8,3\}$ lattice maps exactly onto a model of effective spin-1/2 degrees of freedom and hardcore bosons on the Archimedean $(8,4,8,4)$ lattice.
The latter is the lattice obtained by collapsing the $z$-bond dimers into effective sites, and contains alternating square ($\square$) and octagonal ($\octagon$) plaquettes.
The spin states represent the two ferromagnetic configurations of each dimer, and bosons correspond to excitations out of the low-energy ferromagnetic subspace, with large energy cost $\Delta E/2\:{\approx}\:2J_z$.

To focus on the low-energy physics, we project onto the zero-boson subspace, and obtain the effective spin-1/2 Hamiltonian~\cite{SM}:
\begin{align}\label{main:eq:main-Heff}
    \mathcal{\hat{H}}_\text{eff}=\frac{5}{16}\frac{J_\parallel^4}{J_z^3}\sum_{\square}\hat{W}_{\square}
+\frac{5}{2048}\frac{J_\parallel^8}{J_z^7}\sum_{\octagon}\hat{W}_{\octagon},
\end{align}
where $\hat{W}$ are Wilson loop operators on the $(8,4,8,4)$ lattice, and we have set $J_x\,{=}\,J_y\,{=}\,J_\parallel$ here for simplicity.
The $\hat{W}$ operators all commute with each other, and are in fact equivalent to the plaquette operators $\hat{W}_P$ introduced earlier.
Thus, the positive couplings in \cref{main:eq:main-Heff} imply that $\hat{W}_P\:{=}\:{-}1$ in the ground state, consistent with the exact result Eq.~(\ref{main:eq:GS-flux-config}).
Second, \cref{main:eq:main-Heff} implies that the lowest-energy excitation is a $\mathbb{Z}_2$ vortex with $\hat{W}_{\octagon}\:{=}\:{+}1$ and energy cost ${\sim}\,J_\parallel^8/J_z^7$, much less than the fermion gap $\Delta E/2\:{\approx}\:2J_z$ in that limit.
Finally, the effective model \eqref{main:eq:main-Heff} can be further mapped to a hyperbolic analog of the toric code~\cite{Kitaev:2003} on the $\{8,4\}$ lattice, i.e., a hyperbolic surface code~\cite{Breuckmann:2016,Breuckmann:2017,Lavasani:2019,Jahn:2021,Higgott:2023,Fahimniya:2023}.
This last mapping reveals that the $\square$ and $\octagon$ vortices obey bosonic self-statistics but are mutual semions, establishing that the $\mathsf{G}$ phase is a topologically ordered $\mathbb{Z}_2$ spin liquid~\cite{Wen:1991}.

\begin{figure}[t]
    \subfloat{\label{main:fig:phase-diagram_K=0.1}}
    \subfloat{\label{main:fig:edge-dispersion}}
    \centering
    \includegraphics{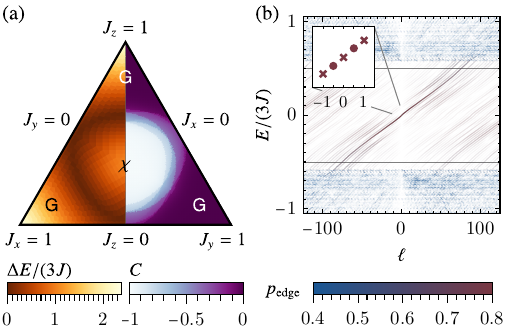}
    \caption{
        (a) Phase diagram for $K/(3J)\:{=}\:0.1$ with the spectral gap $\Delta E$ shown in the left half and the Chern number $C$ computed on a PBC cluster ($2048$ sites) in the right half.
        The chiral ($\chi$) and $\mathbb{Z}_2$ ($\mathsf{G}$) spin liquid phases are separated by a phase transition with a gap closing and an integer change in $C$.
        (b) Energy $E$ vs angular momentum $\ell$ for $J_x\,{=}\,J_y\,{=}\,J_z\,{=}\,J$, $K/(3J)\:{=}\:0.1$, computed on an OBC disk ($896$ sites). Color encodes the degree of edge localization $p_\mathrm{edge}$, and opacity the weight of the corresponding $\ell$. Inset: $\ell\:{\in}\:\mathbb{Z}\:{+}\:\frac{1}{2}$ (red dots) at low energies without a vortex, and $\ell\:{\in}\:\mathbb{Z}$ (red crosses) with a $\mathbb{Z}_2$ vortex at the center of the disk.
    }
    \label{main:fig:chiral-spin-liquid}
\end{figure}

\sectitle{Chiral spin liquid}
A different type of gapped spin liquid is obtained when the emergent Majorana fermions carry a nonzero Chern number. This requires time-reversal symmetry to be broken, which happens for $K\:{\neq}\:0$.
Focusing first on the isotropic point, where for $K\:{=}\:0$ the fermionic spectrum is gapless, a gap opens at infinitesimal $K\:{\neq}\:0$ and subsequently increases with increasing $K$ (\cref{main:fig:gap_vs_K}).
Thus, for finite $K$, a new gapped phase $\chi$ develops around the isotropic point and remains separated from $\mathsf{G}$ by a circular gapless line in parameter space (\cref{main:fig:phase-diagram_K=0.1}, left half). From cuts through the phase diagram for different values of $K$ (\cref{main:fig:gap_vs_Jz_for_K}), we find that the $\chi$ region expands with increasing $K$.

The Chern number $C$ determines the properties of anyonic excitations as well as the existence and character of topologically protected boundary modes~\cite{Kitaev:2006}.
While in Euclidean translation-invariant systems, $C$ can be easily computed in momentum space, we rely here on a real-space formulation~\cite{Kitaev:2006} and compute it on finite PBC clusters~\cite{SM}.
\Cref{main:fig:phase-diagram_K=0.1} shows that the gapped $\chi$ phase around the isotropic point has odd Chern number $C\:{=}\:{-}1$, establishing it as a chiral spin liquid with non-Abelian anyons~\cite{Kitaev:2006}, while $C$ vanishes in the gapped $\mathbb{Z}_2$ spin liquid ($\mathsf{G}$) phase.

Finally, the nonzero Chern number suggests gapless chiral edge modes, which we investigate in a disk-shaped OBC flake at the isotropic point.
For a sufficiently large flake, an approximate continuous rotation symmetry emerges on the edge, allowing us to introduce an approximate angular momentum quantum number $\ell$, as discussed in \cref{App:chiral-edge-states}.
In \cref{main:fig:edge-dispersion}, we show the corresponding angular dispersion together with a measure $p_\mathrm{edge}$ of edge localization defined as the integrated probability density within the outer $10\%$ of the hyperbolic radius of the flake.
Bulk modes (blue) generally do not have sharp angular momentum, but a branch of states sharply peaked at a single $\ell$ and strongly localized on the edge (red) crosses the bulk gap; we identify it with the single dispersive band of chiral edge states expected for the $C\:{=}\:{-}1$ topology. 
In contrast to Euclidean lattices, there is an extensive number of such edge states due to the finite boundary-to-bulk ratio in hyperbolic geometry.

For edge modes described by a chiral Majorana conformal field theory with chiral central charge $c_-\:{=}\:1/2$, we expect a linear low-energy angular dispersion $E\:{\propto}\:\ell$ with half-integer quantization $\ell\;{\in}\;\mathbb{Z}+\frac{1}{2}$~\cite{Read:2000,stone2004}.
The inset in \cref{main:fig:edge-dispersion} (red dots) confirms this expectation, notably the absence of a zero-energy mode with $\ell\:{=}\:0\;{\notin}\;\mathbb{Z}+\frac{1}{2}$.
Inserting a vortex through the center of the disk binds a Majorana zero mode there, shifts $\ell$ by $1/2$ such that $\ell\;{\in}\;\mathbb{Z}$~\cite{Read:2000,stone2004}, and induces a second zero-energy mode on the boundary (red crosses in the inset).

\sectitle{Conclusion}
In summary, we introduced for the first time an exactly solvable model of strongly correlated hyperbolic quantum matter, the hyperbolic Kitaev model (HKM).
The non-Euclidean space-group symmetries of hyperbolic lattices play a crucial role in the model's construction and solution.
In contrast to previous noncrystalline extensions of the Kitaev model, reflection symmetries across geodesics enable an exact analytical determination of the ground-state flux sector via Lieb's lemma, and noncommutative translation symmetries allow for an efficient determination of thermodynamic-limit properties via hyperbolic band theory.
Our detailed study of the HKM on the $\{8,3\}$ lattice reveals both Abelian and non-Abelian gapped topological spin liquids, as well as a gapless spin liquid that, unlike Kitaev's Dirac spin liquid, has a finite low-energy density of states dominated by Majorana non-Abelian Bloch states, a unique feature of hyperbolic space.

Our work opens several vistas for future study.
On the theoretical side, given the degree of analytical control the HKM affords, one should investigate whether the bulk hyperbolic spin liquids found here realize interesting \enquote{holographic spin liquids} on the edge~\cite{asaduzzaman2020,brower2021,basteiro2022,basteiro2023,basteiro2023b,chen2023d,Dey:2024}.
Unlike Kitaev's (unique) honeycomb lattice in 2D, infinitely many $\{p,3\}$ lattices are now open to investigation, as well as other possible extensions of Kitaev physics~\cite{yao2007,yang2007,yao2009,wu2009,barkeshli2015,vaezi2014}.
On the experimental side, the spin-spin interactions in \cref{main:eq:Hamiltonian} could potentially be realized via qubit-photon interactions~\cite{bienias2022} in circuit quantum electrodynamics~\cite{Kollar:2019} with the particular spin interactions implemented using Floquet engineering~\cite{Sun:2023} of Ising-type interactions~\cite{Salathe:2015,Nguyen:2024}.
For applications to quantum error correction, implementing a two-spin interaction in the anisotropic coupling limit $J_z\:{\gg}\:J_x,J_y$ might represent a simpler path towards hyperbolic surface codes than directly engineering the requisite multi-spin interactions~\cite{Breuckmann:2016,Breuckmann:2017,Jahn:2019}.


\emph{Note added.} While finalizing this manuscript, we became aware of an independent work~\cite{Dusel:2024} studying the Kitaev model on the $\{9,3\}$ lattice, where the authors identify a gapless chiral $\mathbb{Z}_2$ spin liquid.

\sectitle{Acknowledgments}
We thank Igor Boettcher for many insightful conversations and suggestions over the course of this project, Julien Vidal for helpful feedback on several aspects of our work, as well as Marin Bukov, Felix Dusel, Juraj Ha\v{s}\'{i}k, Tobias Hofmann, Bastien Lapierre, Ji\v{r}\'{i} Min\'{a}\v{r}, Rémy Mosseri, Benedikt Placke, G. Shankar, Ronny Thomale, and G\"otz S. Uhrig for valuable discussions.
P.~M.~L. acknowledges support by the European Union (ERC, QuSimCtrl, 101113633).
Views and opinions expressed are however those of the authors only and do not necessarily reflect those of the European Union or the European Research Council Executive Agency. Neither the European Union nor the granting authority can be held responsible for them.
S.~D.~was supported by the Faculty of Science at the University of Alberta.
T.~B.~was supported by the Starting Grant No.~211310 by the Swiss National Science Foundation (SNSF).
J.~M.~was supported by NSERC Discovery Grants \#RGPIN-2020-06999 and \#RGPAS-2020-00064; the Canada Research Chair (CRC) Program; and Alberta Innovates.
This research was enabled in part by support provided by the Digital Research Alliance of Canada.

\sectitle{Data availability}
The data that support the findings of this Letter are openly available~\cite{SDC}.

\let\oldaddcontentsline\addcontentsline     
\renewcommand{\addcontentsline}[3]{}        

\pdfbookmark[1]{References}{references}
\bibliography{arXiv_v2.bbl}
\onecolumngrid

\let\addcontentsline\oldaddcontentsline     

\clearpage

\onecolumngrid
\begin{center}
\phantomsection
\large\textbf{End Matter}
\addcontentsline{toc}{chapter}{End Matter}
\end{center}
\twocolumngrid 

\appsection{Symmetric three-edge coloring}\label{App:coloring}%
The three edges coincident on a vertex of the $\{2m,3\}$ lattice are in one-to-one correspondence with the three sides of a face of the dual $\{3,2m\}$ lattice, which is an equilateral triangle (see white/gray triangles in \cref{main:fig:coloring}, for $m=4$).
Those three sides belong to distinct equivalence classes under reflections in the sides of any equilateral triangle, which are bond-cutting reflection symmetries of the original $\{2m,3\}$ lattice.
By coloring the sides of the equilateral triangles according to their equivalence class, we obtain a three-edge coloring which respects those reflection symmetries and allows us to use Lieb's lemma.

Mathematically, the edges of the $\{2m,3\}$ lattice form the right coset space $H\backslash G$ where the hyperbolic triangle group $G\,{=}\,\Delta(2,3,2m)$ is the space group of the $\{2m,3\}$ lattice~\cite{Boettcher:2022,Chen:2023b} and $H$ is the subgroup of $G$ which leaves a given edge invariant (stabilizer subgroup).
Further quotienting out reflections in the sides of the equilateral triangles, which form the subgroup $K\,{=}\,\Delta(m,m,m)$ of $G$, we obtain the double coset space $H\backslash G/K$ which contains only three elements, i.e., three colors.
For computations using hyperbolic band theory the chosen Bravais unit cell or supercell must be compatible with this edge coloring. This is ensured if the corresponding translation group $\Gamma$ is a normal subgroup of both $G$ and $K$.
In Sec.~I of the Supplemental Material (SM)~\cite{SM}, we spell out the above mathematical arguments in more rigor.

\appsection{Majorana hyperbolic band theory}\label{App:HBT-Majorana}%
The generic quadratic Majorana Hamiltonian $\hat{\Ham} = \frac{\i}{4}\sum_{j,k}A_{jk}\hat{c}_j\hat{c}_k$ with $A^\top=-A$ and $\{\hat{c}_j,\hat{c}_k\}=2\delta_{jk}$ defined on a hyperbolic lattice with translation group $\Gamma$ can be written in reciprocal space~\cite{Maciejko:2021,Maciejko:2022} using the generalized Fourier transform~\cite{Canon:2024}
\begin{equation}
    \hat{a}_{\mu\nu,j}^{(K)} = \frac{1}{\sqrt{2\abs{\Gamma}}}\sum_{\gamma\in \Gamma}\sqrt{d_K}\hat{c}_{\gamma,j}D_{\nu\mu}^{(K)}(\gamma),
\end{equation}
where $K$ runs over all irreducible representations (IRs) $D^{(K)}$ and $\mu,\nu$ run from $1$ to $d_K$, the dimension of $D^{(K)}$.
Defining $\i A^{(K)} = \sum_{\gamma\in \Gamma}\i A(\gamma)\otimes D^{(K)}(\gamma)$, we obtain
\begin{equation}
    \hat{\Ham} = \frac{\i}{2}\sum_K\sum_{\mu,\nu,\nu'}\sum_{j,k}A_{\nu,j;\nu',k}^{(K)}
    \adjo{\mbox{$\displaystyle\hat{a}_{\mu\nu,j}^{(K)}$}}\hat{a}_{\mu\nu',k}^{(K)}.
    \label{main:eq:Bloch-decomp}
\end{equation}
Diagonalizing the Hermitian matrices $\i A^{(K)}$, giving operators $\hat{d}_{\mu\lambda,l}^{(K)}$ and eigenvalues $\varepsilon_{\lambda,l}(K)$, finally results in
\begin{equation}
    \hat{\Ham} = \sum_{\substack{K,\lambda,\mu,l\\\varepsilon_{\lambda,l}(K)>0}}\varepsilon_{\lambda,l}(K)\left(\adjo{\mbox{$\displaystyle\hat{d}_{\mu\lambda,l}^{(K)}$}}\hat{d}_{\mu\lambda,l}^{(K)}-\frac{1}{2}\right).
    \label{main:eq:Bloch-Hamiltonian_diagonalized}
\end{equation}
Analogously to the Euclidean case, the sum is constrained to positive energies due to the reality of the Majorana fermions, which relates states corresponding to conjugate IRs.
The derivation of \cref{main:eq:Bloch-decomp,main:eq:Bloch-Hamiltonian_diagonalized} amounts to straightforward algebraic manipulations, see SM~\cite{SM}, Sec.~IV.

\appsection{Extrapolation using the supercell method}\label{App:supercell-extrapolation}%
The supercell method~\cite{Lenggenhager:2023} provides a framework for including the effect of higher-dimensional IRs $D^{(K)}$ by sampling one-dimensional IRs on successively larger unit cells (supercells).
To estimate the \emph{true} value of a given quantity, we compute it for supercell sizes $N\in\{1,4,16,32,128\}$ (see SM~\cite{SM}, Sec.~V) and subsequently extrapolate $N$ to $\infty$.
Below, we discuss the details of this procedure for the ground-state (GS) energy and for the fermionic spectral gap.

\begin{figure}[t]
    \centering
    \includegraphics{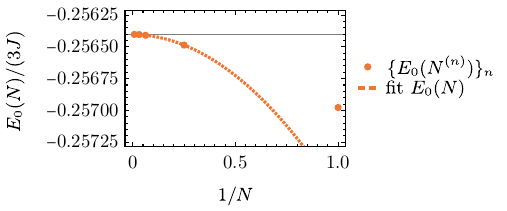}
    \caption{
    Extrapolation of the ground-state energy $E_0(N)$ for the homogeneous $\pi$-flux configuration.
    The data points show the values obtained using \cref{main:eq:ground-state-energy} and the line is the fit according to \cref{main:eq:ground-state-extrapolation:fit-model} (excluding $n=1$).
    We find $E_0/(3J)=-0.25640(1\pm 6)$ from a fit with coefficient of determination $1-R^2\sim 10^{-12}$.
    }
    \label{main:fig:E0-extrapolation}
\end{figure}

The GS energy is given by $E_0 = -\frac{1}{4}\overline{\abs{\varepsilon_{\lambda,l}(K)}}$ with the average running over the full spectrum.
On the $n^\text{th}$ supercell, we randomly sample $N_s$ momenta from the corresponding Abelian Brillouin zone $\mathrm{ABZ}^{(n)}$ and compute
\begin{equation}
    E_0(N^{(n)}) = -\frac{1}{4N_s}\sum_{\vec{k}\in\mathrm{ABZ}^{(n)},l}\abs{\varepsilon_{l}(\vec{k})}.
    \label{main:eq:ground-state-energy}
\end{equation}
We extrapolate using a weighted least-squares fit with model
\begin{equation}
    E_0(N) = E_0 + \frac{u}{N} + \frac{v}{N^2},
    \label{main:eq:ground-state-extrapolation:fit-model}
\end{equation}
and weights $N$, excluding the primitive cell $n=1$.
In \cref{main:fig:E0-extrapolation}, we demonstrate this for the homogeneous $\pi$-flux configuration.
The resulting estimate for $E_0$ is given together with the parameter error reflecting a $95\%$ confidence interval.

\begin{figure*}[t]
    \subfloat{\label{main:fig:spectral-gap:cDOS-vs-E}}
    \subfloat{\label{main:fig:spectral-gap:cDOS-vs-N}}
    \subfloat{\label{main:fig:spectral-gap:slope}}
    \centering
    \includegraphics{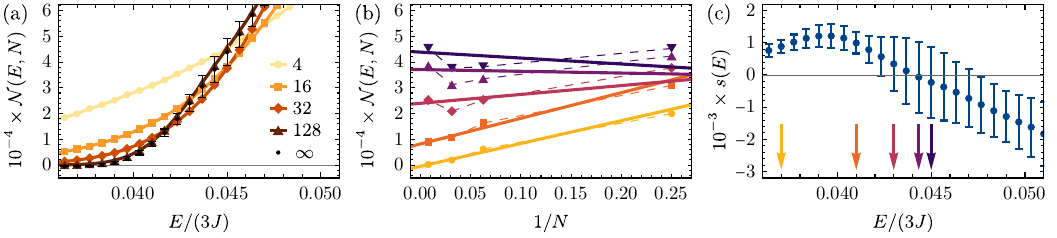}
    \caption{Estimation of the spectral gap.
    Integrated density of states $\mathcal{N}(E,N)$ for $J_x=J_y=J_z=J$ and $K/(3J)=0.1$.
    (a) $\mathcal{N}(E,N)$ as a function of energy $E$ for different supercells (see inset legend for the supercell size $N$), including the extrapolated value $\mathcal{N}_0(E)$ (\enquote{$N=\infty$}) with $95\%$ confidence intervals shown as error bars.
    (b) $\mathcal{N}(E,N)$ as a function of the inverse supercell size $1/N$ for the values of energy indicated by the correspondingly colored arrows in panel (c).
    Dashed lines are guides to the eye and the solid lines are linear maximum-likelihood fits.
    (c) The slope $s(E)$ extracted from the same fits as a function of energy $E$; error bars indicate $95\%$ confidence intervals.}
    \label{main:fig:spectral-gap-method}
\end{figure*}

The spectral gap $\Delta E=2E_g$ is the extent of the interval of energies $[-E_g,E_g]$ with vanishing density of states (DOS).
We estimate $\Delta E$ through the integrated DOS
\begin{equation}
    \mathcal{N}(E,N) = \int_0^E\dd{E'}\rho(E',N),
    \label{main:eq:cDOS}
\end{equation}
obtained by constructing a cumulative histogram of the computed eigenvalues.
It shows a transition between the region $E<E_g$ where $\mathcal{N}(E,N)$ is suppressed for increasing $N$ and the region $E>E_g$ where it is enhanced, see \cref{main:fig:spectral-gap:cDOS-vs-E,main:fig:spectral-gap:cDOS-vs-N}.
The extrapolation of $\mathcal{N}(E,N)$ simultaneously takes into account the effects due to non-Abelian Bloch states and finite sampling of Abelian Bloch states without being overly sensitive to the chosen energy resolution (due to the integration).

Using a maximum-likelihood algorithm (see SM~\cite{SM}, Sec.~V~C), we fit $\mathcal{N}(E,N)$ as a function of $1/N$ (for $N>1$) in the vicinity of $E_g$ using the linearized model
\begin{equation}
    \mathcal{N}(E,N) = \mathcal{N}_0(E) + \frac{s(E)}{N},
    \label{main:eq:cDOS-vs-N_fit-model}
\end{equation}
and weights $N$ to account for the larger weight of non-Abelian states in larger supercells.
Some examples of fits are shown in \cref{main:fig:spectral-gap:cDOS-vs-N}.
From each fit, we extract $\mathcal{N}_0$ and the slope $s$ together with their $95\%$ confidence intervals, see \cref{main:fig:spectral-gap:cDOS-vs-E,main:fig:spectral-gap:slope}, respectively.
The extrapolated integrated DOS $\mathcal{N}_0(E)$ is expected to change from $0$ to a positive value at $E_g$.
On the other hand, the slope $s(E)$ is expected to change its sign from positive below $E_g$ to negative above.
From both datasets, we obtain estimates of $E_g$ including uncertainties due to the confidence intervals.
Results with combined uncertainties are shown in \cref{main:fig:spectral-gap} as a function of~$J_z$~and~$K$ (see also SM~\cite{SM}, Sec.~VII~B).

For sufficiently large $\Delta E$, this reduces to finding the intersection of $\mathcal{N}(E)$ for the largest two supercells.
We employed this computationally more efficient approach to obtain the full phase diagrams in \cref{main:fig:phase-diagram_K=0:triangle,main:fig:phase-diagram_K=0.1}.

\begin{figure}[b]
    \centering
    \includegraphics{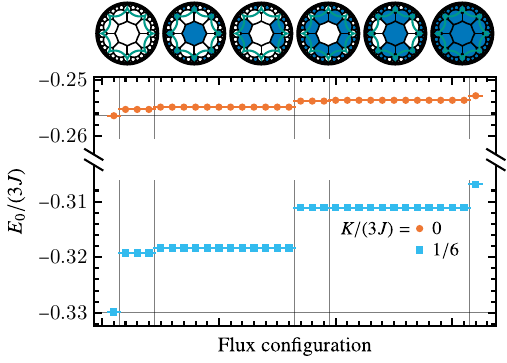}
    \caption{
    Ground-state energy for the $32$ translation-invariant flux configurations with zero net flux per unit cell for $J_x=J_y=J_z=J$ and different values of $K$ (see legend).
    Error bars indicate $95\%$ confidence intervals.
    The flux configurations fall into the six equivalence classes (separated by the vertical gray lines) shown at the top with white (blue shaded) octagons denoting plaquettes with $\pi$ ($0$) flux.
    In both cases, the homogeneous $\pi$-flux configuration has lowest energy.
    }
    \label{main:fig:energy-flux-configs}
\end{figure}

\appsection{Ground-state flux sector}\label{App:GS-flux-sector}%
To verify that the ground-state flux sector is indeed the homogeneous $\pi$-flux configuration both for $K=0$ and $K\neq 0$, we study the ground-state energy $E_0$ in different flux sectors.
For computational reasons, we restrict the analysis to translation-invariant flux configurations with no net flux per unit cell.
Since a primitive cell has six faces and the number of plaquettes with $W_P=-1$ has to be even, there are $2^{6-1}=32$ such configurations, reducing to six equivalence classes by symmetry.
The results are shown in \cref{main:fig:energy-flux-configs} for $K=0$ and $K/(3J)=1/6$ (see also SM~\cite{SM}, Sec.~VII~A for extended figures).

\appsection{Chiral edge states}\label{App:chiral-edge-states}%
To extract the dispersion $E(\ell)$ of the edge states in the chiral phase, plotted in \cref{main:fig:edge-dispersion}, we assign to each eigenstate $\ket{\psi_n}$ of the Majorana Hamiltonian on a circular flake two quantities: (\emph{i}) degree of localization near the edge $p_{n,\textrm{edge}}$, and (\emph{ii})~ angular momentum $\ell_n$.
We define the first as $p_{n,\textrm{edge}}\,{=}\, \sum_{j \in \textrm{edge}} \abs{\psi_n(j)}^2$, where \enquote{$j\,{\in}\,\textrm{edge}$} indicates sites located within the outer $10\,\%$ of the hyperbolic distance to the boundary. 
The computed values exhibit a sharp jump in $p_\textrm{edge}$ at energies $ E/(3J) \approx \pm 0.6$, see~\cref{main:fig:edge-dispersion}.

Due to the discrete rotation symmetry, $\ell_n$ is defined only modulo~$8$.
However, since low-energy states have a wavelength much larger than the lattice spacing, we anticipate the emergence of an unbounded $\ell$ near $E\,{=}\,0$.
In the continuum limit, the eigenvalue $\ell$ is associated with $\e^{\i \ell \varphi}$, where the phase of the wave function grows with angular coordinate $\varphi$. 
Therefore, $c_{n,\ell} = \abs{\sum_{j\in\textrm{sites}}\e^{-2\i \ell \varphi(j)}\psi_n^2(j)} \in [0,1]$ estimates the probability that $\ket{\psi_n}$ carries angular momentum $\ell \in \mathbb{Z}/2$. 
The computed values $c_{n,\ell}$ are dominated by a single branch, $\ell = \ell_n$. 
The branch has approximately linear dispersion $E(\ell)$, with $\ell\,{\in}\,\mathbb{Z}$ ($\ell\,{\in}\,\mathbb{Z}\,{+}\,\tfrac{1}{2}$) in the absence (presence) of a vortex at the center of the disk as shown in \cref{main:fig:edge-dispersion}.
For technical details, see SM~\cite{SM}, Sec.~X.

\clearpage

\setcounter{page}{1}
\setcounter{equation}{0}
\setcounter{section}{0}
\setcounter{figure}{0}

\renewcommand{\theequation}{S\arabic{equation}}
\renewcommand{\thefigure}{S\arabic{figure}}
\renewcommand{\theHfigure}{S\arabic{figure}}
\renewcommand{\thetable}{S\arabic{table}}

\makeatletter 
    
\renewcommand\onecolumngrid{%
    \do@columngrid{one}{\@ne}%
    \def\set@footnotewidth{\onecolumngrid}%
    \def\footnoterule{\kern-6pt\hrule width 1.5in\kern6pt}%
}

\renewcommand\twocolumngrid{%
    \def\footnoterule{
    \dimen@\skip\footins\divide\dimen@\thr@@
    \kern-\dimen@\hrule width.5in\kern\dimen@}
    \do@columngrid{mlt}{\tw@}
}%

\makeatother

\onecolumngrid
\begin{center}
\phantomsection
\large\textbf{Supplemental Material}
\addcontentsline{toc}{chapter}{Supplemental Material}
\end{center}

This document contains the proofs, derivations, and technical/implementation details on the applied methods described below.
\vspace{-1.5cm}

\setcounter{tocdepth}{-1}
\tableofcontents
\setcounter{tocdepth}{2}


\newpage

\labeledsection{Symmetric three-edge coloring on hyperbolic $\{2m,3\}$ lattices}{%
Definitions of the mathematical objects introduced in \cref{App:coloring}; constructive proof of the existence of a symmetry-compatible three-edge coloring for infinite hyperbolic $\{p,3\}$ lattices with even $p=2m$ (\cref{SM:3edge-inf}); compatibility with translation symmetry (\cref{SM:3edge-PBC}).%
}\label{SM:symmetric-3-coloring}

In this section, we give technical details on the systematic procedure for constructing a symmetry-compatible three-edge coloring for hyperbolic $\{p,3\}$ lattices with even $p=2m$ introduced in \cref{App:coloring}.
We define all the relevant mathematical objects and constructively prove the existence of the symmetric coloring under the given conditions.

To define an exactly solvable Kitaev model on a graph, one must first ensure the graph admits a three-edge coloring, that is, a coloring of edges with at most three colors such that no two edges with a common vertex have the same color. 
Infinite $\{2m,3\}$ lattices are bipartite graphs, for which K\H{o}nig's theorem ensures they are three-edge colorable~\cite{Diestel:2017}. 
However, an arbitrary three-edge coloring does not generically exhibit the infinitely many reflection symmetries that are necessary to unambiguously determine the $\mathbb{Z}_2$ flux in each plaquette via Lieb's lemma (see main text). Here, we utilize the space-group symmetries of the $\{2m,3\}$ lattice to construct a three-edge coloring\footnote{Hyperbolic triangle group symmetries were also used to construct three-colorings for hyperbolic tilings in Ref.~\onlinecite{Higgott:2023}.} that exhibits infinitely many reflection symmetries (\cref{SM:3edge-inf}), allowing for a complete determination of the ground-state flux in every plaquette. We also determine a sufficient condition on the translation group $\Gamma$ for the corresponding Bravais unit cell to respect those same symmetries (\cref{SM:3edge-PBC}).

\subsection{Three-edge coloring from space-group symmetries}\label{SM:3edge-inf}

\begin{figure}[t]
    \centering
    \includegraphics{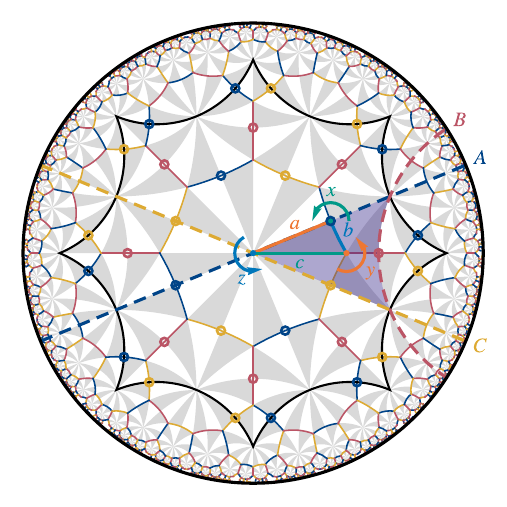}
    \caption{Symmetric three-edge coloring on the hyperbolic $\{2m,3\}$ lattice (depicted here for $m=4$). The space group $\Delta(2,3,2m)$ of this lattice is generated by the reflections $a,b,c$ in the sides of a right triangle, which generate rotations $x=ab$, $y=bc$, $z=ca$ around the corners of this triangle as well as all other space-group operations. The edges of the $\{2m,3\}$ lattice (open circles) are in one-to-one correspondence with the edges of the dual $\{3,2m\}$ lattice, composed of equilateral triangles. Assigning a distinct color to the three sides of a reference equilateral triangle (shaded region), and applying all possible compositions of the reflections $A,B,C$ (dashed lines) in the sides of this triangle (i.e., all operations in the subgroup $\Delta(m,m,m)\nsubg\Delta(2,3,2m)$ generated by $A,B,C$), we generate a three-edge coloring of the entire $\{2m,3\}$ lattice which is invariant under those bond-cutting reflections. A Bravais unit cell (black octagon) preserves all the symmetries of the coloring provided that the corresponding translation group $\Gamma$ is a normal subgroup in both $\Delta(2,3,2m)$ and $\Delta(m,m,m)$. }
    \label{fig:3-coloring}
\end{figure}

The space group $G$ of the infinite $\{2m,3\}$ lattice is the hyperbolic triangle group $\Delta(2,3,2m)$, an infinite discrete group defined by the presentation
\begin{align}
    G\equiv\Delta(2,3,2m)=\langle a,b,c|a^2,b^2,c^2,(ab)^2,(bc)^3,(ca)^{2m}\rangle.
    \label{eq:triangle-group}
\end{align}
In general, one defines the triangle group $\Delta(r,q,p)=\langle a,b,c|a^2,b^2,c^2,(ab)^r,(bc)^q,(ca)^p\rangle$. 
For $\frac{1}{r}+\frac{1}{q}+\frac{1}{p}<1$, as is the case here, this is the symmetry group of a regular tiling of the Poincar\'e disk $\mathbb{D}$ by hyperbolic triangles with interior angles $\frac{\pi}{r},\frac{\pi}{q},\frac{\pi}{p}$. 
The generators $a,b,c$ are reflections with respect to the sides of a reference triangle (highlighted with thick solid edges near the center of \cref{fig:3-coloring}), which is repeated under the action of $\Delta(2,3,2m)$ to tessellate the entire Poincar\'e disk. 
The composite operations $x\equiv ab$, $y\equiv bc$, $z\equiv ca$ are respectively counterclockwise rotations by $\frac{2\pi}{2}=\pi$, $\frac{2\pi}{3}$, $\frac{2\pi}{2m}=\frac{\pi}{m}$ about the vertices (correspondingly labeled $v_x,v_y,v_z$) of the reference triangle.

To construct a three-edge coloring, we first observe that the edges of the $\{2m,3\}$ lattice (open circles in \cref{fig:3-coloring}) are in one-to-one correspondence with all images of the vertex $v_x\in\mathbb{D}$ under the action of $G$, which we simply call ``$x$-vertices''. 
Formally, the set of all $x$-vertices is the orbit $v_x\cdot G$ where we define a right action of $G<\mathrm{PSU}(1,1)$ on $\mathbb{D}$ as:
\begin{align}
    z\cdot g\equiv\frac{\alpha^* z-\beta}{-\beta^*z+\alpha},\hspace{5mm}g\in G,\hspace{5mm}z\in\mathbb{D},
\end{align}
which obeys $(z\cdot g)\cdot g'=z\cdot(gg')$. Here we have used the representation of an element $g\in\Delta(2,3,2m)$ as an $\mathrm{SU}(1,1)$ matrix $\left(\begin{smallmatrix}\alpha & \beta\\ \beta^* & \alpha^*\end{smallmatrix}\right)$ with $|\alpha|^2-|\beta|^2=1$. 
According to the orbit-stabilizer theorem, the orbit $v_x\cdot G$ is in one-to-one correspondence with the right coset space $H\backslash G$, where $H<G$ is the stabilizer of $v_x$ in $G$:
\begin{align}\label{H-Gvx}
    H\equiv G_{v_x}=\{h\in G:v_x\cdot h=v_x\}.
\end{align}
Geometrically, $H$ is the set of elements of the space group $G$ that leave $v_x$ invariant. 
Inspecting \cref{fig:3-coloring}, we see that $H$ is generated by the reflections $a$ and $b$ that pass through $v_x$, and thus inherits from $G$ the following presentation:
\begin{align}\label{stabilizer}
    H=\langle a,b|a^2,b^2,(ab)^2\rangle.
\end{align}
This is a Coxeter group isomorphic to the dihedral group $D_2\cong\mathbb{Z}_2\times\mathbb{Z}_2$ (Klein's Vierergruppe) of order $|H|=4$.

So far, we have found that the set of all edges of the $\{2m,3\}$ lattice is in one-to-one correspondence with the elements of the coset space $H\backslash G$. To construct a three-edge coloring, we wish to partition $H\backslash G$ into three colors, i.e., distinct equivalence classes of $x$-vertices such that any three incident edges of the $\{2m,3\}$ lattice have different colors. 
For this purpose, we first construct a subgroup $K=\langle A,B,C\rangle<G$ as the group generated by the group elements $A,B,C\in G$ defined as
\begin{align}\label{defABC}
    A\equiv a,\hspace{5mm}
    B\equiv yay^{-1}=bcacb,\hspace{5mm}
    C\equiv y^{-1}ay=cbabc.
\end{align}
Geometrically, $A,B,C$ are reflections with respect to the sides of a large equilateral reference triangle with interior angles $\frac{\pi}{m}$, which contains six small triangles of the $(2,3,2m)$ tessellation (dark shaded region in \cref{fig:3-coloring}). 
One can show that $A,B,C$ obey the defining relations of the hyperbolic triangle group $\Delta(m,m,m)$, which tiles the Poincar\'e disk with these equilateral triangles. Thus, we conclude that $K$ is in fact isomorphic to $\Delta(m,m,m)$: 
\begin{align}\label{K}
    K\cong\Delta(m,m,m)=\langle A,B,C|A^2,B^2,C^2,(AB)^m,(BC)^m,(CA)^m\rangle.
\end{align}
The equilateral triangles correspond in fact to faces of the dual $\{3,2m\}$ lattice, and the elements of $K$ are reflections with respect to the edges of this dual lattice. Each of those edges intersects a single $x$-vertex $v\in v_x\cdot G$. Under this correspondence, the three sides of any equilateral triangle of the dual lattice encode a set of three edges of the $\{2m,3\}$ lattice (or $x$-vertices) that meet at a single point. Thus, the problem of finding a three-edge coloring of the $\{2m,3\}$ lattice maps onto the problem of coloring the edges of the dual $\{3,2m\}$ lattice such that the three sides of each equilateral triangle have different colors. A natural solution offers itself: color first the edges of a reference equilateral triangle, and then color the edges of all remaining equilateral triangles by applying all elements of $K$. Thus, algebraically, two $x$-vertices $v,v'\in v_x\cdot G$ belong to the same edge color (denoted $v\sim v'$) if and only if they are related by the right action of $K$, i.e., $v'=v\cdot k$ with $k\in K$. Writing $v'=v_x\cdot(h'g')$ and $v=v_x\cdot(hg)$ with $h,h'\in H$ and $g,g'\in G$, we see that $v\sim v'$ implies $v_x\cdot(h'g')=v_x\cdot(hgk)$ and thus $h'g'=\tilde{h}hgk$ where $\tilde{h}\in H$. Redefining $h^{\prime-1}\tilde{h}h\mapsto h\in H$, this defines an equivalence relation on $G$, namely,
\begin{align}\label{DoubleCosetEquiv}
g\sim g'\text{ iff there exists }h\in H\text{ and }k\in K\text{ such that }hgk=g'.
\end{align}
If $H,K$ are subgroups of a group $G$, the equivalence class $HgK$ of $g$ under equivalence relation (\ref{DoubleCosetEquiv}) is called the $(H,K)$-double coset of $g\in G$, and the set of all $(H,K)$-double cosets is denoted $H\backslash G/K$ (``$G$ mod $K$ mod $H$'')~\cite{Robinson:1996}:
\begin{align}
    H\backslash G/K=\{HK,Hg_2K,\ldots,Hg_NK\},
    \label{eq:double-coset}
\end{align}
where $g_1\equiv 1,g_2,\ldots,g_N\in G$ are double coset representatives, asssuming here a finite number $N\equiv|H\backslash G/K|$ of double cosets. 
As with ordinary cosets, double cosets form a disjoint decomposition of $G$. 
Thus, we have found that the set of distinct edge colors is in one-to-one correspondence with the elements of the double coset space $H\backslash G/K$. 
According to our earlier geometric argument, we expect there should be $N=3$ colors, since $x$-vertices of different colors correspond to inequivalent sides of the $(m,m,m)$ equilateral triangles, i.e., sides that are not related by the action of $K\cong\Delta(m,m,m)$. 
We also see that each site of the $\{2m,3\}$ lattice lies at the center of a unique equilateral $(m,m,m)$ triangle, thus the three edges incident on this site are necessarily assigned different colors by our construction, satisfying the conditions for three-edge coloring.

For consistency, it thus remains to be shown that $|H\backslash G/K|=3$. To show this, we use the formula~\cite{HallBook}
\begin{align}\label{DoubleCosetFormula}
|G:K|=\sum_{HgK\in H\backslash G/K}|H:H\cap gKg^{-1}|,
\end{align}
where $|G:K|$ denotes the index of $K$ in $G$ and the sum is over all double cosets in $H\backslash G/K$. 
We make use of two mathematical facts.
First, $K\nsubg G$; therefore, $gKg^{-1}=K$ and all summands in \cref{DoubleCosetFormula} are equal, implying $|G:K|=|H\backslash G/K||H:H\cap K|$. 
Second, $|G:K|=6$, which corresponds to each triangle of the dual lattice consisting of six white or gray Schwarz triangles of the original lattice (\cref{fig:3-coloring}). 
Combining these two facts, we have
\begin{align}\label{HGK}
    |H\backslash G/K|=\frac{6}{|H:H\cap K|}.
\end{align}
We now consider the intersection $H\cap K$, which is the subset of elements of the stabilizer (\ref{stabilizer}) that are also in the triangle group $\Delta(m,m,m)$. 
Clearly, $a=A\in K$ [see \cref{K}] and thus the intersection $H\cap K$ contains at least the $\mathbb{Z}_2$ subgroup $\{1,a\}< H$. 
Regarding the other elements $\{b,ab\}\subset H$, they clearly cannot be in $K$.
First, the element $b$ is a reflection that passes through the equilateral reference triangle and swaps inequivalent ($B$ and $C$) sides of this triangle (see \cref{fig:3-coloring}), an operation that is not in $\Delta(m,m,m)$. It follows that also $ab\notin K$, as the converse would contradict $b=a^{-1}(ab)\notin K$. Thus, we find that $H\cap K=\{1,a\}\cong\mathbb{Z}_2$. Using Lagrange's theorem~\cite{Robinson:1996}, we have $|H:H\cap K|=|H|/|H\cap K|=2$, thus \cref{HGK} implies:
\begin{align}
    |H\backslash G/K|=3,
\end{align}
predicting an edge coloring of the infinite $\{2m,3\}$ lattice with three colors, as expected. Thus, for any edge associated to an $x$-vertex $v_x\cdot g$ with $g\in G$, we can determine its color algebraically by checking to which double coset in $H\backslash G/K$ the element $g$ belongs, which is easily done in \gap~\cite{GAP4}.

\subsection{Symmetric Bravais unit cells}\label{SM:3edge-PBC}

For reciprocal-space calculations using the supercell method~\cite{Lenggenhager:2023} or real-space calculations using finite clusters with periodic boundary conditions (PBC)~\cite{Maciejko:2022}, we must construct a Bravais unit cell and the corresponding translation group $\Gamma$, which is a torsion-free subgroup of the space group $\Delta(2,3,2m)$. In this section, $\Gamma$ denotes either the primitive translation group~\cite{Chen:2023b} or a supercell translation group~\cite{Lenggenhager:2023}. In particular, for verifications of the ground-state flux sector using the supercell method (\cref{App:GS-flux-sector}), we would like to ensure that the chosen Bravais unit cell preserves the reflection symmetries mandated by Lieb's lemma. In this section, we show this is ensured if the corresponding translation group $\Gamma$ is a normal subgroup of both $G=\Delta(2,3,2m)$ and $K=\Delta(m,m,m)$.

We first observe that on the Bravais unit cell associated to $\Gamma$, the infinite set $v_x\cdot G$ of $x$-vertices is replaced by the finite set of $x$-vertices contained inside the unit cell. Mathematically, the reference vertex $v_x$ should be replaced by the orbit $\tilde{v}_x\equiv v_x\cdot\Gamma$, which is the set of all $\Gamma$-translates of $v_x$. For any $g\in G$ and $\gamma\in\Gamma$, the elements $g$ and $\gamma g$ have the same right action on $\tilde{v}_x$. Thus, the group that acts most naturally on $\tilde{v}_x$ is not $G$ itself but rather the factor group $G/\Gamma$, i.e., the set of cosets $\Gamma g$, which is a finite group of order $|G:\Gamma|$. (If $\Gamma$ is chosen as the torsion-free normal subgroup of smallest possible index in $G$, then it is the primitive translation group of the $\{2m,3\}$ lattice~\cite{Chen:2023b}, and $G/\Gamma$ is the point group of that lattice.) For $G/\Gamma$ to be a group, we require $\Gamma$ to be normal in $G$. Note that since $\Gamma\nsubg G$, right and left cosets are equivalent and $\Gamma\backslash G=G/\Gamma$. Thus, the set of $x$-vertices contained inside the unit cell is the orbit $\tilde{v}_x\cdot(G/\Gamma)$.

Next, we wish to describe the orbit $\tilde{v}_x\cdot(G/\Gamma)$ in purely algebraic terms using the orbit-stabilizer theorem. To do so, we must find the stabilizer of $\tilde{v}_x$ in $G/\Gamma$. We first define the product~\cite{Robinson:1996} of $\Gamma\nsubg G$ and $H<G$ [\cref{H-Gvx}] as
\begin{align}
    \Gamma H=\{\gamma h:\gamma\in\Gamma,h\in H\}=\{\Gamma h:h\in H\}.
\end{align}
Assuming $\gamma_1,\gamma_2\in\Gamma$ and $h_1,h_2\in H$, since $\Gamma$ is normal in $G$, we have $(\gamma_1h_1)(\gamma_2h_2)=\gamma_1h_1\gamma_2h_1^{-1}h_1h_2=\gamma_1\gamma_2'h_1h_2$ with $\gamma_2'\in\Gamma$, thus $\Gamma H<G$. This infinite subgroup of $G$ leaves $\tilde{v}_x$ invariant:
\begin{align}
    \tilde{v}_x\cdot(\gamma h)&=v_x\cdot(\Gamma\gamma h)\nonumber\\
    &=v_x\cdot(h\Gamma)\nonumber\\
    &=\tilde{v}_x,
\end{align}
where we have used the normality of $\Gamma$ and the fact that $H$ stabilizes $v_x$. However, we want the stabilizer of $\tilde{v}_x$ in the finite group $G/\Gamma$, which is the factor group $\Gamma H/\Gamma$, i.e., the set of (right) cosets of $\Gamma$ in $\Gamma H$:
\begin{align}
    \Gamma H/\Gamma&=\{\Gamma,\Gamma\gamma_2 h_2,\ldots,\Gamma\gamma_M h_M\}\nonumber\\
    &=\{\Gamma,\Gamma h_2,\ldots,\Gamma h_M\},
\end{align}
assuming $M$ cosets. This can be viewed as another generalization of coset space (i.e., one cannot directly write $H/\Gamma$ since $\Gamma$ is not a subgroup of $H$). That $\Gamma H/\Gamma$ is a group in this case follows from the fact that $\Gamma\nsubg\Gamma H$. Indeed, first, $\Gamma\subset\Gamma H$ since $\Gamma=\Gamma\cdot 1$ with $1\in H$. Second, $\Gamma<\Gamma H$ since $\Gamma$ is a group. Third, $\Gamma$ is normal in $\Gamma H$ since it is normal in $G>\Gamma H$. Using the second isomorphism theorem~\cite{Robinson:1996}, we have
\begin{align}
    \Gamma H/\Gamma\cong H/(\Gamma\cap H)\cong H\cong\mathbb{Z}_2\times\mathbb{Z}_2,
\end{align}
using the fact that $\Gamma\cap H=\{1\}$ since $\Gamma$ is torsion-free and $H$ contains only elements of finite order [\cref{stabilizer}]. Finally, $\Gamma H/\Gamma$ is a subgroup of $G/\Gamma$: it can be viewed as a set of (right) cosets of $\Gamma$ in $G$, and it is a group. Thus, the stabilizer of $\tilde{v}_x$ in $G/\Gamma$ is $\Gamma H/\Gamma$, which is a finite group isomorphic to $H$.

Having identified $\tilde{G}\equiv G/\Gamma$ and $\tilde{H}\equiv\Gamma H/\Gamma$ as the unit cell equivalents of the groups $G$ and $H$ for the infinite lattice, we further define $\tilde{K}\equiv K/\Gamma$ as the third object required to construct a symmetric three-edge coloring on the unit cell. For $\tilde{K}$ to be a group, $\Gamma$ must be a normal subgroup of $K=\Delta(m,m,m)$. Assuming this condition, by the third isomorphism theorem~\cite{Robinson:1996}, we have $\tilde{K}\nsubg\tilde{G}$ and also $\tilde{G}/\tilde{K}\cong G/K$ which implies $|\tilde{G}:\tilde{K}|=|G:K|=6$. Thus, applying \cref{DoubleCosetFormula} to the groups $\tilde{G}$, $\tilde{H}$, and $\tilde{K}$, we have
\begin{align}
    |\tilde{H}\backslash\tilde{G}/\tilde{K}|=\frac{6}{|\tilde{H}:\tilde{H}\cap\tilde{K}|}.
\end{align}
Next, we compute $\tilde{H}\cap\tilde{K}$. Using the correspondence theorem~\cite{Humphreys}, since $\Gamma H,K<G$ and $\Gamma\nsubg\Gamma H,K$, we have
\begin{align}
    \tilde{H}\cap\tilde{K}=(\Gamma H/\Gamma)\cap(K/\Gamma)=(\Gamma H\cap K)/\Gamma.
\end{align}
Finally, we invoke Dedekind's modular law~\cite{Robinson:1996}, which implies that $\Gamma H\cap K=\Gamma(H\cap K)$. Using once more the second isomorphism theorem, we have
\begin{align}
    \tilde{H}\cap\tilde{K}=\Gamma(H\cap K)/\Gamma\cong(H\cap K)/(\Gamma\cap H\cap K)=(H\cap K)/\{1\}=H\cap K\cong\mathbb{Z}_2.
\end{align}
Using again Lagrange's theorem, we have $|\tilde{H}:\tilde{H}\cap\tilde{K}|=|\tilde{H}|/|\tilde{H}\cap\tilde{K}|=|H|/|H\cap K|=2$, thus
\begin{align}
    |\tilde{H}\backslash\tilde{G}/\tilde{K}|=3,
\end{align}
giving us again a (symmetric) three-edge coloring, but this time on a finite unit cell.

Under the condition $\Gamma\nsubg G,K$, the unit cell is manifestly compatible with the reflection symmetries relevant for Lieb's lemma, encoded in $K$. Indeed, on such a symmetric unit cell, the space group $G$ descends to the quotient $\tilde{G}=G/\Gamma$, and the reflection symmetries $K\nsubg G$ descend to the quotient $\tilde{K}=K/\Gamma\nsubg\tilde{G}$. An example of symmetric unit cell is provided by the $\{8,8\}$ unit cell on the $\{8,3\}$ lattice (black octagon in \cref{fig:3-coloring}). In general, given any finite-index translation group $\Gamma\nsubg G$ not necessarily normal in $K$, one can always construct a symmetry-compatible translation group via $\Gamma'\equiv\Gamma\cap K$. The resulting $\Gamma'$ also has finite index in $G$ (and $K$), implying that the corresponding unit cell contains finitely many sites.

\labeledsection{Reflection symmetry and the three-spin term}{%
Clarification of notation used in \cref{main:eq:Hamiltonian} and proof that the reflection symmetries of the coloring are preserved.%
}\label{SM:K-term}

In this section, we clarify our notation for the three-spin term in \cref{main:eq:Hamiltonian} and demonstrate that our choice of coefficient ensures that the reflection-symmetries of the coloring are preserved.

\begin{figure}[t]
    \centering
    \includegraphics{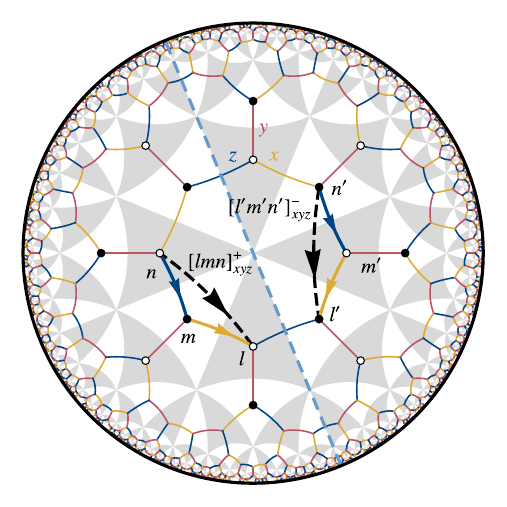}
    \caption{
        Illustration of the triplets $[lmn]^+_{\alpha\beta\gamma}$ involved in the definition of the $K$-term on the $\{8,3\}$ lattice with symmetric coloring.
        A three-spin interaction is defined by a triplet $(n,m,l)$ of sites, where the superscript \enquote{$+$} in the label $[lmn]^+_{xyz}$ denotes counterclockwise (positive) orientation around the octagonal plaquette the path $n\to m\to l$ is part of.
        This is illustrated with the sites $n,m,l$ to the left of the blue dashed mirror line.
        In the Majorana representation, this becomes a next-nearest-neighbor hopping (dashed black arrow) from $n$ to $l$, inheriting the counterclockwise orientation with respect to going around the center of the plaquette.
        The corresponding bond operators $\hat{u}_{mn}$ ($z$-bond from $n$ to $m$) and $\hat{u}_{lm}$ ($x$-bond from $m$ to $l$) whose eigenvalues determine the hopping amplitude are indicated as bold solid arrows.
        Note that the reflected triplet is indeed oriented oppositely, i.e., clockwise, justifying the label $[l'm'n']^-_{xyz}$.
    }
    \label{fig:K-term-illustration}
\end{figure}

The three-spin term in \cref{main:eq:Hamiltonian} in the Letter contains the totally antisymmetric tensor $\varepsilon_{\alpha\beta\gamma}$ (which satisfies $\varepsilon_{xyz}\:{=}\:{+}1$):
\begin{equation}
    \hat{\Ham}_K = - K\sum_{[lmn]_{\alpha\beta\gamma}^+}\varepsilon_{\alpha\beta\gamma}\hat{\sigma}_l^\alpha\hat{\sigma}_m^\beta\hat{\sigma}_n^\gamma.
    \label{eq:spin-K-term}
\end{equation}
The two sublattices (filled/empty disks in \cref{main:fig:coloring} in the Letter and in \cref{fig:K-term-illustration}) of the infinite $\{p,3\}$ hyperbolic lattice are related by reflection symmetry.
This manifests as filled (empty) sites always lying in white (gray) Schwarz triangles of $\Delta(p/2,p/2,p/2)$.
Our symmetric coloring is, by construction, compatible with $\Delta(p/2,p/2,p/2)$ which implies that the colors of the bonds around sites from different sublattices are related to each other by an odd number of reflections and are thus oriented oppositely, see \cref{fig:3-coloring}.

We now argue that \cref{eq:spin-K-term} is reflection-symmetric.
Any reflection operator $\hat{\vartheta}$ acts on the spin operators as $\hat{\vartheta}\hat{\sigma}_j^\alpha\hat{\vartheta}^{-1}=-\hat{\sigma}_{j'}^\alpha$, where $j'$ is the reflected partner of $j$ and we use the fact that the coloring is invariant under reflections.
Considering a single term in $\hat{\Ham}_K$, associated with the triplet $[lmn]_{\alpha\beta\gamma}$, we first note that the reflected triplet has opposite orientation: $[l'm'n']_{\alpha\beta\gamma}^-$ as can be seen in \cref{fig:K-term-illustration}.
Then,
\begin{align}
    \hat{\vartheta}\varepsilon_{\alpha\beta\gamma}\hat{\sigma}_l^\alpha\hat{\sigma}_m^\beta\hat{\sigma}_n^\gamma\hat{\vartheta}^{-1}
    &= -\varepsilon_{\alpha\beta\gamma}\hat{\sigma}_{l'}^\alpha\hat{\sigma}_{m'}^\beta\hat{\sigma}_{n'}^\gamma,\nonumber\\
    &= \varepsilon_{\gamma\beta\alpha}\hat{\sigma}_{n'}^\gamma\hat{\sigma}_{m'}^\beta\hat{\sigma}_{l'}^\alpha,
\end{align}
where we used the antisymmetry of $\varepsilon_{\alpha\beta\gamma}$, the fact that the spin operators on different sites commute, and rearranged the operators such that they correspond to the triplet $[n'm'l']_{\gamma\beta\alpha}^+$ oriented as it appears in the sum in \cref{eq:spin-K-term}.
Thus, we arrive at
\begin{equation}
    \hat{\vartheta}\hat{\Ham}_K\hat{\vartheta}^{-1} = \hat{\Ham}_K.
\end{equation}

For the coloring used in Ref.~\onlinecite{Kitaev:2006}, the factor $\varepsilon_{\alpha\beta\gamma}$ is unnecessary, because for any positively oriented triplet $[lmn]_{\alpha\beta\gamma}^+$, $\varepsilon_{\alpha\beta\gamma}$ always takes the same value.
We also note that, although reminiscent of the scalar spin chiralities $\hat{\chi}_{lmn}\equiv\hat{\boldsymbol{\sigma}}_l\cdot(\hat{\boldsymbol{\sigma}}_m\times\hat{\boldsymbol{\sigma}}_n)$~\cite{Wen:1989}, the operators appearing in \cref{eq:spin-K-term} are not the same as $\hat{\chi}_{lmn}$ because there is no free sum over $\alpha,\beta,\gamma$; rather, the spin components are fixed by the colors $\alpha,\beta,\gamma$ of the bonds incident on $l,m,n$. Therefore, just like the two-spin term in the Hamiltonian, the three-spin term (\ref{eq:spin-K-term}) completely breaks the $\mathrm{SU}(2)$ spin rotation symmetry.

\labeledsection{Majorana representation of the Hamiltonian}{%
Derivation of the Majorana Hamiltonian, \cref{main:eq:Majorana-Hamiltonian}, starting from the spin Hamiltonian given in \cref{main:eq:Hamiltonian}.%
}\label{SM:H_Majorana}

In this section, we derive the Majorana representation, i.e., \cref{main:eq:Majorana-Hamiltonian} in the Letter, for the Hamiltonian given in \cref{main:eq:Hamiltonian}:
\begin{equation}
    \hat{\Ham} = -\sum_{\nn{j,k}_\alpha}J_\alpha\hat{\sigma}_j^\alpha\hat{\sigma}_k^\alpha - K\sum_{[lmn]_{\alpha\beta\gamma}^+}\varepsilon_{\alpha\beta\gamma}\hat{\sigma}_l^\alpha\hat{\sigma}_m^\beta\hat{\sigma}_n^\gamma.
    \label{eq:spin-Hamiltonian}
\end{equation}
We follow Ref.~\onlinecite{Kitaev:2006} and introduce the Majorana fermions $\hat{b}_j^\alpha$, $\alpha\in\{x,y,z\}$, and $\hat{c}_j$, at each site $j$, such that $\hat{\sigma}_j^\alpha = \i\hat{b}_j^\alpha\hat{c}_j$.
Further, for the nearest-neighbor bond $\nn{j,k}_\alpha$, we define the bond variable $\hat{u}_{jk}=\i\hat{b}_j^\alpha\hat{b}_k^\alpha$.
Recall that this extends the Hilbert space and comes with the gauge transformation operators $\hat{D}_j=\hat{b}_j^x\hat{b}_j^y\hat{b}_j^z\hat{c}_j$ that define the physical subspace as their common $+1$ eigenspace.
The first term in the Hamiltonian then becomes
\begin{align}
    \hat{\Ham}_J
    &= -\sum_{\nn{j,k}_\alpha}J_\alpha\left(\i\hat{b}_j^\alpha\hat{c}_j\right)\left(\i\hat{b}_k^\alpha\hat{c}_k\right)\nonumber\\
    &= \sum_{\nn{j,k}_\alpha}J_\alpha\left(\i\hat{b}_j^\alpha\hat{b}_k^\alpha\right)\i\hat{c}_j\hat{c}_k\nonumber\\
    &= \sum_{\nn{j,k}_\alpha}J_\alpha\hat{u}_{jk}\i\hat{c}_j\hat{c}_k,
\end{align}
and the second
\begin{align}
    \hat{\Ham}_K
    &= -K\sum_{[lmn]_{\alpha\beta\gamma}^+}\varepsilon_{\alpha\beta\gamma}
    \left(\i\hat{b}_l^\alpha\hat{c}_l\right)
    \left(\i\hat{b}_m^\beta\hat{c}_m\right)
    \left(\i\hat{b}_n^\gamma\hat{c}_n\right)\nonumber\\
    &= K\sum_{[lmn]_{\alpha\beta\gamma}^+}\varepsilon_{\alpha\beta\gamma}
    \left(\i\hat{b}_l^\alpha\right)
    \left(\i\hat{b}_m^\beta\hat{c}_m\right)
    \hat{b}_n^\gamma\i\hat{c}_l\hat{c}_n\nonumber\\
    &= K\sum_{[lmn]_{\alpha\beta\gamma}^+}\varepsilon_{\alpha\beta\gamma}
    \left(\i\hat{b}_l^\alpha\hat{b}_m^\alpha\hat{b}_m^\alpha\right)
    \hat{b}_m^\beta\hat{c}_m\hat{b}_m^\gamma
    \left(\i\hat{b}_m^\gamma\hat{b}_n^\gamma\right)
    \i\hat{c}_l\hat{c}_n\nonumber\\
    &= K\sum_{[lmn]_{\alpha\beta\gamma}^+}\varepsilon_{\alpha\beta\gamma}
    \hat{b}_m^\alpha\hat{b}_m^\beta\hat{b}_m^\gamma\hat{c}_m
    \hat{u}_{lm}\hat{u}_{mn}\i\hat{c}_l\hat{c}_n\nonumber\\
    &= K\sum_{[lmn]_{\alpha\beta\gamma}^+}
    \hat{b}_m^x\hat{b}_m^y\hat{b}_m^z\hat{c}_m
    \hat{u}_{lm}\hat{u}_{mn}\i\hat{c}_l\hat{c}_n\nonumber\\
    &= K\sum_{[lmn]_{\alpha\beta\gamma}^+}\hat{D}_m\hat{u}_{lm}\hat{u}_{mn}\i\hat{c}_l\hat{c}_n.
\end{align}
Since $D_m=+1$ in the physical subspace, we can drop it, such that we arrive at
\begin{equation}
    \hat{\Ham} = \sum_{\nn{j,k}_\alpha}J_\alpha\hat{u}_{jk}\i\hat{c}_j\hat{c}_k + K\sum_{[lmn]_{\alpha\beta\gamma}^+}\hat{u}_{lm}\hat{u}_{mn}\i\hat{c}_l\hat{c}_n.
    \label{eq:Hamiltonian_Majorana-rep:full}
\end{equation}
This can be written compactly as
\begin{equation}
    \hat{\Ham} = \frac{\i}{4}\sum_{j,k}\hat{A}_{jk}\hat{c}_j\hat{c}_k,
    \label{eq:Hamiltonian_Majorana-rep}
\end{equation}
with
\begin{equation}
    \hat{A}_{jk} =
    \begin{cases}
        2J_\alpha\hat{u}_{jk},&\text{for }\nn{j,k}_\alpha,\\
        \pm 2K\hat{u}_{jl}\hat{u}_{lk},&\text{for }[jlk]_{\alpha\beta\gamma}^\pm,
    \end{cases}
    \label{eq:A-operator}
\end{equation}
where the additional factor of $1/2$ compensates the double counting of each bond and triplet.
Note that $\hat{A}_{kj}=-\hat{A}_{jk}$, because $\hat{u}_{kj}=-\hat{u}_{jk}$ and $[klj]_{\gamma\beta\alpha}^\pm\equalhat[jlk]_{\alpha\beta\gamma}^\mp$, see also \cref{fig:K-term-illustration}.

As discussed in the Letter, in a fixed flux sector, we can replace the bond operators $\hat{u}_{jk}$ by their expectation values $u_{jk}$ and thus $\hat{A}_{jk}$ by the skew-symmetric real matrix $A_{jk}$, resulting in the Hamiltonian
\begin{equation}
    \hat{\Ham} = \frac{\i}{4}\sum_{j,k}A_{jk}\hat{c}_j\hat{c}_k,
    \label{eq:Majorana-Hamiltonian}
\end{equation}
acting only on the fermionic degrees of freedom in the specified flux sector.
Note that $A$ in \cref{eq:Majorana-Hamiltonian} is gauge dependent.
Projection to the physical (gauge-invariant) subspace is important for exact computations of quantities in the original spin model like the many-body ground-state energy and the $\mathbb{Z}_2$ vortex gap, especially in a finite system with periodic boundary conditions~\cite{Pedrocchi:2011}.
However, as in earlier work~\cite{Kitaev:2006,Kamfor:2010}, we focus here only on the fermionic excitation spectrum in the thermodynamic limit, which is correctly given by the (positive) eigenvalues of the Hermitian matrix $\BlochHam = \i A$ in a fixed flux sector.
We further make the simplifying assumption that the $K=0$ flux sector determined by Lieb's lemma persists for $K\neq 0$, which is correct for infinitesimal $K$ but ignores the possibility of flux phase transitions at finite $K$.
The assumption is consistent with our results in \cref{App:GS-flux-sector}, where we find that among all the translation-invariant flux configurations with zero net flux per primitive cell, the homogeneous $\pi$-flux configuration has lowest energy even for relatively large $K$.
We leave an analysis of the stability of the $\pi$-flux sector for future work.

We finally comment on time-reversal symmetry. The antiunitary time-reversal operator $\hat{\TRS}$ acts on the spin operators as $\hat{\TRS}\hat{\sigma}_j^\alpha\hat{\TRS}^{-1}=-\hat{\sigma}_j^\alpha$, thus the three-spin term in \cref{eq:spin-K-term} is odd under time reversal:
\begin{align}
    \hat{\TRS}\hat{\mathcal{H}}_K\hat{\TRS}^{-1}=-\hat{\mathcal{H}}_K.
\end{align}
In the Majorana representation, \cref{eq:Majorana-Hamiltonian}, it is useful to consider time reversal composed with a gauge transformation to compensate for the change in sign of $\hat{u}_{jk}=\i\hat{b}_j^\alpha\hat{b}_k^\alpha$ due to the antiunitary nature of $\hat{\TRS}$~\cite{Kitaev:2006}.
The corresponding symmetry $\hat{\TRS}'$, which leaves $\hat{u}_{jk}$ invariant but flips the sign of $\hat{c}_j$ on one of the sublattices, enforces couplings only between the two sublattices.
Correspondingly, $K\:{\neq}\:0$ turns on a coupling within each sublattice, analogous to the next-nearest neighbor hopping in the Haldane model~\cite{Haldane:1988}, suggesting a nonvanishing Chern number (see Sec.~\ref{SM:Chern}).

\labeledsection{Hyperbolic band theory for Majorana Hamiltonians}{%
Derivation of \cref{main:eq:Bloch-decomp,main:eq:Bloch-Hamiltonian_diagonalized} in \cref{App:HBT-Majorana}.%
}

In this section, we derive \cref{main:eq:Bloch-decomp,main:eq:Bloch-Hamiltonian_diagonalized}, extending the concept of hyperbolic band theory and hyperbolic Bloch Hamiltonians~\cite{Maciejko:2021,Maciejko:2022} including the supercell method~\cite{Lenggenhager:2023} to generic quadratic Majorana Hamiltonians of the form of \cref{eq:Majorana-Hamiltonian}, i.e.,
\begin{equation}
    \hat{\Ham} = \frac{\i}{4}\sum_{j,k}A_{jk}\hat{c}_j\hat{c}_k,
    \label{eq:generic-Majorana-Hamiltonian}
\end{equation}
where $A$ is a skew-symmetric real matrix, $A^\top=-A$, and $\{\hat{c}_j,\hat{c}_k\}=2\delta_{jk}$.

Following the ideas in Ref.~\onlinecite{Lenggenhager:2023}, we assume $\hat{\Ham}$ in \cref{eq:generic-Majorana-Hamiltonian} to be defined on a large (but finite) cluster of a hyperbolic lattice with periodic boundary conditions (PBC cluster).
At the very end, we formally take the limit~\cite{Lux:2022} of an infinite PBC cluster, recovering a description of the infinite lattice.
The large PBC cluster is defined~\cite{Maciejko:2022} by a normal subgroup $\Gamma_\mathrm{PBC}$ of the full translation group $\Gamma$ and of the triangle group $\Delta$.
Further, we assume a division of the PBC cluster into $N$ translated copies of a \mbox{(super-)cell}, characterized by $\Gamma_\mathrm{sc}\nsubg\Delta$, $\Gamma_\mathrm{sc}\nsubgeq\Gamma$, such that $\Gamma_\mathrm{PBC}\nsubg\Gamma_\mathrm{sc}$.
These copies are labeled by the corresponding translations from the supercell at the origin to the copy, modulo translations in $\Gamma_\mathrm{PBC}$.
Thus, the different copies of supercells are labeled by cosets $\eta\in G=\Gamma_\mathrm{sc}/\Gamma_\mathrm{PBC}$.
This, in turn, allows us to enumerate all sites in the PBC cluster as tuples $(\eta,j)$, where $j$ now labels the corresponding Wyckoff position in the supercell rather than the site in the PBC cluster; there are $M=\gindex{\Delta}{\Gamma_\mathrm{sc}}$ such Wyckoff positions.
Writing $\hat{c}_{(\eta,j)}=\hat{c}_{\eta,j}$, the Hamiltonian can be written as
\begin{equation}
    \hat{\Ham} = \frac{\i}{4}\sum_{\eta,\eta'\in G}\sum_{j,k=1}^M A_{(\eta',j),(\eta,k)}\hat{c}_{\eta',j}\hat{c}_{\eta,k}.
\end{equation}

By translation invariance, the skew-symmetric matrix satisfies $A_{(\eta',j),(\eta,k)}=A_{jk}(\eta'\eta^{-1})$, i.e., it depends only on the relative translation $\gamma=\eta'\eta^{-1}\in G$ between the two copies of the supercell and not their absolute position.
Recognizing that $\eta'=\gamma\eta$, we arrive at
\begin{equation}
    \hat{\Ham} = \frac{\i}{4}\sum_{\eta,\gamma\in G}\sum_{j,k} A_{jk}(\gamma)\hat{c}_{\gamma\eta,j}\hat{c}_{\eta,k}
    \label{eq:Majorana-Ham_translation-inv}
\end{equation}
where $j,k$ implicitly run from $1$ to $M$.

To rewrite \cref{eq:Majorana-Ham_translation-inv} in hyperbolic reciprocal space, we use the formalism recently developed in Ref.~\onlinecite{Canon:2024}, which defines a Fourier transform on PBC clusters.
Let $f(\gamma)$ be a function (potentially operator-valued) on $G$ and denote the irreducible representations of $G$ by $D^{(K)}$.
Recall that because $G$ is generally non-Abelian, $D^{(K)}$ can be matrix valued and we denote its dimension by $d_K$; $K$ takes the role of momentum in hyperbolic reciprocal space.
Then, the Fourier transform is defined as
\begin{subequations}
    \begin{align}
        f(\gamma) &= \frac{1}{\sqrt{N}}\sum_{K,\mu,\nu}\sqrt{d_K} f_{\mu\nu}^{(K)}\cconj{D_{\nu\mu}^{(K)}(\gamma)},\\
        f_{\mu\nu}^{(K)} &= \frac{1}{\sqrt{N}}\sum_{\gamma\in G}\sqrt{d_K}f(\gamma)D_{\nu\mu}^{(K)}(\gamma),
\end{align}
\end{subequations}
where $N=\abs{G}=\gindex{\Gamma_\mathrm{PBC}}{\Gamma_\mathrm{sc}}$, $K$ runs over all irreducible representations and $\mu,\nu$ run from $1$ to $d_K$.
Here, the unitary matrix $D_{\nu\mu}^{(K)}(\gamma)$ plays a role akin to the usual phase factor $e^{i\boldsymbol{K}\cdot\boldsymbol{R}}$ on Euclidean lattices.

This allows us to define the following reciprocal-space representation $\hat{a}_{\mu\nu,j}^{(K)}$ for the Majorana operators $\hat{c}_{\eta,j}$,
\begin{subequations}
    \begin{align}
    \hat{c}_{\eta,j} &= \sqrt{\frac{2}{N}}\sum_{K,\mu,\nu}\sqrt{d_K} \hat{a}_{\mu\nu,j}^{(K)}\cconj{D_{\nu\mu}^{(K)}(\eta)},\\
    \hat{a}_{\mu\nu,j}^{(K)} &= \frac{1}{\sqrt{2N}}\sum_{\eta\in G}\sqrt{d_K}\hat{c}_{\eta,j}D_{\nu\mu}^{(K)}(\eta),
\end{align}
\end{subequations}
which satisfy canonical anticommutation relations:
\begin{align}
    \acomm{\hat{a}_{\mu\nu,j}^{(K)}}{\adjo{\mbox{$\displaystyle\hat{a}_{\mu'\nu',k}^{(K')}$}}}
    &= \frac{d_K}{2N}\sum_{\eta,\eta'\in G}D_{\nu\mu}^{(K)}(\eta)\cconj{D_{\nu'\mu'}^{(K')}(\eta)}\acomm{\hat{c}_{\eta,j}}{\hat{c}_{\eta',k}}\nonumber\\
    &= \frac{d_K}{N}\delta_{jk}\sum_{\eta\in G}D_{\nu\mu}^{(K)}(\eta)\cconj{D_{\nu'\mu'}^{(K')}(\eta)}\nonumber\\
    &= \delta_{KK'}\delta_{\mu\mu'}\delta_{\nu\nu'}\delta_{jk},
\end{align}
where we used $\acomm{\hat{c}_{\eta,j}}{\hat{c}_{\eta',k}}=2\delta_{\eta\eta'}\delta_{jk}$ as well as Schur's orthogonality relation.
However, the fact that $\adjo{\hat{c}_{\eta,j}}=\hat{c}_{\eta,j}$ implies that $\hat{a}_{\mu\nu,j}^{(K)}$ and $\adjo{\mbox{$\hat{a}_{\mu\nu,k}^{(K)}$}}$ are \emph{not} independent:
\begin{align}
    \adjo{\mbox{$\displaystyle\hat{a}_{\mu\nu,k}^{(K)}$}}
    &= \frac{1}{\sqrt{2N}}\sum_{\eta\in G}\sqrt{d_K}\hat{c}_{\eta,j}\cconj{D_{\nu\mu}^{(K)}(\eta)}\nonumber\\
    &= \frac{1}{\sqrt{2N}}\sum_{\eta\in G}\sqrt{d_K}\hat{c}_{\eta,j}{D_{\nu\mu}^{(-K)}(\eta)}\nonumber\\
    &= \hat{a}_{\mu\nu,k}^{(-K)},
    \label{eq:aK-dependent}
\end{align}
where $D^{(-K)}$ is the representation conjugate to $D^{(K)}$.
Thus, self-conjugate (real) representations $D^{(-K)}=D^{(K)}$, if any, generalize to hyperbolic lattices the notion of time-reversal invariant momenta.

Rewriting \cref{eq:Majorana-Ham_translation-inv}, we thus find
\begin{align}
    \hat{\Ham} &= \frac{\i}{2N}\sum_{\eta,\gamma\in G}\sum_{j,k}
    A_{jk}(\gamma)
    \sum_{K,\mu,\nu}\sqrt{d_{K}} \adjo{\mbox{$\displaystyle\hat{a}_{\mu\nu,j}^{(K)}$}}{D_{\nu\mu}^{(K)}(\gamma\eta)}
    \sum_{K',\mu',\nu'}\sqrt{d_{K'}} \hat{a}_{\mu'\nu',k}^{(K')}\cconj{D_{\nu'\mu'}^{(K')}(\eta)}\nonumber\\
    &= \frac{\i}{2N}\sum_{\gamma\in G}\sum_{j,k}
    A_{jk}(\gamma)
    \sum_{\substack{K,\mu,\nu\\K',\mu',\nu'}}\sqrt{d_{K}d_{K'}}
    \adjo{\mbox{$\displaystyle\hat{a}_{\mu\nu,j}^{(K)}$}}\hat{a}_{\mu'\nu',k}^{(K')}\sum_{\lambda}D_{\nu\lambda}^{(K)}(\gamma)
    \sum_{\eta\in G}D_{\lambda\mu}^{(K)}(\eta)\cconj{D_{\nu'\mu'}^{(K')}(\eta)},\nonumber\\
    \intertext{where we used that $D^{(K)}(\gamma\eta)=D^{(K)}(\gamma)D^{(K)}(\eta)$. Using Schur's orthogonality relation, this becomes}
    &= \frac{\i}{2N}\sum_{\gamma\in G}\sum_{j,k}
    A_{jk}(\gamma)
    \sum_{\substack{K,\mu,\nu\\K',\mu',\nu'}}\sqrt{d_{K}d_{K'}}
    \adjo{\mbox{$\displaystyle\hat{a}_{\mu\nu,j}^{(K)}$}}\hat{a}_{\mu'\nu',k}^{(K')}D_{\nu\mu}^{(K)}(\gamma)
    \frac{N}{d_K}\delta_{KK'}\delta_{\mu\mu'}\delta_{\lambda\nu'},\nonumber\\
    &= \frac{\i}{2}\sum_{K,\lambda,\mu,\nu}\sum_{j,k}
    \sum_{\gamma\in G}A_{jk}(\gamma)D_{\nu\lambda}^{(K)}(\gamma)
    \adjo{\mbox{$\displaystyle\hat{a}_{\mu\nu,j}^{(K)}$}}\hat{a}_{\mu\lambda,k}^{(K)}\nonumber\\
    &= \frac{\i}{2}\sum_K\sum_{\mu,\nu,\nu'}\sum_{j,k}A_{\nu,j;\nu',k}^{(K)}
    \adjo{\mbox{$\displaystyle\hat{a}_{\mu\nu,j}^{(K)}$}}\hat{a}_{\mu\nu',k}^{(K)},
    \label{eq:Bloch-decomp}
\end{align}
where in the last step, we defined the reciprocal-space representation of the matrix $A$:
\begin{equation}
    A_{\nu,j;\nu',k}^{(K)} = \sum_{\gamma\in G}A_{jk}(\gamma)D_{\nu\nu'}^{(K)}(\gamma).
    \label{eq:Bloch-A-matrix}
\end{equation}

Since $A(\gamma)$ is skew-symmetric, we can consider the Hermitian matrix $\i A(\gamma)$, such that
\begin{equation}
    \BlochHam(K) = \i A^{(K)} = \sum_{\gamma\in G}\i A(\gamma)\otimes D^{(K)}(\gamma),
    \label{eq:Majorana-Bloch-Ham}
\end{equation}
is a Bloch Hamiltonian as defined in Ref.~\onlinecite{Lenggenhager:2023}.
Let $U^{(K)}$ be the unitary matrix diagonalizing $\BlochHam(K)$, such that
\begin{equation}
    H_{\nu,j;\nu',k}(K) = \sum_{\lambda,l} (\adjo{U^{(K)}})_{\nu,j;\lambda,l}\varepsilon_{\lambda,l}(K)U_{\lambda,l;\nu',k}^{(K)}
\end{equation}
and define
\begin{equation}
    \hat{d}_{\mu\lambda,l}^{(K)} = \sum_{\nu,j}U_{\lambda,l;\nu,j}^{(K)}\hat{a}_{\mu\nu,j}^{(K)},
\end{equation}
which are again canonically normalized, i.e., $\acomm{\hat{d}_{\mu\lambda,l}^{(K)}}{\adjo{\mbox{$\displaystyle\hat{d}_{\mu'\lambda',m}^{(K')}$}}}=\delta_{KK'}\delta_{\mu\mu'}\delta_{\lambda\lambda'}\delta_{lm}$.
This diagonalizes the Hamiltonian in \cref{eq:Bloch-decomp}:
\begin{align}
    \hat{\Ham}
    &= \frac{1}{2}\sum_K\sum_{\mu,\nu,\nu',\lambda}\sum_{j,k,l}\adjo{\mbox{$\displaystyle\hat{a}_{\mu\nu,j}^{(K)}$}}(\adjo{U^{(K)}})_{\nu,j;\lambda,l}\varepsilon_{\lambda,l}U_{\lambda,l;\nu',k}^{(K)}
    \hat{a}_{\mu\nu',k}^{(K)}\nonumber\\
    &= \frac{1}{2}\sum_{K,\lambda,\mu,l}\varepsilon_{\lambda,l}(K)\adjo{\mbox{$\displaystyle\hat{d}_{\mu\lambda,l}^{(K)}$}}\hat{d}_{\mu\lambda,l}^{(K)}.
    \label{eq:diagonal-Ham}
\end{align}

Note that \cref{eq:Majorana-Bloch-Ham} is Hermitian, $\adjo{\BlochHam(K)}=\BlochHam(K)$ and thus satisfies
\begin{equation}
    \BlochHam(K)^\top
    = \cconj{\BlochHam(K)}
    = -\sum_{\gamma\in G}\i A(\gamma)\otimes \cconj{D^{(K)}(\gamma)}
    = -\BlochHam(-K),
\end{equation}
where we used that $\cconj{D^{(K)}(\gamma)}=D^{(-K)}(\gamma)$.
Since $\BlochHam(K)^\top$ and $\BlochHam(K)$ have the same spectrum, it is always possible to choose appropriate bases (for the indices $\lambda,l$) in the $K$ and $-K$ sectors, such that
\begin{equation}
    \varepsilon_{\lambda,l}(-K) = -\varepsilon_{\lambda,l}(K).
    \label{eq:double-spectrum-property}
\end{equation}

The above allows us to rewrite the sum in \cref{eq:diagonal-Ham} by splitting it according to the sign of $\varepsilon_{\lambda,l}(K)$:
\begin{equation}
    \hat{\Ham} = \frac{1}{2}\sum_{\substack{K,\lambda,\mu,l\\\varepsilon_{\lambda,l}(K)>0}}\varepsilon_{\lambda,l}(K)\adjo{\mbox{$\displaystyle\hat{d}_{\mu\lambda,l}^{(K)}$}}\hat{d}_{\mu\lambda,l}^{(K)} + \frac{1}{2}\sum_{\substack{K,\lambda,\mu,l\\\varepsilon_{\lambda,l}(K)<0}}\varepsilon_{\lambda,l}(K)\adjo{\mbox{$\displaystyle\hat{d}_{\mu\lambda,l}^{(K)}$}}\hat{d}_{\mu\lambda,l}^{(K)}.
    \label{eq:Ham_split-BZ}
\end{equation}
In the second term, we rewrite the sum over $K$ as a sum over $K'=-K$, such that
\begin{align}
    \frac{1}{2}\sum_{\substack{K,\lambda,\mu,l\\\varepsilon_{\lambda,l}(K)<0}}\varepsilon_{\lambda,l}(K)\adjo{\mbox{$\displaystyle\hat{d}_{\mu\lambda,l}^{(K)}$}}\hat{d}_{\mu\lambda,l}^{(K)}
    &= \frac{1}{2}\sum_{\substack{K',\lambda,\mu,l\\\varepsilon_{\lambda,l}(K')>0}}\varepsilon_{\lambda,l}(-K')\adjo{\mbox{$\displaystyle\hat{d}_{\mu\lambda,l}^{(-K')}$}}\hat{d}_{\mu\lambda,l}^{(-K')}\nonumber\\
    &= -\frac{1}{2}\sum_{\substack{K',\lambda,\mu,l\\\varepsilon_{\lambda,l}(K')>0}}\varepsilon_{\lambda,l}(K')\hat{d}_{\mu\lambda,l}^{(K')}\adjo{\mbox{$\displaystyle\hat{d}_{\mu\lambda,l}^{(K')}$}}\nonumber\\
    &= \frac{1}{2}\sum_{\substack{K',\lambda,\mu,l\\\varepsilon_{\lambda,l}(K')>0}}\varepsilon_{\lambda,l}(K')\left(\adjo{\mbox{$\displaystyle\hat{d}_{\mu\lambda,l}^{(K')}$}}\hat{d}_{\mu\lambda,l}^{(K')}-1\right).
\end{align}
Substituting back into \cref{eq:Ham_split-BZ}, we find
\begin{align}
    \hat{\Ham} &= \frac{1}{2}\sum_{\substack{K,\lambda,\mu,l\\\varepsilon_{\lambda,l}(K)>0}}\varepsilon_{\lambda,l}(K)\adjo{\mbox{$\displaystyle\hat{d}_{\mu\lambda,l}^{(K)}$}}\hat{d}_{\mu\lambda,l}^{(K)} + \frac{1}{2}\sum_{\substack{K,\lambda,\mu,l\\\varepsilon_{\lambda,l}(K)>0}}\varepsilon_{\lambda,l}(K)\left(\adjo{\mbox{$\displaystyle\hat{d}_{\mu\lambda,l}^{(K)}$}}\hat{d}_{\mu\lambda,l}^{(K)}-1\right)\nonumber\\
    &= \sum_{\substack{K,\lambda,\mu,l\\\varepsilon_{\lambda,l}(K)>0}}\varepsilon_{\lambda,l}(K)\left(\adjo{\mbox{$\displaystyle\hat{d}_{\mu\lambda,l}^{(K)}$}}\hat{d}_{\mu\lambda,l}^{(K)}-\frac{1}{2}\right).
    \label{eq:Bloch-Hamiltonian_diagonalized}
\end{align}

\Cref{eq:Bloch-Hamiltonian_diagonalized} implies that the spectrum of \cref{eq:generic-Majorana-Hamiltonian} can be found by diagonalizing the Bloch Hamiltonian \cref{eq:Majorana-Bloch-Ham} for different $K$.
However, since we are dealing with hyperbolic lattices, $K$ labels not simply ordinary Bloch states, but also non-Abelian Bloch states transforming in higher-dimensional irreducible representations of the translation group.
In order to deal with those, we apply the supercell method~\cite{Lenggenhager:2023} as discussed in the next section.

\labeledsection{Computations based on the supercell method}{%
Technical details and parameter choices for computations based on the supercell method: supercell sequence and sampling parameters (\cref{SM:SC:sequence}); explicit model definition and gauge choices (\cref{SM:SC:model}); details on the fitting algorithm employed for the extrapolation described in \cref{App:supercell-extrapolation} (\cref{SM:SC:intDOS-fit}).%
}\label{SM:SC-method}

The supercell method gives access to non-Abelian Bloch states on hyperbolic lattices by considering Abelian Bloch states on a sequence of supercells increasing in size~\cite{Lenggenhager:2023}.
In the context of our work, we apply the method to the Bloch Hamiltonian given in \cref{eq:Majorana-Bloch-Ham}.
In this section, we provide all the necessary details on the computations performed to obtain the data presented in the Letter.
In \cref{SM:SC:sequence}, we specify the particular supercell-sequence that we use, the detailed parameters used for the random sampling in the Abelian Brillouin zones and for constructing the density of states.
This is followed by \cref{SM:SC:model} with explicit expressions for the matrices defining the model.
Finally in \cref{SM:SC:intDOS-fit}, we describe the maximum-likelihood fitting algorithm employed in the extrapolation of the spectral gap as described in \cref{App:supercell-extrapolation}.

\subsection{Supercell sequence and state sampling}\label{SM:SC:sequence}
\subsubsection*{Supercell sequence}
We consider the hyperbolic Kitaev model on the $\{8,3\}$ lattice which has symmetry group $\Delta(2,3,8)$, see \cref{eq:triangle-group}.
The symmetric three-edge coloring discussed in \cref{SM:symmetric-3-coloring}, breaks that symmetry down to the normal subgroup $\Delta(4,4,4)\nsubg\Delta(2,3,8)$, see \cref{main:fig:coloring,fig:3-coloring}.
The primitive unit cell of the model is thus given by the torsion-free normal subgroup $\Gamma\nsubg\Delta(4,4,4),\Delta(2,3,8)$ of lowest index.
Such normal subgroups are tabulated, e.g., in Ref.~\onlinecite{Conder:2007}, in terms of the quotients of the proper triangle group $\Delta^+$, i.e., $\Delta^+/\Gamma$ labeled in the form $\tgquot{\genus}{n}$.
Here, $\genus$ is the genus of the Riemann surface the group $\Delta$ acts on, i.e., on which the corresponding cell can be embedded with periodic boundary conditions, and $n$ is a running index for given $\genus$.
In our case, we find that $\Gamma$ corresponds to the quotient $\tgquot{2}{1}$ which, given in terms of the generators of $\Delta(2,3,8)$, is
\begin{equation}
    \Delta(2,3,8)/\Gamma = \Delta(2,3,8)/\Gamma_{\tgquot{2}{1}}
    = \gpres{a,b,c}{a^2,b^2,c^2,x^2,y^3,z^8,zyxz(zy)^{-1}xz}
    \label{eq:quotient-T2.1}
\end{equation}
with $x = ab$, $y = bc$, $z = ca$.
Using the implementation of this library of normal subgroups in the \textsc{HyperCells} package~\cite{HyperCells}, we compute the kernel of the quotient homomorphism, which gives the corresponding translation group with the simplified presentation
\begin{equation}\label{TG21-pres}
    \Gamma = \Gamma_{\tgquot{2}{1}}
    = \gpres{\gamma_1,\gamma_2,\gamma_3,\gamma_4}{\gamma_1\gamma_2^{-1}\gamma_3\gamma_4^{-1}\gamma_1^{-1}\gamma_2\gamma_3^{-1}\gamma_4}
\end{equation}
and
\begin{equation}
    \gamma_i = z^{5-i}yz^4y^{-1}z^{i-1}.
\end{equation}
In \gap{}~\cite{GAP4}, we can easily check that $\Gamma$ is a subgroup of (and thus normal in) $\Delta(4,4,4)$.
Indeed, with $A=a$, $B=yay^{-1}$, and $C=y^{-1}ay$ the generators of $\Delta(4,4,4)$ and the corresponding rotations $X=AB$, $Y=BC$, and $Z=CA$, $\gamma_i$, $i=1,2,3,4$, can be written as
\begin{subequations}
    \begin{align}
        \gamma_1 &= ZY^{-1}X,\\
        \gamma_2 &= ZYX^{-1},\\
        \gamma_3 &= Y^{-1}XZ,\\
        \gamma_4 &= YX^{-1}Z,
    \end{align}
\end{subequations}
implying $\gamma_i\in\Delta(4,4,4)$.

\begin{figure}[t]
    \centering
    \subfloat{\label{fig:supercell-sequences:T2.1}}
    \subfloat{\label{fig:supercell-sequences:T5.1}}
    \subfloat{\label{fig:supercell-sequences:T17.2}}
    \subfloat{\label{fig:supercell-sequences:T33.1}}
    \includegraphics{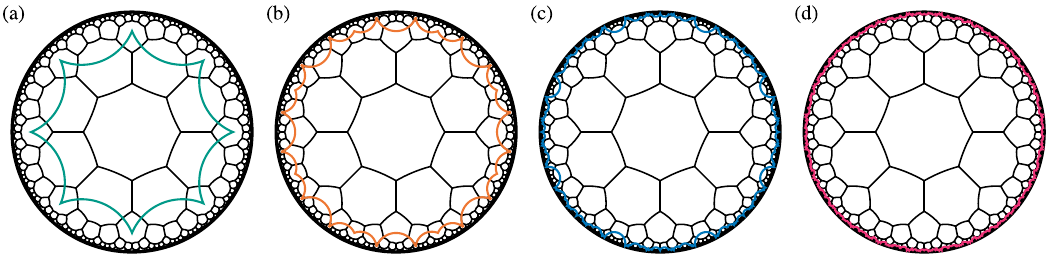}
    \caption{
        Supercell sequence used in the computations and defined by the quotients in \cref{eq:quotient-T2.1,eq:supercell-sequence:238}:
        (a) primitive cell $\tgquot{2}{1}$ with $16$ sites and $6$ plaquettes;
        (b) first supercell $\tgquot{5}{1}$ with $64$ sites and $24$ plaquettes;
        (c) second supercell $\tgquot{17}{2}$ with $256$ sites and $96$ plaquettes;
        (d) third supercell $\tgquot{33}{1}$ with $512$ sites and $192$ plaquettes.
        The largest supercell (see text) used in the computations is not shown.
    }
    \label{fig:supercell-sequence}
\end{figure}

To apply the supercell method, we construct a sequence of normal translation subgroups $\HTGsc{m}\nsubg\Delta(2,3,8)$, $\HTGsc{m+1}\nsubg\HTGsc{m}$ with $\HTGsc{1}=\Gamma$ using \textsc{HyperCells}~\cite{HyperCells}.
In terms of their quotient groups $\Delta(2,3,8)/\HTGsc{m}$, they are given as
\begin{subequations}
    \label{eq:supercell-sequence:238}
    \begin{align}
        \Delta(2,3,8)/\Gamma_{\tgquot{5}{1}} &= \gpres{a,b,c}{a^2,b^2,c^2,x^2,y^3,z^8,z^3yz^{-1}xzy^{-1}xy^{-1}z^{-2}x},\\
        \Delta(2,3,8)/\Gamma_{\tgquot{17}{2}} &= \gpres{a,b,c}{a^2,b^2,c^2,x^2,y^3,z^8,(z^2yx)^2(zy^{-1}z^{-1}x)^2},\\
        \Delta(2,3,8)/\Gamma_{\tgquot{33}{1}} &= \gpres{a,b,c}{a^2,b^2,c^2,x^2,y^3,z^8,xz^2(zyx)^3z^2y^2z^{-2}xy^{-1}z^{-2}},\\
        \Delta(2,3,8)/\Gamma_{\tgquot{33}{1}\cap\tgquot{65}{1}} &= \gpres{a,b,c}{a^2,b^2,c^2,x^2,y^3,z^8,
        (xzy^{-1}z^{-2})^2(xz^2yz^{-1})^2, (xz^2y^{-1}z^{-2})^2(xz^3yz^{-1})^2}.
        \label{eq:supercell-5}
    \end{align}
\end{subequations}
The first three were obtained from the library of quotients~\cite{HyperCells,Conder:2007}, while the fourth was constructed using the procedure of intersections of normal subgroups~\cite{Tummuru:2024,Lenggenhager:PhDThesis}.
In particular, $\Gamma_{\tgquot{33}{1}\cap\tgquot{65}{1}}=\Gamma_{\tgquot{33}{1}}\cap\Gamma_{\tgquot{65}{1}}$
with $\Delta(2,3,8)/\Gamma_{\tgquot{65}{1}}$ from Ref.~\onlinecite{Conder:2007}.
The resulting supercell has genus $129$ and as such corresponds to $128$ copies of the primitive cell resulting in $2048$ sites and $768$ plaquettes.
The connected and symmetric (super)cells for $m=1,2,3,4$ are illustrated in \cref{fig:supercell-sequence}.

\subsubsection*{Sampling of Abelian Bloch states}

Given that $\HTGsc{m}<\Gamma<\Delta(4,4,4)<\Delta(2,3,8)$ and $\HTGsc{m}\nsubg\Delta(2,3,8)$, we immediately find that $\HTGsc{m}\nsubg\Delta(4,4,4)$, as required for compatibility with the symmetric coloring, cf.~\cref{SM:3edge-PBC}.
Using the \textsc{HyperBloch} package~\cite{HyperBloch}, we then extend the tight-binding model, defined by \cref{eq:A-operator}, with a choice of gauge $\vec{u}=\{u_{jk}\}$ on the primitive cell to the supercells defined by \cref{eq:supercell-sequence:238}, following \cref{eq:Majorana-Ham_translation-inv}.
Next, we define the corresponding Abelian Bloch Hamiltonian $\BlochHam(\vec{k}^{(n)})$, \cref{eq:Majorana-Bloch-Ham}, for the one-dimensional irreducible representations $D^{(\vec{k}^{(n)})}(\gamma_j^{(n)})=\e^{\i k_j^{(n)}}$ defined on the generators $\gamma_j^{(n)}$ of $\HTGsc{m}$.
Finally, we diagonalize $\BlochHam(\vec{k}^{(n)})$ for $10^5$ (in some cases $10^6$) randomly sampled momenta $\vec{k}^{(n)}$ in the Abelian Brillouin zone $\torus{2\genus{}^{(n)}}$, where $\genus{}^{(n)}$ is the genus of the surface on which the corresponding supercell is compactified.
Based on the resulting sample of eigenvalues, we compute the ground-state energy, density of states, and the spectral gap as detailed below.

While the Hamiltonian in \cref{eq:Majorana-Hamiltonian} is gauge dependent through the choice of gauge $\vec{u}=\{u_{jk}\}$ in the definition of the matrix $A$ according to \cref{eq:A-operator}, the single-particle energies, i.e., the eigenvalues of $\i A$, are explicitly gauge-invariant.
They only depend on the corresponding flux sector.
However, as we comment in the Letter, for a Hamiltonian defined on a PBC cluster with $F$ plaquettes, $\vec{u}$ is \emph{not} fully determined by the $F-1$ independent fluxes through the plaquettes.
Instead, the expectation values of the Wilson loop operators along the $2\genus{}$ noncontractible loops need to be specified too; we call them the \emph{global fluxes}.
The differences in the eigenvalues for different global flux configurations are finite-size effects and vanish in the thermodynamic limit.

On the other hand, hyperbolic band theory explicitly describes eigenstates and energies on the infinite lattice, where there are no global fluxes.
Consequently, the flux sector is uniquely determined by the $F-1$ fluxes through the plaquettes in the unit cell and the single-particle energies $\varepsilon_{\lambda,l}(K)$ only depend on $\{W_P\}_{P=1}^{F-1}$.
Operationally, however, the supercell method considers Abelian Bloch Hamiltonians $\BlochHam(\vec{k})$ defined on finite clusters with particular (twisted) boundary conditions specified by $\vec{k}$.
To reconcile these two perspectives, note that each global flux precisely corresponds to the shift of momentum $\vec{k}$ by $\pi$ in a particular direction (there are $2\genus{}$ directions in the Abelian Brillouin zone).
The global fluxes thus only rearrange the blocks $\BlochHam(\vec{k})$ without affecting the full spectrum.

\subsubsection*{Density of states}

The density of states is obtained by constructing a histogram from the computed eigenvalues.
Here, we use an energy-bin-width of $0.002J$ when sampling $10^5$ momenta (applicable to most computations) and $0.0002$ for $10^6$ momenta (when higher resolution was required for the data presented in \cref{fig:gap_vs_K}).
When plotting the density of states, we smooth the data with a moving average with a window of five data points, resulting in an effective energy resolution of $0.01J\approx 0.0033\times 3J$ and $0.001J\approx 0.00033\times 3J$, respectively.
Some example data at different points in the phase diagram are shown in \cref{SM:ext-figs:spectrum}; note that only states with $E\:{>}\:0$ are physical excitations. The convergence of the density of states obtained from the supercell method is studied in Ref.~\onlinecite{Lenggenhager:2023}.

\subsection{Model definition}\label{SM:SC:model}

\begin{figure}[t]
    \subfloat{\label{fig:gauge-choice:hom-Pi}}
    \subfloat{\label{fig:gauge-choice:hom-0}}
    \centering
    \includegraphics{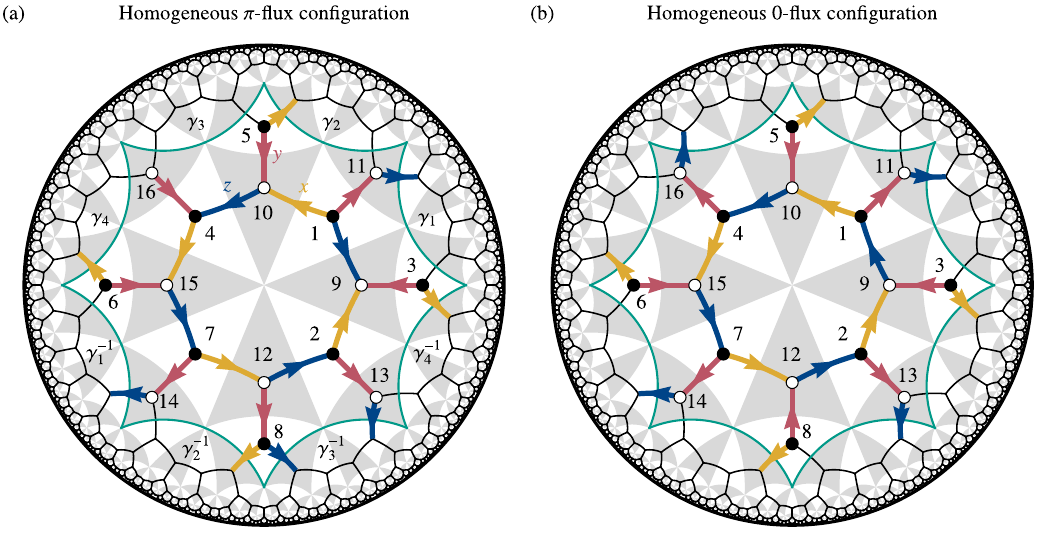}
    \caption{
    Gauge choices and site ordering from the homogeneous (a) $\pi$- and (b) $0$-flux configurations.
    The choice of gauge, $\{u_{jk}\}$ is indicated by arrows pointing from $k$ to $j$ such that $u_{jk}=+1$.
    This allows us to easily read off the corresponding eigenvalues of the plaquette operators $W_P$ by going around the plaquette counterclockwise and counting the number of arrows that do not follow that orientation.
    The numbers by the sites indicate their index in our choice of basis for \cref{eq:A-primitive-cell_explicit}.
    Finally, the translations associated with the boundaries are shown in terms of the translation generators $\gamma_j$, $j=1,2,3,4$.
    }
    \label{fig:gauge-choice}
\end{figure}

We define the hyperbolic Kitaev model on the primitive cell and then use \cref{eq:Majorana-Ham_translation-inv} to extend the definition to any of the supercells (or a compatible PBC cluster or even an OBC flake, see \cref{SM:CF}).
First, we construct the primitive cell defined by $\Gamma_{\tgquot{2}{1}}$; its symmetry group is $G=\Delta(2,3,8)/\Gamma_{\tgquot{2}{1}}$.
Following the labeling convention of Ref.~\onlinecite{Lenggenhager:2023}, the set $V$ of sites of the $\{8,3\}$ lattice is given by the right coset $G_{v_y}\backslash G$, where $G_{v_y}=\gpres{b,c}{b^2,c^2,(bc)^3}$ is the stabilizer of the vertex $v_y$ of the reference triangle, cf.~\cref{fig:3-coloring}.
Analogously, the set $E$ of edges is given by $G_{v_x}\backslash G$, where $G_{v_x}=H=\gpres{a,b}{a^2,b^2,(ab)^2}$ is the stabilizer of the vertex $v_x$ of the reference triangle.
Colors can then be assigned to the edges through partitioning of $H\backslash G$ into the double cosets, $H\backslash G/K$ with $K=\Delta(4,4,4)$, following \cref{eq:double-coset}.

Second, we fix a gauge $\vec{u}=\{u_{jk}\}_{(j,k)\in E}$ on the set of edges that is compatible with a chosen flux configuration $\{W_P\}_{P\in F}$, where $F$ is the set of faces, $G_{v_z}\backslash G$.
This results in a system of $\abs{F}-1=5$ equations with $\abs{E}=24$ unknowns, for which we select an arbitrary solution.
Our choices for the homogeneous $0$- and $\pi$-flux configurations are depicted in \cref{fig:gauge-choice}.
Note that while there are $2^{\abs{V}-1}=2^{15}$ gauge-equivalent choices of $\vec{u}$ for a given flux configuration, the fact that we do not specify the $2\genus{}=4$ global fluxes implies that the above system of equations has $2^{19}$ possible solutions and the resulting $\vec{u}$ will not be equivalent for finite systems.
However, as discussed above, the $2^4$ configurations of global fluxes are equivalent in the thermodynamic limit, which we are studying explicitly with the supercell method.

Third, we construct $A(\gamma)$, cf.~\cref{eq:A-operator} for our choice of $\vec{u}$.
In the basis indicated by the numbering of the sites in \cref{fig:gauge-choice}, we find
\begin{subequations}\label{eq:A-primitive-cell_explicit}
\begin{equation}
    A(1) = 
    \begin{pmatrix}
    A_K(1) & A_J(1)\\
    -A_J(1) & -A_K(1)
    \end{pmatrix},
\end{equation}
with
\begin{equation}
\begin{alignedat}{4}
    A_J(1) &=
    \begin{pmatrix}
     -J_{z} & -J_{x} & -J_{y} & 0 & 0 & 0 & 0 & 0 \\
     -J_{x} & 0 & 0 & J_{z} & -J_{y} & 0 & 0 & 0 \\
     -J_{y} & 0 & 0 & 0 & 0 & 0 & 0 & 0 \\
     0 & J_{z} & 0 & 0 & 0 & 0 & -J_{x} & J_{y} \\
     0 & -J_{y} & 0 & 0 & 0 & 0 & 0 & 0 \\
     0 & 0 & 0 & 0 & 0 & 0 & -J_{y} & 0 \\
     0 & 0 & 0 & -J_{x} & 0 & -J_{y} & J_{z} & 0 \\
     0 & 0 & 0 & J_{y} & 0 & 0 & 0 & 0
    \end{pmatrix},&\qquad
    A_K(1) &= K
    \begin{pmatrix}
     0 & -1 & 1 & -1 & -1 & 0 & 0 & 0 \\
     1 & 0 & -1 & 0 & 0 & 0 & 1 & 1 \\
     -1 & 1 & 0 & 0 & 0 & 0 & 0 & 0 \\
     1 & 0 & 0 & 0 & -1 & -1 & -1 & 0 \\
     1 & 0 & 0 & 1 & 0 & 0 & 0 & 0 \\
     0 & 0 & 0 & 1 & 0 & 0 & 1 & 0 \\
     0 & -1 & 0 & 1 & 0 & -1 & 0 & 1 \\
     0 & -1 & 0 & 0 & 0 & 0 & -1 & 0 
    \end{pmatrix},
\end{alignedat}
\end{equation}
and
\begin{align}
    A(\gamma_j) &= 
    \begin{pmatrix}
 0 & 0 & 0 & 0 & 0 & 0 & 0 & 0 & 0 & 0 & 0 & 0 & 0 & 0 & 0 & 0 \\
 0 & 0 & 0 & 0 & K \delta _{{j3}} & K \delta _{{j4}} & 0 & 0 & 0 & 0 & 0 & 0 & 0 & 0 & 0 & 0 \\
 0 & 0 & 0 & K \delta _{{j4}} & 0 & 0 & 0 & 0 & 0 & 0 & 0 & 0 & 0 & 0 & 0 & -J_{x} \delta _{{j4}} \\
 0 & 0 & 0 & 0 & 0 & 0 & 0 & 0 & 0 & 0 & 0 & 0 & 0 & 0 & 0 & 0 \\
 0 & 0 & 0 & 0 & 0 & 0 & 0 & 0 & 0 & 0 & 0 & 0 & 0 & 0 & 0 & 0 \\
 -K \delta _{{j1}} & 0 & 0 & 0 & 0 & 0 & 0 & 0 & 0 & 0 & J_{z} \delta _{{j1}} & 0 & 0 & 0 & 0 & 0 \\
 0 & 0 & K \delta _{{j1}} & 0 & K \delta _{{j2}} & 0 & 0 & 0 & 0 & 0 & 0 & 0 & 0 & 0 & 0 & 0 \\
 -K \delta _{{j2}} & 0 & 0 & -K \delta _{{j3}} & 0 & 0 & 0 & 0 & 0 & 0 & -J_{x} \delta _{{j2}} & 0 & 0 & 0 & 0 & -J_{z} \delta _{{j3}} \\
 0 & 0 & 0 & 0 & 0 & 0 & 0 & 0 & 0 & 0 & 0 & 0 & 0 & 0 & 0 & -K \delta _{{j4}} \\
 0 & 0 & 0 & 0 & 0 & 0 & 0 & 0 & 0 & 0 & 0 & 0 & 0 & 0 & 0 & 0 \\
 0 & 0 & 0 & 0 & 0 & 0 & 0 & 0 & 0 & 0 & 0 & 0 & 0 & 0 & 0 & 0 \\
 0 & 0 & 0 & 0 & 0 & 0 & 0 & 0 & 0 & 0 & K \delta _{{j2}} & 0 & 0 & 0 & 0 & -K \delta _{{j3}} \\
 0 & 0 & 0 & 0 & -J_{z} \delta _{{j3}} & J_{x} \delta _{{j4}} & 0 & 0 & 0 & K \delta _{{j3}} & 0 & 0 & 0 & 0 & K \delta _{{j4}} & 0 \\
 0 & 0 & -J_{z} \delta _{{j1}} & 0 & J_{x} \delta _{{j2}} & 0 & 0 & 0 & K \delta _{{j1}} & K \delta _{{j2}} & 0 & 0 & 0 & 0 & 0 & 0 \\
 0 & 0 & 0 & 0 & 0 & 0 & 0 & 0 & 0 & 0 & -K \delta _{{j1}} & 0 & 0 & 0 & 0 & 0 \\
 0 & 0 & 0 & 0 & 0 & 0 & 0 & 0 & 0 & 0 & 0 & 0 & 0 & 0 & 0 & 0
    \end{pmatrix},\\
    \intertext{with $j=1,2,3,4$,}
    A(\gamma_j\gamma_{j+1}^{-1}) &= 
    \begin{pmatrix}
     0 & 0 & 0 & 0 & 0 & 0 & 0 & 0 & 0 & 0 & 0 & 0 & 0 & 0 & 0 & 0 \\
     0 & 0 & 0 & 0 & 0 & 0 & 0 & 0 & 0 & 0 & 0 & 0 & 0 & 0 & 0 & 0 \\
     0 & 0 & 0 & 0 & K \delta _{{j5}} & 0 & 0 & 0 & 0 & 0 & 0 & 0 & 0 & 0 & 0 & 0 \\
     0 & 0 & 0 & 0 & 0 & 0 & 0 & 0 & 0 & 0 & 0 & 0 & 0 & 0 & 0 & 0 \\
     0 & 0 & 0 & 0 & 0 & K \delta _{{j7}} & 0 & 0 & 0 & 0 & 0 & 0 & 0 & 0 & 0 & 0 \\
     0 & 0 & 0 & 0 & 0 & 0 & 0 & K \delta _{{j1}} & 0 & 0 & 0 & 0 & 0 & 0 & 0 & 0 \\
     0 & 0 & 0 & 0 & 0 & 0 & 0 & 0 & 0 & 0 & 0 & 0 & 0 & 0 & 0 & 0 \\
     0 & 0 & -K \delta _{{j3}} & 0 & 0 & 0 & 0 & 0 & 0 & 0 & 0 & 0 & 0 & 0 & 0 & 0 \\
     0 & 0 & 0 & 0 & 0 & 0 & 0 & 0 & 0 & 0 & 0 & 0 & 0 & 0 & 0 & 0 \\
     0 & 0 & 0 & 0 & 0 & 0 & 0 & 0 & 0 & 0 & 0 & 0 & 0 & 0 & 0 & 0 \\
     0 & 0 & 0 & 0 & 0 & 0 & 0 & 0 & 0 & 0 & 0 & 0 & 0 & 0 & 0 & -K \delta _{{j6}} \\
     0 & 0 & 0 & 0 & 0 & 0 & 0 & 0 & 0 & 0 & 0 & 0 & 0 & 0 & 0 & 0 \\
     0 & 0 & 0 & 0 & 0 & 0 & 0 & 0 & 0 & 0 & K \delta _{{j4}} & 0 & 0 & 0 & 0 & 0 \\
     0 & 0 & 0 & 0 & 0 & 0 & 0 & 0 & 0 & 0 & 0 & 0 & K \delta _{{j2}} & 0 & 0 & 0 \\
     0 & 0 & 0 & 0 & 0 & 0 & 0 & 0 & 0 & 0 & 0 & 0 & 0 & 0 & 0 & 0 \\
     0 & 0 & 0 & 0 & 0 & 0 & 0 & 0 & 0 & 0 & 0 & 0 & 0 & K \delta _{{j8}} & 0 & 0
    \end{pmatrix},
\end{align}
with $j=1,2,\dotsc,8$ and $\gamma_{4+i}=\gamma_i^{-1}$.
The remaining twelve matrices are defined via
\begin{equation}
    A(\gamma^{-1}) = -A(\gamma)^\top.
\end{equation}
\end{subequations}
Note that the actual calculations~\cite{SDC} have been performed in the gauge obtained from the above choice by performing a single gauge transformation on site $16$.

\subsection{Fitting the integrated density of states}\label{SM:SC:intDOS-fit}

To fit the dependence of the integrated density of states, \cref{main:eq:cDOS}, $\mathcal{N}(E,N) = \int_0^E\dd{E'}\rho(E',N)$, on the supercell-size $N$ according to the linearized model given in \cref{main:eq:cDOS-vs-N_fit-model}, $\mathcal{N}(E,N) = \mathcal{N}_0(E) + \frac{s(E)}{N}$, care must be taken with the statistics.
This is due to the nonnormal distribution of the estimates for $\mathcal{N}$ obtained from sampling.

We assume the $N_s$ randomly sampled states, characterized by energy $E$, to be independent samples from the probability distribution $\rho(E)$.
Then, the probability of finding a state in the energy range $[0,E]$ is precisely $\mathcal{N}(E)$.
The probability of observing $m$ of them within an energy window $[0,E]$ is then given by the binomial distribution
\begin{equation}
    \mathrm{P}(m)=\binom{N_s}{m}\mathcal{N}(E)^{m}(1-\mathcal{N}(E))^{N_s-m},
\end{equation}
with estimator for its mean $\hat{\mathcal{N}}(E) = \frac{m(E)}{N_s}$, where $m(E)$ is the number of numerically observed states in the window.
The best fit to the model is thus found using maximum-likelihood estimation, which we have done using the \texttt{GeneralizedLinearModelFit} function in \textsc{Mathematica} with the exponential family option set to \texttt{"QuasiLikelihood"}, the variance function to $\mu\mapsto\mu(1-\mu)$ (giving the variance as a function of the mean), the response domain to $[0,1]$, and the link function to the identity.
The result of such a fit for several choices of energy is shown in \cref{main:fig:spectral-gap-method} in \cref{App:supercell-extrapolation} in the Letter.

\labeledsection{Computations based on the continued-fraction method}{%
Technical details and parameter choices for computations based on the continued-fraction method: choice of clusters with open (\cref{SM:CF-OBC}) and periodic (\cref{SM:CF-PBC}) boundary conditions; details on the algorithm itself, in particular on the modifications necessary to capture gaps and a benchmark-comparison to the supercell method (\cref{SM:CF-method}).%
}\label{SM:CF}

The supercell method discussed in Sec.~\ref{SM:SC-method} provides an efficient way to study hyperbolic lattices directly in the thermodynamic limit.
In practice, it is useful, because relatively small cutoffs on the supercell sequence and the momentum sampling produce good approximations.
However, near band edges and band degeneracies, the presence of such cutoffs can lead to the sampling of Bloch states that are strongly suppressed in the thermodynamic limit~\cite{Lenggenhager:2023,Tummuru:2024}. Thus, it is helpful to have an alternative method to verify our results.
In this section, we provide details on the computation of the single-particle density of states (DOS) using the continued fraction method~\cite{Haydock:1972,Haydock:1975}, which was applied recently to nearest-neighbor hyperbolic tight-binding models~\cite{Mosseri:2023}. To the difference of Ref.~\onlinecite{Mosseri:2023}, which only considered clusters with open boundary conditions (OBC), here we consider clusters with either OBC (\cref{SM:CF-OBC}) or periodic boundary conditions (PBC, \cref{SM:CF-PBC}). We provide details of the continued-fraction method itself in \cref{SM:CF-method}, which also introduces modifications to the algorithm of Ref.~\onlinecite{Mosseri:2023} that are needed to capture the presence of energy gaps~\cite{Gaspard:1973,Turchi:1982}.

\subsection{Clusters with open boundary conditions (OBC)}\label{SM:CF-OBC}

To construct large hyperbolic lattices with OBC, we use the procedure detailed in Appendix~A of Ref.~\onlinecite{Boettcher:2022}. We use the fact that the $\{8,3\}$ lattice can be viewed as an $\{8,8\}$ Bravais lattice with a 16-site basis~\cite{Boettcher:2022}. In other words, the $\{8,3\}$ lattice can be generated by acting on a reference unit cell containing 16 sites (black and white dots in \cref{main:fig:coloring} in the Letter) with all elements $\gamma$ of the Fuchsian translation group $\Gamma$ of the $\{8,8\}$ lattice. The latter group is a finitely presented infinite group with the presentation already given in \cref{TG21-pres},
\begin{align}
    \Gamma=\langle\gamma_1,\gamma_2,\gamma_3,\gamma_4|\gamma_1\gamma_2^{-1}\gamma_3\gamma_4^{-1}\gamma_1^{-1}\gamma_2\gamma_3^{-1}\gamma_4\rangle.
\end{align}
The elements $\gamma\in\Gamma$ are expressed as words in the generators $\gamma_1,\ldots,\gamma_4,\gamma_1^{-1},\ldots,\gamma_4^{-1}$. To generate a finite, rotationally symmetric flake centered around the reference unit cell, we apply to this unit cell all words in $\Gamma$ of length up to a maximal length $p$. To eliminate redundancy between different words that correspond to the same element of $\Gamma$, we work with the $\mathrm{PSU}(1,1)$ matrix representation of the generators~\cite{Maciejko:2021}, which gives a faithful representation of $\Gamma$. This also allows us to compute the complex coordinate $z_\gamma\equiv\gamma(0)$ in the Poincar\'e disk $\mathbb{D}$ of the center of a unit cell translated by $\gamma$ from the reference cell centered at $z=0$. We enumerate all words of length up to $p=8$, corresponding to a finite OBC cluster with $N=7\,579\,465$ unit cells and thus $V=16N=121\,271\,440$ sites.

Once all distinct translations $\gamma$ and corresponding $\{8,8\}$ cell coordinates $z_\gamma$ have been identified, to construct the Kitaev model on an OBC cluster, we also need to identify nearest-neighbor (NN) and next-nearest-neighbor (NNN) pairs. (Once those are known, NN and NNN pairs on the $\{8,3\}$ lattice are easily identified.) The hyperbolic distance $d(z,z')$ between two points $z,z'\in\mathbb{D}$, measured in units of the radius of curvature of $\mathbb{D}$, is defined by~\cite{Balazs:1986}
\begin{align}
    \cosh d(z,z')=1+\frac{2|z-z'|^2}{(1-|z|^2)(1-|z'|^2)}.
\end{align}
Using the $\mathrm{PSU}(1,1)$ matrix representation of the generators, the NN and NNN distances on the $\{8,8\}$ lattice are found to be
\begin{align}\label{NN-NNN-dist}
    \cosh d_\text{NN}=5+4\sqrt{2},\hspace{5mm}\cosh d_\text{NNN}=17+12\sqrt{2}.
\end{align}
For each pair $z_\gamma,z_{\gamma'}$ of cell coordinates on the flake, we first compute $d(z_\gamma,z_{\gamma'})$ and find the NN and NNN pairs that satisfy \cref{NN-NNN-dist} within numerical error. Second, for each such pair, we identify the Fuchsian translation $\gamma^{-1}\gamma'$ that relates the two cells, which is either a single generator (for a NN pair) or a product of two generators (for a NNN pair). From the knowledge of those translations, we construct the $16N\times 16N$ matrix $A_{jk}$ appearing in the Majorana Hamiltonian (\ref{eq:Majorana-Hamiltonian}).

\subsection{Clusters with periodic boundary conditions (PBC)}\label{SM:CF-PBC}

PBC clusters correspond to finite hyperbolic lattices where a certain subset of translations within the primitive translation group $\Gamma$ are set to the identity, i.e., unit cells differing by those translations are identified, producing a closed lattice without boundaries. Algebraically, a PBC cluster with $N$ unit cells is defined by a choice of normal subgroup $\Gamma'$ of $\Gamma$ with (finite) index $N$, and the unit cells on the cluster are labeled by the cosets $\Gamma/\Gamma'$. Translation symmetry on the finite cluster can be viewed as the action of the finite group $\Gamma/\Gamma'$ on itself via the (left-)regular representation~\cite{Maciejko:2022}. For our calculations on the $\{8,3\}$ lattice, we generate a PBC cluster with $N=13\,063\,680$ unit cells ($V=16N=209\,018\,880$ sites) by utilizing the method described in the Supplemental Material of Ref.~\onlinecite{Tummuru:2024}. In this method, we compute $\Gamma'$ as the intersection of several normal subgroups $\Gamma^{(i)}$ of index $|\Gamma:\Gamma^{(i)}|\leq 25$, previously obtained using the low-index normal subgroups procedure~\cite{Maciejko:2022}. To mitigate finite-size effects, we impose a $\pi$ flux in each plaquette---given that this is the lowest-energy configuration in the thermodynamic limit (see main text)---even though the particular PBC cluster we work with does not respect all bond-cutting symmetries (i.e., $\Gamma'$ is not constructed as a normal subgroup of $G$ or $K$).

For both OBC and PBC clusters, the $16\times 16$ submatrices appearing in $A_{jk}$ are given explicitly in \cref{eq:A-primitive-cell_explicit}.

\subsection{The continued-fraction method}\label{SM:CF-method}

The single-particle spectrum is controlled by the eigenvalues of the Hermitian antisymmetric matrix $H_{jk}\equiv iA_{jk}$ in \cref{eq:Majorana-Hamiltonian}.
From the single-particle Green's function in the site basis $\{\ket{i}\}$,
\begin{align}
    \mathcal{G}_{ij}(z)=\langle i|(z-H)^{-1}|j\rangle,
\end{align}
we obtain the local DOS on site $i$, $\rho_i(E)=-\frac{1}{\pi}\Im\mathcal{G}_{ii}(E+i0^+)$. For an infinite $\{p,q\}$ lattice, because of lattice symmetries, the local DOS is the same on every site, and we obtain the (global) DOS as
\begin{align}
    \rho(E)=-\frac{1}{\pi}\Im\mathcal{G}_{00}(E+i0^+),
\end{align}
computed on some reference site $i=0$. In practice, $H$ is the Hamiltonian matrix for a finite cluster with periodic boundary conditions (PBC) or open boundary conditions (OBC), and we choose the reference site to be at the center of the cluster, to mitigate finite-size artefacts introduced by the boundary conditions.

\begin{figure}[t]
    \centering
    \includegraphics[width=0.4\columnwidth]{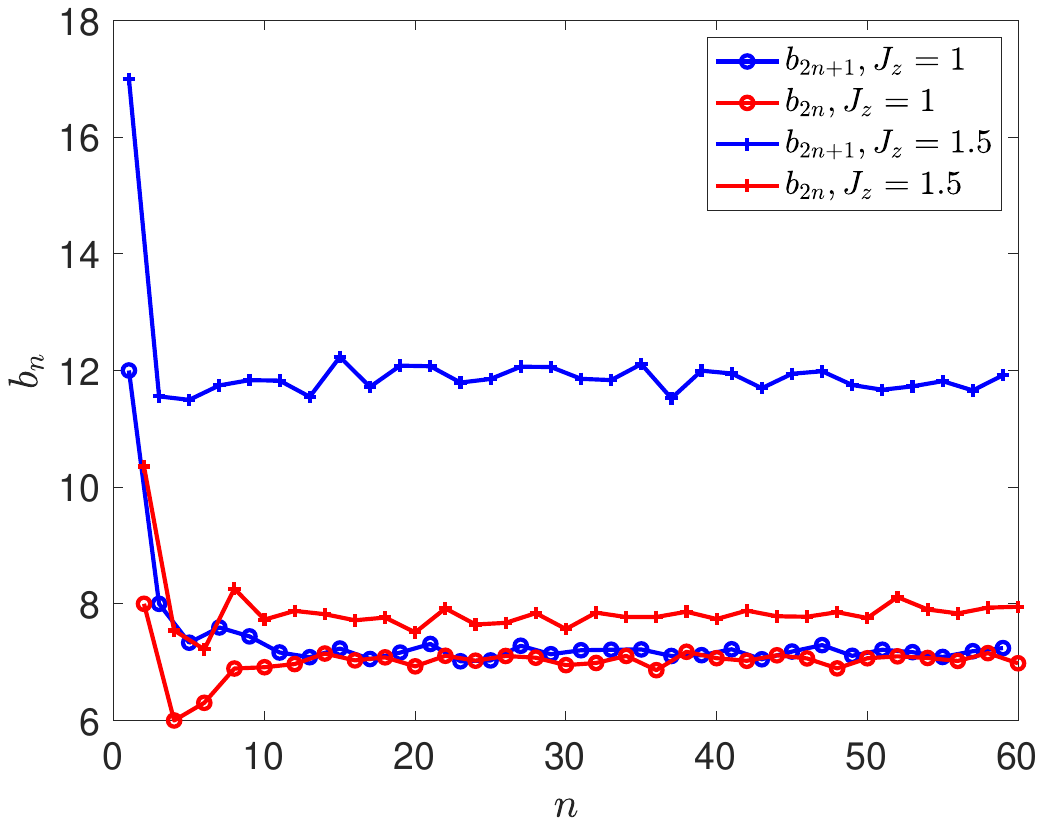}\hspace{10mm}
    \includegraphics[width=0.4\columnwidth]{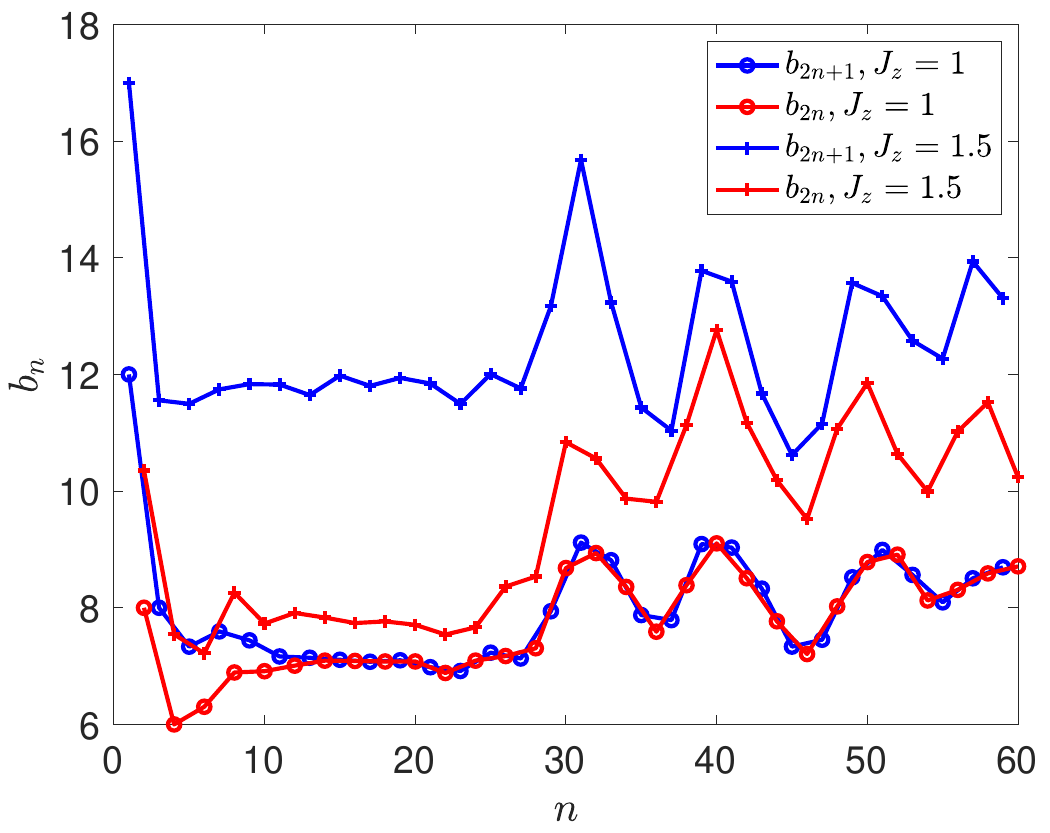}
    \caption{Continued-fraction coefficients $b_n$ computed on finite hyperbolic lattices for PBC with $V=209\,018\,880$ sites (left panel); and OBC with $V=121\,271\,440$ sites (right panel). Here, $J_x=J_y=1$ and $K=0$. In the isotropic limit ($J_z=1$), the odd and even coefficients converge to approximately the same value for $n\geq 15$, indicative of a gapless spectrum. In OBC, the large oscillations for $n\geq 30$ indicate finite-size artefacts from danglings bonds at the boundary, and we estimate $b_\infty$ by averaging over the interval $n\in[15,20]$. In PBC, we estimate $b_\infty$ by averaging over the interval $n\in[30,60]$. Over those respective intervals, for $J_z\neq 1$, the odd and even coefficients converge to a different value, indicative of a single spectral gap.  }
    \label{fig:bn}
\end{figure}

The continued fraction method is a Lanczos-type method which, starting from a normalized state vector $|1\}\equiv\ket{0}$ localized on the reference site, constructs an entire new basis $|n\}$ via a recursion relation,
\begin{align}\label{SM:Lanczos1}
    |n+1\}=H|n\}-a_n|n\}-b_{n-1}|n-1\},\hspace{5mm}n=1,2,3,\ldots,
\end{align}
with coefficients $a_n$ and $b_n$ given as
\begin{align}\label{SM:Lanczos2}
    a_n=\frac{\{n|H|n\}}{\{n|n\}},\hspace{5mm}b_n=\frac{\{n+1|n+1\}}{\{n|n\}},\hspace{5mm}n=1,2,3,\ldots,
\end{align}
with $b_0\equiv 0$. The coefficients are chosen to make the basis orthogonal, and an orthonormal basis is obtained as $|n)\equiv|n\}/\sqrt{\{n|n\}}$. In this orthonormal basis, the Hamiltonian is tridiagonal,
\begin{align}\label{SM:tridiagonal}
    (n|H|n')=\left(\begin{array}{cccc}
    a_1 & \sqrt{b_1} & & \\
    \sqrt{b_1} & a_2 & \sqrt{b_2} & \\
    & \sqrt{b_2} & a_3 & \ddots \\
    & & \ddots & \ddots
    \end{array}\right).
\end{align}
For models with a particle-hole symmetric DOS $\rho(E)=\rho(-E)$, as is the case here, all the $a_n$ coefficients vanish identically. $\mathcal{G}_{00}(z)$ is obtained as the $11$ element of the Green's function matrix in this basis:
\begin{align}
    \mathcal{G}_{00}(z)=(1|(z-H)^{-1}|1)=\displaystyle\frac{1}{z-\displaystyle\frac{b_1}{z-\displaystyle\frac{b_2}{z-\cdots\displaystyle\frac{b_{N-1}}{z-b_N t(z)}}}},
\end{align}
for a given $N>1$, where we define the remainder $t(z)$ as the infinite continued fraction
\begin{align}
    t(z)=\frac{1}{z-\displaystyle\frac{b_{N+1}}{z-\displaystyle\frac{b_{N+2}}{z-\cdots}}}.
\end{align}

In a gapless system, the sequence $b_n$ converges to a well-defined value $b_\infty\equiv\lim_{n\rightarrow\infty}b_n$~\cite{Turchi:1982}. We compute the $b_n$ coefficients via Lanczos iteration (\ref{SM:Lanczos1}-\ref{SM:Lanczos2}) on a finite hyperbolic lattice with PBC or OBC up to a value $n=N$ until the desired degree of convergence is reached. However, on a finite hyperbolic lattice with $V$ sites, only the first $n_\text{max}$ coefficients capture the true bulk physics where $n_\text{max}$ grows logarithmically with $V$~\cite{Mosseri:2023}. In practice, we compute the coefficients until approximate convergence is reached within a certain range $n_1<n<n_2$ (see \cref{fig:bn}). We then obtain an estimate of $b_\infty$ as the average of the $b_n$ over this range. We then set $b_n=b_\infty$ for all $n>N$ for some chosen $N<n_1,n_2$. Thus, the remainder can be calculated exactly:
\begin{align}\label{remainder-gapless}
    t(z)=\frac{1}{z-\displaystyle\frac{b_\infty}{z-\displaystyle\frac{b_\infty}{z-\cdots}}}=\frac{1}{z-b_\infty t(z)}
    =\frac{1}{2b_\infty}\left(z-\sqrt{z^2-4b_\infty}\right),
\end{align}
upon solving the quadratic equation $(z-b_\infty t)t=1$ for $t$, where the negative square root is chosen to ensure $t(z)$ has the correct analytic properties of a Green's function (vanishes as $z\rightarrow\infty$). Since $t(z)$ develops a nonzero imaginary part for $|\Re z|<2\sqrt{b_\infty}$, the DOS has nonzero support in the interval $-2\sqrt{b_\infty}<E<2\sqrt{b_\infty}$.

\begin{table}[t]
\begin{tabular}{c|rrr} 
 \hline\hline
 \multicolumn{1}{c|}{$k$} & \multicolumn{1}{c}{Supercell} & \multicolumn{1}{c}{Continued-fraction (PBC)} & \multicolumn{1}{c}{Continued-fraction (OBC)} \\ 
 \hline
 2 & 3 & 3 & 3 \\ 
 4 & 15 & 15 & 15 \\
 6 & 87 & 87 & 87 \\
 8 & 537 & 537 & 537 \\
 10 & 3\,423 & 3\,423 & 3\,423 \\ 
 12 & 22\,239 & 22\,239 & 22\,239 \\
 14 & 146\,289 & 146\,289 & 146\,289 \\
 16 & 970\,677 & 970\,677 & 970\,677 \\
 18 & 6\,482\,175 & 6\,482\,175 & 6\,482\,175 \\
 20 &  43\,502\,59\textcolor{orange}{6} & 43\,502\,59\textcolor{orange}{5} & 43\,502\,59\textcolor{orange}{5} \\
 22 & 293\,107\,9\textcolor{orange}{27} & 293\,107\,9\textcolor{orange}{16} & 293\,107\,9\textcolor{orange}{16} \\
 24 & 1\,981\,314\,\textcolor{orange}{254} & 1\,981\,314\,\textcolor{orange}{166} & 1\,981\,314\,\textcolor{orange}{176} \\
 26 & 13\,429\,78\textcolor{orange}{9\,221} & 13\,429\,78\textcolor{orange}{8\,112} & 13\,429\,78\textcolor{orange}{8\,608} \\
 \hline\hline
\end{tabular}
\caption{Comparison of (normalized) $k$th DOS moments $\mu_k/2^k$ obtained from the supercell method ($2048$ sites, $10^6$ momenta, energy resolution $0.0002$; rounded to the nearest integer) and the continued-fraction method with PBC and OBC clusters, in the isotropic limit $J_x=J_y=J_z=1$, $K=0$. Digits that disagree across methods are highlighted in orange.}\label{tab:moments}
\end{table}

In a system with a (single) gap, the odd coefficients ($b_{2n+1}$) and even coefficients ($b_{2n}$) asymptote to two different limits as $n\rightarrow\infty$~\cite{Turchi:1982}. In this case, we choose $N$ to be an even integer, such that the remainder becomes
\begin{align}\label{SM:remainder-gap}
    t(z)=\frac{1}{z-\displaystyle\frac{b_\infty^{(o)}}{z-\displaystyle\frac{b_\infty^{(e)}}{z-\displaystyle\frac{b_\infty^{(o)}}{z-\displaystyle\frac{b_\infty^{(e)}}{\cdots}}}}}
    =\frac{1}{z-\displaystyle\frac{b_\infty^{(o)}}{z-b_\infty^{(e)}t(z)}},
\end{align}
having defined $b_\infty^{(o)}\equiv\lim_{n\rightarrow\infty}b_{2n+1}$ and $b_\infty^{(e)}\equiv\lim_{n\rightarrow\infty}b_{2n}$. In practice, we again estimate $b_\infty^{(o)}$ and $b_\infty^{(e)}$ as averages over a suitable interval of (odd/even) $n$ where approximate convergence is observed, together with error bars from the maximum/minimum values of $b_n$ over this convergence interval. The quadratic equation for $t(z)$ in \cref{SM:remainder-gap} can again be solved, yielding
\begin{align}\label{remainder-gapped}
    t(z)=\frac{1}{2b_\infty^{(e)}z}\left(z^2-\Delta_\infty-\sqrt{(z^2-\Delta_\infty)^2-4b_\infty^{(e)}z^2}\right),
\end{align}
where we define $\Delta_\infty\equiv b_\infty^{(o)}-b_\infty^{(e)}$. Inspecting the branch points of $t(z)$ as before, we find that the gap in the DOS is given by
\begin{align}
    \Delta E=2\left(\sqrt{b_\infty^{(o)}}-\sqrt{b_\infty^{(e)}}\right).
\end{align}

In \cref{fig:gap_vs_Jz,fig:gap_vs_K}, we plot the gap $\Delta E$ together with estimated error bars coming from the maximum/minimum gap over the convergence interval discussed previously. For both PBC and OBC clusters, within the estimated error, we find that the spectrum is gapless in the isotropic limit $J_x=J_y=J_z=J$ and for $K=0$. Thus, in the Letter, we present results in this case that are calculated using the remainder formula (\ref{remainder-gapless}) for gapless systems. For $J_z\neq 1$ or $K\neq 0$, we use the two-sided remainder formula  (\ref{remainder-gapped}).

To compare results obtained from the supercell method (\cref{SM:SC-method}) vs the continued-fraction method, we compare in Table~\ref{tab:moments} the moments of the DOS, defined as
\begin{align}\label{moments}
    \mu_k=\int_{-\infty}^\infty dE\,\rho(E)E^k.
\end{align}
Due to the bipartite nature of the $\{8,3\}$ lattice, the model has a particle-hole symmetry, $\rho(-E)=\rho(E)$, and thus only the even moments are nonvanishing. We compare the even moments from the supercell method, obtained via numerical integration of the DOS, to those obtained from the continued-fraction method (in the isotropic, $K=0$ limit). For the latter method, we use the fact that for a translation invariant system with $V$ sites,
\begin{align}
    \mu_k=\frac{1}{V}\sum_{i}\langle i|H^k|i\rangle=\langle 0|H^k|0\rangle=(1|H^k|1),
\end{align}
with $|0\rangle=|1)$ denoting the localized state at the center of the cluster. The $(1,1)$ element of the $k$th power of $H$ is easily computed using the tridiagonal representation of $H$ in \cref{SM:tridiagonal}, and only involves $b_n$ coefficients up to $n=k/2$. For the PBC cluster, translation symmetry implies that $\langle i|H^k|i\rangle$ is truly independent of $i$, while for the OBC cluster, since there is no true translation invariance, $i=0$ is chosen to mitigate boundary effects. The computation of moments can also be used to benchmark our finite-size clusters. For the simple nearest-neighbor tight-binding model on the $\{8,3\}$ lattice~\cite{Mosseri:2023}, we find that our PBC (OBC) cluster captures the correct moments up to $k=22$ ($k=30$).

\begin{figure}[p]
    \subfloat{\label{fig:flux-configs:homPi}}
    \subfloat{\label{fig:flux-configs:2-nnns}}
    \subfloat{\label{fig:flux-configs:2-nns}}
    \subfloat{\label{fig:flux-configs:4-nnns}}
    \subfloat{\label{fig:flux-configs:4-nns}}
    \subfloat{\label{fig:flux-configs:hom0}}
    \centering
    \includegraphics{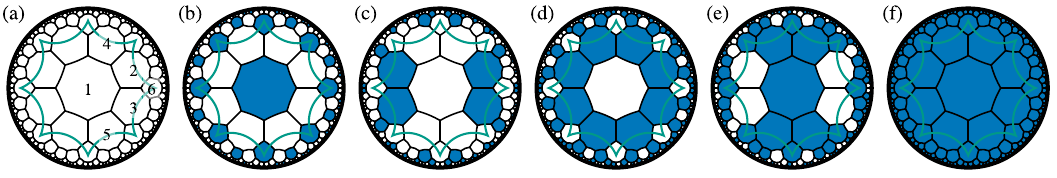}
    \caption{The six equivalence classes of translation-invariant flux configurations. White (blue shaded) octagons denote plaquettes with $\pi$ ($0$) flux, $W_P=-1$ ($W_P=+1$).
    The following representative flux configurations are shown:
    (a) $(\pi,\pi,\pi,\pi,\pi,\pi)$,
    (b) $(0,\pi,\pi,\pi,\pi,0)$,
    (c) $(\pi,0,0,\pi,\pi,\pi)$,
    (d) $(\pi,0,0,0,0,\pi)$,
    (e) $(0,\pi,\pi,0,0,0)$, and
    (f) $(0,0,0,0,0,0)$,
    where the plaquette fluxes are given in the order specified by the numbers in panel (a).
    }
    \label{fig:flux-configs}
\end{figure}

\begin{figure}[p]
    \subfloat{\label{fig:energy-flux-configs:K=0:GSE}}
    \subfloat{\label{fig:energy-flux-configs:K=0:energy-per-vortex}}
    \centering
    \includegraphics{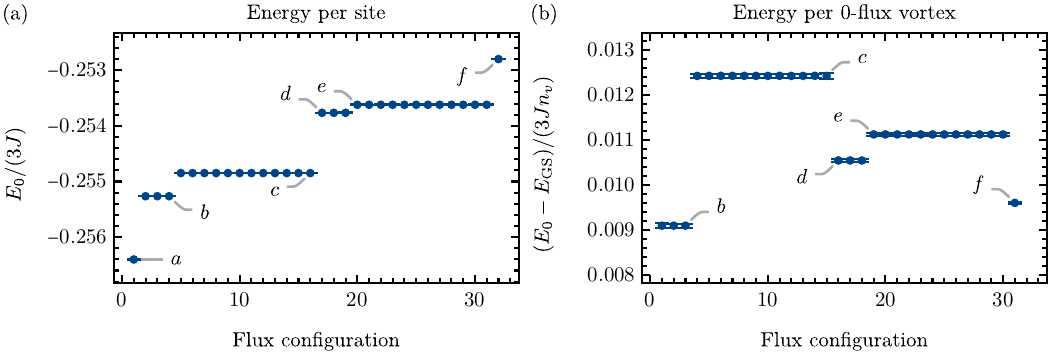}
    \caption{
    Ground-state energy for the $32$ translation-invariant flux configurations with zero net flux per primitive cell for $J_x=J_y=J_z=J$ and $K=0$ with error bars indicating $95\%$ confidence intervals.
    The flux configurations fall into the six equivalence classes shown in \cref{fig:flux-configs} and are labeled by the corresponding panel label.
    (a) Estimated ground-state energy per site $E_0$ with errors from the fit discussed in \cref{App:supercell-extrapolation}.
    (b) Energy per $0$-flux vortex obtained by computing the energy difference $E_0-E_\mathrm{GS}$ between the ground-state energy $E_0$ of the given flux configuration and $E_\mathrm{GS}$, the minimal one (homogeneous $\pi$ flux), and dividing by the density of vortices $n_v$ in the given configuration.
    }
    \label{fig:energy-flux-configs:K=0}
\end{figure}

\begin{figure}[p]
    \subfloat{\label{fig:energy-flux-configs:K=0.5:GSE}}
    \subfloat{\label{fig:energy-flux-configs:K=0.5:energy-per-vortex}}
    \centering
    \includegraphics{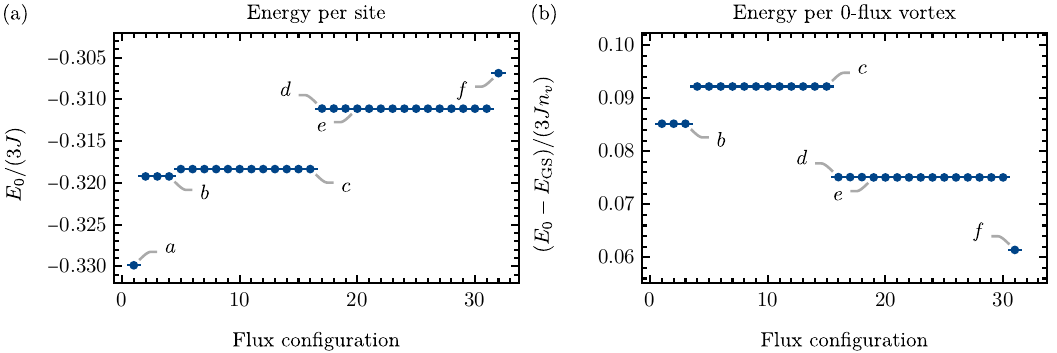}
    \caption{Ground-state energy for the $32$ translation-invariant flux configurations for $J_x=J_y=J_z=J$ and $K=0.5J$. See \cref{fig:energy-flux-configs:K=0}.}
    \label{fig:energy-flux-configs:K=0.5}
\end{figure}

\labeledsection{Extended data figures}{%
Extended figures of data already presented in the Letter: \cref{SM:ext-figs:flux-sector} contains extensions of \cref{main:fig:energy-flux-configs}, and \cref{SM:ext-figs:spectrum} of \cref{main:fig:gap_vs_Jz,main:fig:gap_vs_K} as well as full density-of-states plots, extending \cref{main:fig:DOS_K=0}.%
}

In this section, we provide extended figures of data already presented in the Letter, e.g., with some variation in the precise quantities plotted or with full ranges where only close-ups are shown in the Letter.
\Cref{SM:ext-figs:flux-sector} contains extensions of \cref{main:fig:energy-flux-configs} comparing not only the ground-state energy \emph{per site} but also the energy \emph{per vortex} for different flux configurations.
\Cref{SM:ext-figs:spectrum}, on the other hand, contains plots of the density of states over the full range of energies for different points in the phase diagram and with both $K=0$ and $K\neq 0$, as well as plots of the spectral gap (cf.~\cref{main:fig:gap_vs_Jz,main:fig:gap_vs_K}) for larger ranges of $J_z$ and $K$, respectively.

\subsection{Ground-state flux sector}\label{SM:ext-figs:flux-sector}

In \cref{App:GS-flux-sector}, we have studied the ground-state energy in different flux sectors.
We have used this to verify the prediction that the flux configuration with homogeneous $\pi$ flux has lowest energy for $K=0$, as implied by reflection positivity, and to demonstrate that even for finite $K$, the homogeneous-$\pi$-flux configuration remains the lowest.

For $K=0$, we find that the configuration with homogeneous $\pi$ flux is separated by a gap of $0.00113(7\pm 6)\times 3J$ from the configuration with next higher energy, in this case the one with two neighboring vortices of $0$ flux, cf.~\cref{fig:flux-configs,fig:energy-flux-configs:K=0:GSE}.
Further insights can be gained by computing the energy per vortex (corresponding to a plaquette with flux $0$) $(E_0-E_\mathrm{GS})/n_v$ for each of the configurations, see \cref{fig:energy-flux-configs:K=0:energy-per-vortex}.
Here $E_\mathrm{GS}$ is the ground-state energy $E_0$ of the homogeneous $\pi$-flux configuration and $n_v$ is the vortex density, i.e., the number of vortices per site.
We observe a nontrivial dependence on the distance between the vortices involved in each configuration.
The lowest energy per vortex is attained for configurations of the type shown in \cref{fig:flux-configs:2-nnns}, which have two vortices per unit cell spread out symmetrically.
This is closely followed by the homogenous $0$-flux configuration, \cref{fig:flux-configs:hom0}.
Curiously, the other configuration with two vortices per unit cell, \cref{fig:flux-configs:2-nns}, where the vortices are closer to each other, has highest energy.

We repeat the same analysis for $K=0.5J$ and show the results in \cref{fig:energy-flux-configs:K=0.5}.
The gap to the next translation-invariant configuration, which is still \cref{fig:flux-configs:2-nnns} as for $K=0$, increases by an order of magnitude to $0.01064(84\pm 18)\times 3J$.
Interestingly, there is a rearrangement of the configurations in terms of energy per vortex.
While configurations of the class shown in \cref{fig:flux-configs:2-nnns} also had the lowest energy per vortex for $K=0$, now the homogeneous $0$-flux configuration drops significantly below it.
This suggests a change in the vortex-vortex interaction potential when increasing $K$.

\subsection{Fermionic spectrum in the homogeneous $\pi$-flux sector}\label{SM:ext-figs:spectrum}

Here, we present the full plots on the fermionic spectrum in the homogeneous $\pi$-flux sector going beyond the close-ups of interesting features shown in \cref{main:fig:DOS_K=0,main:fig:gap_vs_Jz,main:fig:gap_vs_K}.
In particular, in \cref{SM:gap-anisotropy}, we show how the gap opens when departing from the isotropic point $J_x=J_y=J_z$ ($K=0$) in the phase diagram and in \cref{SM:gap-Kterm}, we consider the gap induced by $K\neq0$.
In both cases, we comment on the complications arising from each of the two methods.

\subsubsection{Gapping by anisotropy}\label{SM:gap-anisotropy}

\begin{figure}[b]
    \subfloat{\label{fig:gap_vs_Jz}}
    \subfloat{\label{fig:DOS:K=0_Jz=1}}
    \subfloat{\label{fig:DOS:K=0_Jz=2}}
    \centering
    \includegraphics{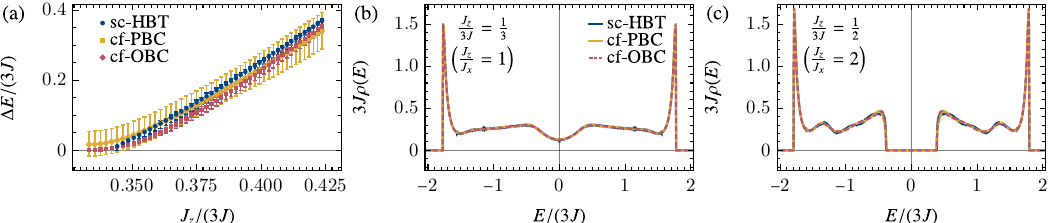}
    \caption{
    (a) Spectral gap $\Delta E$ including uncertainties shown as error bars as a function of $J_z/(3J)$ for $J_x=J_y$ and $J_x+J_y+J_z=3J$.
    The gap opens immediately when departing from the isotropic point $J_z/(3J)=1/3$.
    (b,c) Density of states $\rho$ as a function of energy $E$ at (b) the isotropic point (gapless) and (c) $J_z/(3J)=2$ (gapped).
    Data obtained using the supercell method (sc-HBT; $2048$ sites, moving average window $0.01J$), the continued-fraction method applied to clusters with periodic boundary conditions (cf-PBC; $\approxq 10^8$ sites) and to flakes with open boundary conditions (cf-OBC; $\approxq 10^8$ sites) is shown, see legend.
    The density of states obtained from the continued-fraction method explicitly assumes the presence of (b) no and (c) a single gap.
    }
    \label{fig:gap-opening-Jz}
\end{figure}

\Cref{fig:gap-opening-Jz} shows data for $K=0$ starting at the isotropic point $J_x=J_y=J_z$ and going along a vertical cut through the parameter space from the center of the phase diagram towards the top corner: $J_x=J_y$, $J_x+J_y+J_z=3J$, i.e., $J_x=J_y=3J(1-J_z/(3J))/2$.
Extending \cref{main:fig:gap_vs_Jz}, \cref{fig:gap_vs_Jz} shows the spectral gap $\Delta E$ as a function of $J_z$ near the isotropic point $J_z/(3J)=1/3$.
Note that no data points for the gap obtained from the supercell method are shown very close to the isotropic point.
For those, the algorithm introduced in \cref{App:supercell-extrapolation} fails to detect a gap, which either implies a truly gapless spectrum (in the thermodynamic limit) or insufficient convergence in the supercell size for extrapolation to work accurately.
Based on our data, we thus cannot distinguish whether the gapless phase is restricted to the isotropic point or a small region around it.
However, we can deduce an upper bound to its extent: $(J_z-J)/(3J)<0.013$.
Overall, the results obtained from the three methods agree within their respective error bars.

\Cref{fig:DOS:K=0_Jz=1,fig:DOS:K=0_Jz=2} show the density of states for the two choices $J_z/(3J)=1/3$ and $J_z/(3J)=1/2$, respectively, covering the full energy range in contrast to \cref{main:fig:DOS_K=0}.
We again observe very good agreement between the three methods.
Note that in the case of the continued-fraction method, an assumption on the number of gaps needs to be made.
This is done based on \cref{fig:gap_vs_Jz}, i.e., in \cref{fig:DOS:K=0_Jz=1} no gap while in \cref{fig:DOS:K=0_Jz=2} a single gap is assumed.

\subsubsection{Gapping by time-reversal-symmetry breaking}\label{SM:gap-Kterm}

\begin{figure}[t]
    \subfloat{\label{fig:gap_vs_K}}
    \subfloat{\label{fig:DOS:Jz=1_K=0.002}}
    \subfloat{\label{fig:DOS:Jz=1_K=0.01}}
    \centering
    \includegraphics{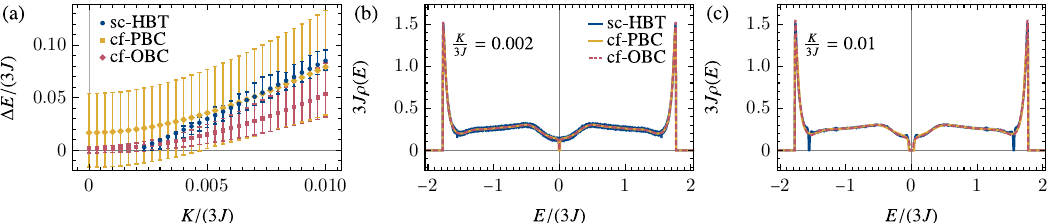}
    \caption{
    Spectral gap $\Delta E$ and density of states at the isotropic point $J_x=J_y=J_z=J$ for different values of $K$. See caption to \cref{fig:gap-opening-Jz} for more details.
    Note the two gaps at $E/(3J)\approx \pm 1.55$ that start developing at (b) $K/(2J)\approx 0.002$ and have fully formed (as far as can be discerned from the supercell method at the given resolution) at (c) $K/(3J)=0.01$.
    The density of states obtained from the continued-fraction method explicitly assumes the presence of a single gap in both (b) and (c).
    }
    \label{fig:gap-opening-K}
\end{figure}

Similarly, we have demonstrated the effect of the time-reversal-breaking term, i.e., $K\neq0$, in \cref{main:fig:gap_vs_K}.
\Cref{fig:gap_vs_K} shows an extension of that figure, i.e., the spectral gap as a function of $K$ at the isotropic point.
As expected from symmetry, we observe that any $K\neq 0$ opens the gap.
However, here, we find increasing deviations in the results obtained from the three methods with increasing $K$.
Considering the density of states obtained from the supercell method, \cref{fig:DOS:Jz=1_K=0.002,fig:DOS:Jz=1_K=0.01}, we recognize that additional gaps develop away from zero energy.
These gaps are \emph{not} captured by our implementation of the continued-fraction method, which assumes at most a single gap.
Multiple gaps can in principle be captured by the method~\cite{Turchi:1982}; however, the formalism becomes increasingly complicated and would require even larger system sizes, which are currently not accessible to us.
As long as the additional gaps are small enough, we do not expect them to significantly influence the extracted main gap $\Delta E$, but the influence grows with $K$.
In fact, even the density of states away from the gaps is barely affected in \cref{fig:DOS:Jz=1_K=0.01}.
Thus, the continued-fraction method allows us to extract the gap at small $K$, where the spectrum obtained from the supercell method is not converged to the necessary resolution in energy, while the supercell method becomes more reliable at larger $K$.

\labeledsection{Gapped $\mathbb{Z}_2$ spin liquid in the anisotropic coupling limit}{%
Mapping of the Kitaev model to a spin-boson model on the $(8,4,8,4)$ lattice (\cref{sec:spinboson}); derivation of \cref{main:eq:main-Heff}, the effective Hamiltonian in the anisotropic coupling limit, using perturbative methods (\cref{sec:PCUT}) and its mapping to a hyperbolic surface code on the $\{8,4\}$ lattice (\cref{sec:48code}).
}

In this section, we provide derivations of the statements made in the section \emph{$\mathbb{Z}_2$ spin liquid} of the Letter, in particular of the hyperbolic Kitaev model [\cref{main:eq:Hamiltonian} in the Letter] in the anisotropic coupling limit $J_x,J_y\ll J_z$ with $K=0$.
From the phase diagram obtained via the free-fermion solution of the model [\cref{main:fig:phase-diagram_K=0:triangle} in the Letter], we know that in this limit, the fermion sector is gapped. By studying the model directly in the spin-1/2 representation but treating $J_x/J_z$, $J_y/J_z$ as small parameters, we show here that the model maps in that limit onto an analog of the toric code~\cite{Kitaev:2003} but on the hyperbolic $\{8,4\}$ lattice. We find that the latter model is again exactly solvable, and use it to show that the low-energy spectrum contains two types of gapped anyons ($e$ and $m$ particles) with mutual semionic statistics. By adiabatic continuity, this establishes that the entire phase denoted $\mathsf{G}$ in \cref{main:fig:phase-diagram_K=0:triangle} in the Letter is a gapped spin liquid with $\mathbb{Z}_2$ topological order.

In \cref{sec:spinboson}, we rewrite the $\{8,3\}$ Kitaev model as a model of effective $s=1/2$ spins and hard-core bosons on the Archimedean $(8,4,8,4)$ lattice. In \cref{sec:PCUT}, we formulate a perturbative approach to systematically compute an effective Hamiltonian as a power series in the small parameters $J_x/J_z$, $J_y/J_z$ and derive \cref{main:eq:Hamiltonian} in the Letter. In \cref{sec:48code}, we focus on the limit of low energies $\omega\ll J_z$ (i.e., energies much less than the Majorana fermion gap $E_g=\Delta E/2\approx 2J_z$) and show that the effective Hamiltonian reduces to a hyperbolic surface code on the $\{8,4\}$ lattice with (static) anyonic excitations. Throughout this section, we rely heavily on the perturbative methods developed in Ref.~\onlinecite{Knetter:2000} and applied in Refs.~\onlinecite{Vidal:2008,Schmidt:2008} to the original Kitaev model on the honeycomb lattice.

\subsection{Mapping to spin-boson model on $(8,4,8,4)$ lattice}\label{sec:spinboson}

In the limit $J_x=J_y=0$ with $J_z>0$, the model consists of decoupled dimers with ferromagnetic interactions on the $z$-bonds of the three-edge colored $\{8,3\}$ lattice. The many-body ground state in this limit is macroscopically degenerate, each dimer being in either the $\uparrow\uparrow$ or $\downarrow\downarrow$ configuration. The lowest excited state above this degenerate ground state corresponds to flipping a single spin on a single dimer, with an energy cost $2J_z$. In the Majorana fermion language, this corresponds to the single-particle gap $\Delta E=4J_z$ obtained at the corners of the phase diagram in \cref{main:fig:phase-diagram_K=0:triangle} in the Letter [the physical many-body gap is $\Delta E/2$, since only Majorana excitations with energy $\varepsilon>0$ are physical---see \cref{eq:Bloch-Hamiltonian_diagonalized}]. Here, we are interested in how the macroscopic ground-state degeneracy is lifted upon turning on small exchange couplings $J_x,J_y\ll J_z$. As we will see, this results in a much smaller many-body gap $\Delta\propto J_x^4J_y^4/J_z^7\ll J_z$, corresponding to $\mathbb{Z}_2$ vortex (or vison) excitations.

\begin{figure}[t!]
    \centering
    \subfloat{\label{fig:anis:a}}
    \subfloat{\label{fig:anis:b}}
    \includegraphics[width=0.85\linewidth]{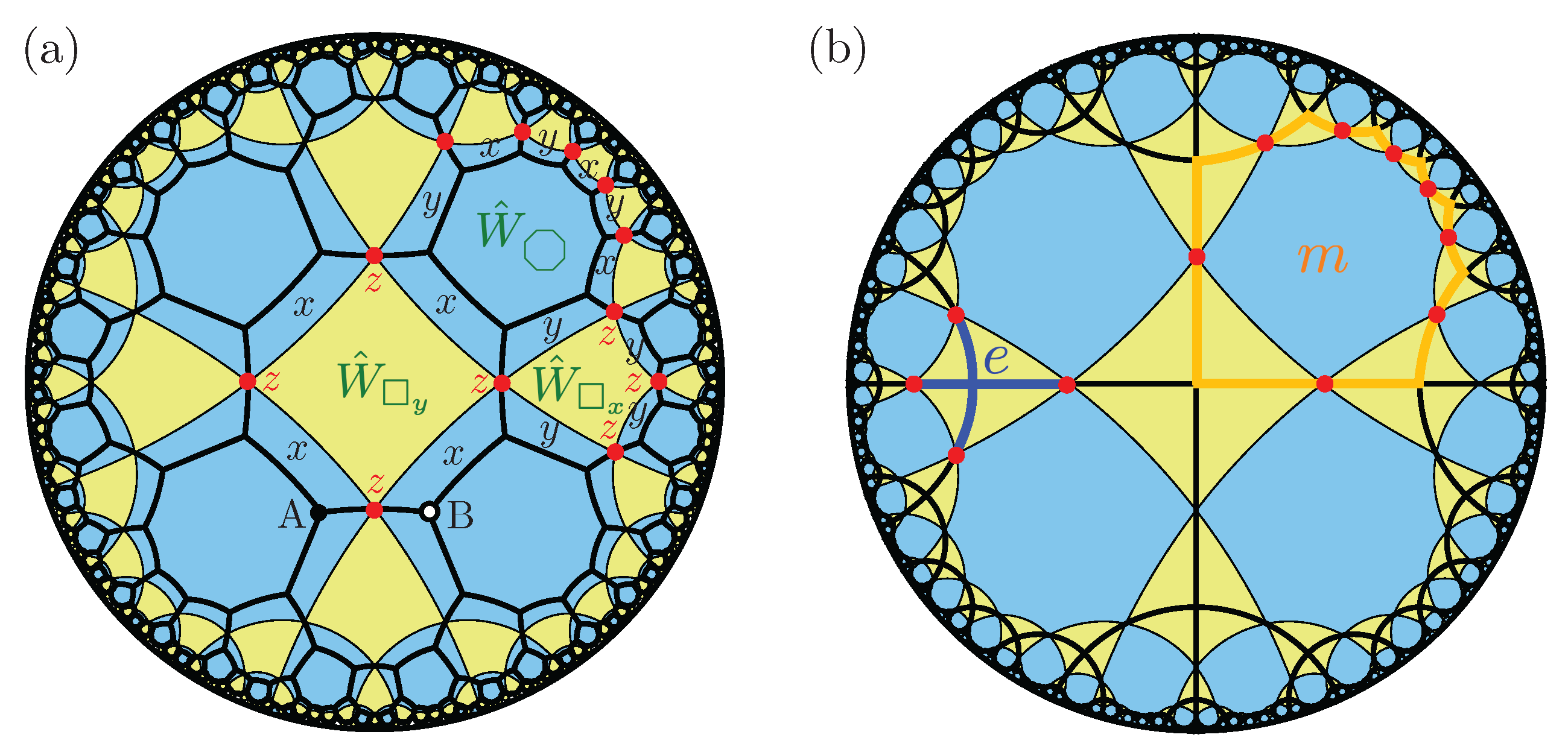}
    \caption[]{In the anisotropic coupling limit $J_x,J_y\ll J_z$, the Kitaev model on the $\{8,3\}$ lattice [black lines in (a)] maps onto a model of effective $s=1/2$ spins and hard-core bosons on the hyperbolic Archimedean $(8,4,8,4)$ lattice [tiling by blue octagons and yellow squares in (a) and (b), with lattice sites indicated as red dots]. At low energies, this effective model reduces to a hyperbolic analog of the toric code, where spins (red dots) live on the edges of the $\{8,4\}$ lattice [black lines in (b)]. Low-energy $e$ ($m$) excitations correspond to star (plaquette) excitations on the $\{8,4\}$ lattice [blue and orange in (b), respectively] and obey mutual semionic statistics.}
\label{fig:anis}
\end{figure}

The idea~\cite{Vidal:2008} is to represent the four possible states of a dimer, including the (ferromagnetic) low-energy states $\ket{\uparrow\uparrow},\ket{\downarrow\downarrow}$ and the antiferromagnetic (high-energy) states $\ket{\uparrow\downarrow},\ket{\downarrow\uparrow}$, in terms of an effective spin-1/2 moment $\hat{\boldsymbol{\tau}}$ (with $\hat{\tau}^z$ eigenstates denoted as $\Uparrow,\Downarrow$) and a hard-core boson $\hat{b},\hat{b}^\dag$ (with occupation number states denoted as $0,1$):
\begin{align}\label{Hilberteff}
\ket{\uparrow\uparrow}=\ket{\Uparrow 0},\hspace{5mm}
\ket{\downarrow\downarrow}=\ket{\Downarrow 0},\hspace{5mm}
\ket{\uparrow\downarrow}=\ket{\Uparrow 1},\hspace{5mm}
\ket{\downarrow\uparrow}=\ket{\Downarrow 1}.
\end{align}
As the local Hilbert space dimension on a given dimer is still four, the mapping (\ref{Hilberteff}) does not necessitate the imposition of any local gauge constraint. Here, we use the bipartite property of the $\{8,3\}$ lattice to assign sublattice (A/B) labels to the two physical spins on any $z$-bond dimer [black and white dots in \cref{fig:anis:a}]. The corresponding spin operators can be written as follows in the new representation:
\begin{align}\label{spinbosonops}
\begin{array}{lllll}
\hat{\sigma}_A^x=\hat{\tau}^x(\hat{b}^\dag+\hat{b}), & &
\hat{\sigma}_A^y=\hat{\tau}^y(\hat{b}^\dag+\hat{b}), & &
\hat{\sigma}_A^z=\hat{\tau}^z, \\
\hat{\sigma}_B^x=\hat{b}^\dag+\hat{b}, & &
\hat{\sigma}_B^y=i\hat{\tau}^z(\hat{b}^\dag-\hat{b}), & &
\hat{\sigma}_B^z=\hat{\tau}^z(1-2\hat{b}^\dag \hat{b}).
\end{array}
\end{align}
Although this representation explicitly breaks the bipartite symmetry of the lattice, as will be seen, the symmetry will be restored in the effective Hamiltonian.

Collapsing each $z$-bond to a single effective site on which those new degrees of freedom live, the original Kitaev model is now mapped exactly onto a model of interacting effective spins and hard-core bosons on the hyperbolic Archimedean $(8,4,8,4)$ lattice [\cref{fig:anis:a}]. An Archimedean $(n_1^{a_1},n_2^{a_2},n_3^{a_3},\ldots)$ lattice is a lattice composed of more than one type of regular polygons (here squares and regular octagons), where each vertex is surrounded by an identical sequence of $a_1$ adjacent $n_1$-gons, followed by $a_2$ adjacent $n_2$-gons, followed by $a_3$ adjacent $n_3$-gons, $\ldots$, as one goes clockwise around the vertex~\cite{Grunbaum:1987}. (For example, the Euclidean honeycomb and kagome lattices are $(6^3)$ and $(6,3,6,3)$ Archimedean lattices, respectively.) Using \cref{spinbosonops}, the $\{8,3\}$ Kitaev model can be written as:
\begin{align}\label{Heff1}
\hat{\mathcal{H}}=-N_zJ_z+2J_z\hat{Q}+\hat{T}_0+\hat{T}_2+\hat{T}_{-2},
\end{align}
where $N_z$ is the total number of $z$-bonds,
\begin{align}
\hat{Q}=\sum_R \hat{b}_R^\dag \hat{b}_R,
\end{align}
is the total number operator for hard-core bosons, with $R$ denoting the sites of the $(8,4,8,4)$ lattice,
\begin{align}\label{T0}
\hat{T}_0=-J_x\sum_{\langle RR'\rangle_x}\hat{b}_R^\dag \hat{b}_{R'}\hat{\tau}_{R'}^x-J_y\sum_{\langle RR'\rangle_y}\hat{b}_R^\dag \hat{b}_{R'}i\hat{\tau}^z_R\hat{\tau}^y_{R'}
+\mathrm{H.c.},
\end{align}
is a boson hopping term, and
\begin{align}
\hat{T}_2&=-J_x\sum_{\langle RR'\rangle_x}\hat{b}_R^\dag \hat{b}_{R'}^\dag\hat{\tau}_{R'}^x-J_y\sum_{\langle RR'\rangle_y}\hat{b}_R^\dag \hat{b}_{R'}^\dag i\hat{\tau}^z_R\hat{\tau}^y_{R'},\label{T2}\\
\hat{T}_{-2}&=\hat{T}_2^\dag,\label{T2dag}
\end{align}
are boson pair-creation/annihilation terms. (Recall that hard-core boson operators obey anticommutation relations for $R=R'$ and commutation relations for $R\neq R'$.) Here, $\langle RR'\rangle_x$ ($\langle RR'\rangle_y$) denote nearest-neighbor $x$-bonds ($y$-bonds) on the $(8,4,8,4)$ lattice [see bonds labeled by $x$ ($y$) between nearest-neighbor red sites in \cref{fig:anis:a}]. In \cref{Heff1}, $-N_zJ_z$ is the (macroscopically degenerate) ground-state energy in the $J_x=J_y=0$ limit. As is clear from \cref{Hilberteff}, the term proportional to $\hat{Q}$ indicates that creating a boson corresponds to flipping a physical spin on a single dimer, which costs energy $2J_z$. To investigate vison excitations with a gap much less than $2J_z$, we wish to derive an effective Hamiltonian valid in the $Q=0$ (zero-boson) sector. To achieve this, the strategy~\cite{Vidal:2008} is to perform a unitary transformation to a Hamiltonian $\hat{\mathcal{U}}^\dag\hat{\mathcal{H}}\hat{\mathcal{U}}$ which conserves the boson number $\hat{Q}$, and evaluate this Hamiltonian in the low-energy sector $Q=0$.

\subsection{Effective Hamiltonian from perturbation theory}
\label{sec:PCUT}

The desired unitary transformation cannot be performed exactly, but it can be performed order by order as a perturbative expansion~\cite{Knetter:2000}, here in powers of $J_x/J_z$ and $J_y/J_z$. We refer the reader to the original reference~\cite{Knetter:2000} for the details of the method and outline here only the key steps. The method works for Hamiltonians of the type
\begin{align}\label{Hknetter}
\hat{\mathcal{H}}=\hat{\mathcal{H}}_0+\sum_{n=-N}^N \hat{T}_n,
\end{align}
where the unperturbed Hamiltonian $\hat{\mathcal{H}}_0=-N_zJ_z+2J_z\hat{Q}$ is proportional (up to an additive constant term, here $-N_zJ_z$) to a ``charge'' $\hat{Q}$ with nonnegative integer spectrum $0,1,2,\ldots$, and $\hat{T}_n$ increases this charge by $n$, i.e., $[\hat{Q},\hat{T}_n]=n\hat{T}_n$. The operators $\hat{T}_n$ are assumed to be proportional to a common small parameter which controls the perturbative expansion. The outcome of the method is a unitarily equivalent Hamiltonian expressed as a perturbative series,
\begin{align}\label{UdagHU}
\hat{\mathcal{U}}^\dag\hat{\mathcal{H}}\hat{\mathcal{U}}=\hat{\mathcal{H}}_0+\sum_{k=1}^\infty\sum_{|\vec{m}|=k}C(\vec{m})\hat{T}(\vec{m}),
\end{align}
where $k$ is the order in perturbation theory. For each $k$, the second sum is a sum over all $k$-component vectors $\vec{m}=(m_1,m_2,\ldots,m_k)$ (indicated by the notation $|\vec{m}|=k$) where the entries $m_i\in\{-2,0,2\}$, which are the allowed values of $n$ in \cref{Hknetter} for our case [compare with \cref{Heff1}]. Furthermore, one only keeps in the sum the vectors that obey $\sum_{i=1}^km_i=0$. Finally, the operators $\hat{T}(\vec{m})$ are products of $k$ operators $\hat{T}_n$ specified by the vector $\vec{m}$,
\begin{align}
\hat{T}(\vec{m})=\hat{T}_{m_1}\hat{T}_{m_2}\cdots \hat{T}_{m_k},
\end{align}
and the numbers $C(\vec{m})$ are coefficients that are computed using a recursive method~\cite{SDC}. Note that the condition $\sum_{i=1}^km_i=0$ implies that the transformed Hamiltonian commutes with the boson number $\hat{Q}$ as desired. Once \cref{UdagHU} has been evaluated up to a desired order $k_\text{max}\geq k\geq 1$ in perturbation theory, we project it onto the $Q=0$ sector, i.e., we compute its expectation value in the boson vacuum $\ket{0}$ defined by $\hat{b}_R\ket{0}=0$ for all $R$:
\begin{align}\label{Heff2}
\hat{\mathcal{H}}_\text{eff}\equiv\langle 0|\hat{\mathcal{U}}^\dag\hat{\mathcal{H}}\hat{\mathcal{U}}|0\rangle=-N_zJ_z+\sum_{k=1}^{k_\text{max}}\sum_{|\vec{m}|=k}C(\vec{m})\langle 0|\hat{T}(\vec{m})|0\rangle.
\end{align}
Note that $\hat{\mathcal{H}}_\text{eff}$ is still an operator acting in the effective spin Hilbert space, because only the boson degrees of freedom have been projected out.

\subsection{Hyperbolic $\{8,4\}$ surface code and anyonic excitations}
\label{sec:48code}

In the $Q=0$ sector, the only nonconstant contributions to $\hat{\mathcal{H}}_\text{eff}$ in \cref{Heff2} come from closed loops, i.e., plaquette operators~\cite{Vidal:2008}. In general, $\hat{\mathcal{H}}_\text{eff}$ contains both single-plaquette terms and multi-plaquette interactions.
For simplicity, we ignore multi-plaquette interactions here, and focus on the single-plaquette terms which give a toric-code-like Hamiltonian.
The $(8,4,8,4)$ lattice itself contains only two types of plaquettes, square and octagonal, but in the general case $J_x\neq J_y$ the effective model distinguishes three types of plaquettes: two square ($\square_x,\square_y$) and one octagonal ($\octagon$), see \cref{fig:anis:a}. For single-plaquette terms, the computation is done by considering separately a single plaquette of each type~\cite{SDC}. Starting with the $\square_x$ plaquette, we see from \cref{fig:anis:a} that it involves four $z$-sites connected by four $y$-bonds. To compute the contribution to the effective Hamiltonian (\ref{Heff2}) from such a plaquette, it is sufficient to truncate the full Hamiltonian (\ref{Heff1}) to a four-site Hilbert space ($R=1,2,3,4$) with effective spins $\hat{\boldsymbol{\tau}}_1,\ldots,\hat{\boldsymbol{\tau}}_4$ and hard-core bosons $\hat{b}^{(\dag)}_1,\ldots,\hat{b}^{(\dag)}_4$. Thus, for this calculation, only terms involving sites $R,R'\in\{1,2,3,4\}$ need to be kept in the $\hat{T}_n$ operators in Eqs.~(\ref{T0}-\ref{T2dag}). To project out the products $\hat{T}(\vec{m})$ in \cref{Heff2} onto the boson vacuum $\ket{0}$, we work with a $2^4\times 2^4$ matrix representation of the $\hat{b}^{(\dag)}_1,\ldots,\hat{b}^{(\dag)}_4$ operators in the Fock basis. In practice, this is conveniently done by viewing hard-core boson Fock states as the binary representation of the integers $0,\ldots,2^4-1$~\cite{Sandvik2010}. Since the effective spin degrees of freedom are not projected out, but remain as noncommuting operators obeying the Pauli algebra, we handle them using the \texttt{DiracQ} \textsc{Mathematica} package~\cite{Wright:2013}. Since the $\square_x$ plaquette contains four bonds, we expect the first nonconstant term to appear at fourth order in perturbation theory, and indeed we find, omitting a constant correction to the ground-state energy,
\begin{align}\label{Heff_x}
\hat{\mathcal{H}}_\text{eff}^{(\square_x)}=-\frac{5}{16}\frac{J_y^4}{J_z^3}\hat{\tau}_1^x\hat{\tau}_2^x\hat{\tau}_3^x\hat{\tau}_4^x.
\end{align}
Repeating the calculation for the $\square_y$ plaquette with four $z$-sites connected by four $x$-bonds [\cref{fig:anis:a}], we obtain
\begin{align}\label{Heff_y}
\hat{\mathcal{H}}_\text{eff}^{(\square_y)}=-\frac{5}{16}\frac{J_x^4}{J_z^3}\hat{\tau}_1^x\hat{\tau}_2^x\hat{\tau}_3^x\hat{\tau}_4^x,
\end{align}
i.e., the same result but with $J_y$ replaced by $J_x$. Finally, for the octagonal plaquette with eight $z$-sites connected by alternating $x$- and $y$-bonds, we expect a nonconstant contribution $\propto J_x^4J_z^4/J_z^7$ at eighth order in perturbation theory,\footnote{At the same order in perturbation theory, we also expect an interaction term between neighboring $\square_x$ and $\square_y$ plaquettes, but we neglect such terms here for simplicity.} and indeed obtain,
\begin{align}\label{Heff_z}
\hat{\mathcal{H}}_\text{eff}^{(\octagon)}=-\frac{5}{2048}\frac{J_x^4J_y^4}{J_z^7}\hat{\tau}_1^z\hat{\tau}_2^z\hat{\tau}_3^z
\hat{\tau}_4^z\hat{\tau}_5^z\hat{\tau}_6^z\hat{\tau}_7^z\hat{\tau}_8^z,
\end{align}
working with a $2^8$-dimensional boson Hilbert space. Summing over all plaquettes, we obtain the effective Hamiltonian
\begin{align}\label{Heff3}
\hat{\mathcal{H}}_\text{eff}=-\frac{5}{16}\frac{J_y^4}{J_z^3}\sum_{\square_x}\prod_{R\in\square_x}\hat{\tau}^x_R
-\frac{5}{16}\frac{J_x^4}{J_z^3}\sum_{\square_y}\prod_{R\in\square_y}\hat{\tau}^x_R
-\frac{5}{2048}\frac{J_x^4J_y^4}{J_z^7}\sum_{\octagon}\prod_{R\in\octagon}\hat{\tau}_R^z.
\end{align}

We next show that the plaquette operators appearing in \cref{Heff3} are nothing but the Wilson loop operators $\hat{W}_P=\prod_{\langle j,k\rangle_\alpha\in P}\hat{\sigma}_j^\alpha\hat{\sigma}_k^\alpha$ introduced in the Letter. Going back to the original $\{8,3\}$ lattice, our choice of three-edge coloring defines three different types of plaquette operators (as for dimer covering III in Ref.~\onlinecite{Kamfor:2010}), corresponding to the parent octagons for the $\square_x$, $\square_y$, and $\octagon$ plaquettes introduced earlier. Numbering the sites around an $\{8,3\}$ plaquette as $j=1,\ldots,8$, we have
\begin{align}
\hat{W}_{\square_x}&=\hat{\sigma}_1^y\hat{\sigma}_2^y\hat{\sigma}_2^z\hat{\sigma}_3^z\hat{\sigma}_3^y\hat{\sigma}_4^y\hat{\sigma}_4^z\hat{\sigma}_5^z\hat{\sigma}_5^y\hat{\sigma}_6^y\hat{\sigma}_6^z\hat{\sigma}_7^z\hat{\sigma}_7^y\hat{\sigma}_8^y\hat{\sigma}_8^z\hat{\sigma}_1^z
=-\hat{\sigma}_1^x\hat{\sigma}_2^x\hat{\sigma}_3^x\hat{\sigma}_4^x\hat{\sigma}_5^x\hat{\sigma}_6^x\hat{\sigma}_7^x\hat{\sigma}_8^x,\\
\hat{W}_{\square_y}&=\hat{\sigma}_1^z\hat{\sigma}_2^z\hat{\sigma}_2^x\hat{\sigma}_3^x\hat{\sigma}_3^z\hat{\sigma}_4^z\hat{\sigma}_4^x\hat{\sigma}_5^x\hat{\sigma}_5^z\hat{\sigma}_6^z\hat{\sigma}_6^x\hat{\sigma}_7^x\hat{\sigma}_7^z\hat{\sigma}_8^z\hat{\sigma}_8^x\hat{\sigma}_1^x
=-\hat{\sigma}_1^y\hat{\sigma}_2^y\hat{\sigma}_3^y\hat{\sigma}_4^y\hat{\sigma}_5^y\hat{\sigma}_6^y\hat{\sigma}_7^y\hat{\sigma}_8^y,\\
\hat{W}_{\octagon}&=\hat{\sigma}_1^x\hat{\sigma}_2^x\hat{\sigma}_2^y\hat{\sigma}_3^y\hat{\sigma}_3^x\hat{\sigma}_4^x\hat{\sigma}_4^y\hat{\sigma}_5^y\hat{\sigma}_5^x\hat{\sigma}_6^x\hat{\sigma}_6^y\hat{\sigma}_7^y\hat{\sigma}_7^x\hat{\sigma}_8^x\hat{\sigma}_8^y\hat{\sigma}_1^y
=-\hat{\sigma}_1^z\hat{\sigma}_2^z\hat{\sigma}_3^z\hat{\sigma}_4^z\hat{\sigma}_5^z\hat{\sigma}_6^z\hat{\sigma}_7^z\hat{\sigma}_8^z.
\end{align}
Next, we map these operators to the spin-boson Hilbert space using the definitions (\ref{spinbosonops}). In the following, $R=1,\dots,8$ once again denote sites of the $(8,4,8,4)$ lattice, as in Eqs.~(\ref{Heff_x}-\ref{Heff_z}). Since the A/B sublattices alternate as we go around each octagon, we obtain
\begin{align}
\hat{W}_{\square_x}&=-\hat{\sigma}_{1,A}^x\hat{\sigma}_{1,B}^x\hat{\sigma}_{2,A}^x\hat{\sigma}_{2,B}^x\hat{\sigma}_{3,A}^x\hat{\sigma}_{3,B}^x\hat{\sigma}_{4,A}^x\hat{\sigma}_{4,B}^x=-\prod_{R\in\square_x}\hat{\tau}_R^x,\label{WL1}\\
\hat{W}_{\square_y}&=-\hat{\sigma}_{1,A}^y\hat{\sigma}_{1,B}^y\hat{\sigma}_{2,A}^y\hat{\sigma}_{2,B}^y\hat{\sigma}_{3,A}^y\hat{\sigma}_{3,B}^y\hat{\sigma}_{4,A}^y\hat{\sigma}_{4,B}^y=-(-1)^{\sum_{R\in\square_y}\hat{Q}_R}\prod_{R\in\square_y}\hat{\tau}_R^x,\\
\hat{W}_{\octagon}&=-\hat{\sigma}_{1,A}^z\hat{\sigma}_{2,B}^z\hat{\sigma}_{3,A}^z\hat{\sigma}_{4,B}^z\hat{\sigma}_{5,A}^zy\hat{\sigma}_{6,B}^z\hat{\sigma}_{7,A}^z\hat{\sigma}_{8,B}^z=-(-1)^{\sum_{R=2,4,6,8}\hat{Q}_R}\prod_{R\in\octagon}\hat{\tau}_R^z.\label{WL3}
\end{align}
Here, $\hat{Q}_R\equiv\hat{b}_R^\dag\hat{b}_R$ is the boson number operator on site $R$. We see that $\hat{W}_{\square_y}$ involves the total boson number on plaquette $\square_y$, while $\hat{W}_{\octagon}$ involves only the sum of boson numbers on every other site of plaquette $\octagon$. After projecting to the boson vacuum, we see that the Wilson loop operators (\ref{WL1}-\ref{WL3}) become precisely the plaquette operators appearing in the effective Hamiltonian (\ref{Heff3}), such that we can write
\begin{align}\label{Heff4}
\mathcal{\hat{H}}_\text{eff}=\frac{5}{16}\frac{J_y^4}{J_z^3}\sum_{\square_x}\hat{W}_{\square_x}
+\frac{5}{16}\frac{J_x^4}{J_z^3}\sum_{\square_y}\hat{W}_{\square_y}
+\frac{5}{2048}\frac{J_x^4J_y^4}{J_z^7}\sum_{\octagon}\hat{W}_{\octagon}.
\end{align} 
Note that all the (projected) plaquette operators commute with each other.\footnote{This model can thus be viewed as a hyperbolic analog of the Wen plaquette model~\cite{Wen:2003}.} We make two observations. Firstly, since all the couplings in \cref{Heff4} are positive, the ground state is obtained by setting all plaquette operators $\hat{W}_P$ to $-1$, which is consistent with the argument based on Lieb's lemma in the Letter. Secondly, setting $J_x=J_y\equiv J_\parallel\ll J_z$ for simplicity, we see that the lowest-energy excitation is a $\mathbb{Z}_2$ vortex on a single $\octagon$ plaquette ($\hat{W}_{\octagon}=+1$), which costs an energy $E_{\octagon}/J_z\propto(J_\parallel/J_z)^8$ that is much lower than a vortex excitation on a square plaquette, $E_\square/J_z\propto(J_\parallel/J_z)^4$.

Finally, we show that the Hamiltonian (\ref{Heff4}) can be mapped to a hyperbolic surface code on the $\{8,4\}$ lattice [\cref{fig:anis:b}], i.e., a hyperbolic analog of Kitaev's toric code on the square lattice~\cite{Kitaev:2003}. To achieve this, we simply view the $(8,4,8,4)$ lattice as the medial lattice of the $\{8,4\}$ lattice (i.e., the lattice obtained by placing sites at the mid-points of the edges of the original lattice). For simplicity, we also set $J_x=J_y\equiv J_\parallel$ as in the previous paragraph, which allows us to ignore the difference between $x$-bonds and $y$-bonds. Then, we can reinterpret the Hamiltonian (\ref{Heff4}) as a model on the $\{8,4\}$ lattice,
\begin{align}\label{toriccode}
\hat{\mathcal{H}}_\text{eff}=-J_e\sum_v \hat{A}_v-J_m\sum_p \hat{B}_p,
\end{align}
where $J_e=5J_\parallel^4/(16J_z^3)$ and $J_m=5J_\parallel^8/(2048J_z^7)$ are positive coupling constants, the sum over $v$ ($p$) is over all vertices (plaquettes/faces) of the $\{8,4\}$ lattice, and the (commuting) $\hat{A}_v$ and $\hat{B}_p$ operators are defined as
\begin{align}
\hat{A}_v=\prod_{\ell\in +_v}\hat{\tau}_\ell^x,\hspace{5mm}
\hat{B}_p=\prod_{\ell\in\partial p}\hat{\tau}_\ell^z,
\end{align}
with $+_v$ denoting the ``star'' of $v$ (i.e., the four edges incident on $v$) and $\partial p$ the perimeter of the octagonal plaquette $p$~\cite{Kitaev:2003}. We note in passing that in the context of quantum information processing, such hyperbolic surface codes have been the object of much interest in recent years~\cite{Breuckmann:2016,Breuckmann:2017,Lavasani:2019,Jahn:2021,Higgott:2023,Fahimniya:2023}.

\begin{figure}[t!]
\centering
    \includegraphics[width=0.4\linewidth]{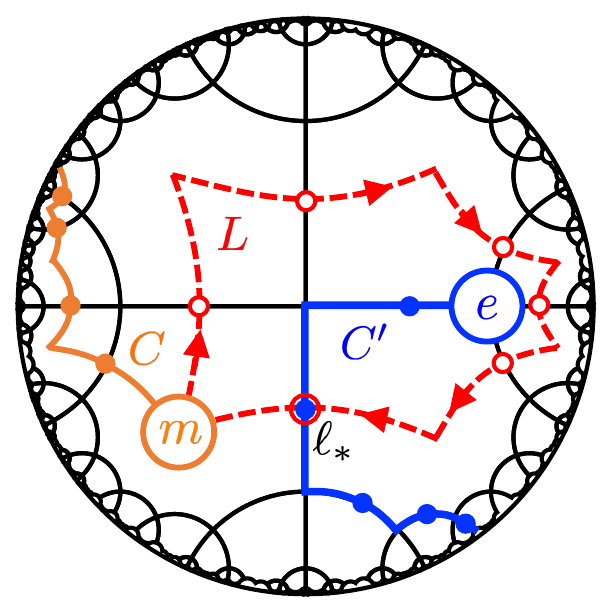}
    \caption[]{Mutual anyonic statistics in the $\{8,4\}$ hyperbolic code. A $\mathbb{Z}_2$ flux $m$ (charge $e$) is initially created by a string operator $\hat{X}$ ($\hat{Z}$) supported on the links (solid dots) crossed by the semi-infinite contour $C$ ($C'$). The flux is then adiabatically braided around the charge along the closed contour $L$, which is implemented by a loop operator $\hat{X}'$ with support on the red open circles. The contour $L$ must cross the ``Dirac string'' $C'$ an odd number of times (here once, on link $\ell_*$), resulting in an overall phase factor of $-1$ which indicates mutual semionic statistics.}
\label{fig:braiding}
\end{figure}

As in the original toric code, the hyperbolic code (\ref{toriccode}) can be viewed as a fixed-point Hamiltonian for the deconfined phase of a $\mathbb{Z}_2$ gauge theory on the $\{8,4\}$ lattice, with $\hat{\tau}_\ell^z$ and $\hat{\tau}_\ell^x$ corresponding to the $\mathbb{Z}_2$ gauge and electric fields on link $\ell$, respectively. The ground state has $\hat{A}_v=+1$ for all $v$ and $\hat{B}_p=+1$ for all $p$. The low-energy excitations are static $\mathbb{Z}_2$ charges (``$e$ particles'') with $\hat{A}_v=-1$ on some vertex $v$, that cost energy $2J_e$, and static $\mathbb{Z}_2$ fluxes or vortices (``$m$ particles'') with $\hat{B}_p=-1$ on some plaquette $p$, that cost energy $2J_m\ll 2J_e$. Fluxes and charges obey bosonic self-statistics but are mutual semions, which can be seen as follows~\cite{Vidal:2008}. Assuming an infinite lattice, consider a nonlocal operator $\hat{X}$ defined as the product of $\hat{\tau}^x_\ell$ operators over all links $\ell$ crossed by a semi-infinite contour $C$ living on the dual lattice (orange contour in \cref{fig:braiding}):
\begin{align}
\hat{X}=\prod_{\ell\in C}\hat{\tau}^x_\ell.
\end{align}
When applied to the ground state $\ket{\text{GS}}$ of Hamiltonian (\ref{toriccode}), this operator flips the sign of $\hat{\tau}^z_\ell$ on every link crossed, thus it creates a single $m$ flux ($\hat{B}_p=-1$) on the plaquette $p$ where $C$ terminates. If we create two such excitations on different plaquettes, their corresponding string operators obviously commute (being made of products of only $\hat{\tau}^x$ operators), thus the $m$ particles have bosonic self-statistics. Likewise, we define a nonlocal operator $\hat{Z}$ as the product of $\hat{\tau}^z_\ell$ operators over all links $\ell$ traversed by a semi-infinite contour $C'$ living on the $\{8,4\}$ lattice (blue contour in \cref{fig:braiding}):
\begin{align}
\hat{Z}=\prod_{\ell\in C'}\hat{\tau}^z_\ell.
\end{align}
This operator flips the sign of $\hat{\tau}^x_\ell$ on every link traversed, thus it creates a single $e$ charge ($\hat{A}_v=-1$) on the vertex $v$ where $C'$ terminates. Again, the string operators creating two separate $\mathbb{Z}_2$ charges commute, thus those particles also obey bosonic self-statistics.

Now, begin with the state $\ket{\Psi}=\hat{Z}\hat{X}\ket{\text{GS}}$ containing both $m$ and $e$ particles and consider braiding $m$ around $e$ along a closed contour $L$ (dashed red contour in \cref{fig:braiding}). Moving $m$ along this contour is accomplished by the operator
\begin{align}
\hat{X}'=\prod_{\ell\in L}\hat{\tau}^x_\ell,
\end{align}
where the product is now only over the red open circles in \cref{fig:braiding}. We obtain the state $\ket{\Psi'}=\hat{X}'\hat{Z}\hat{X}\ket{\text{GS}}$, but $\hat{X}'\hat{Z}=-\hat{Z}\hat{X}'$ because there is necessarily one link $\ell_*$ (more generally, an odd number of links) where $L$ and $C'$ cross and thus where $\hat{\tau}^x_{\ell_*}$ from $\hat{X}'$ and $\hat{\tau}^z_{\ell_*}$ from $\hat{Z}$ must be anticommuted. Thus, we obtain $\ket{\Psi'}=-\hat{Z}\hat{X}\hat{X}'\ket{\text{GS}}$, using also that $\hat{X}$ and $\hat{X}'$ obviously commute. Finally, $\hat{X}'$ is the product over all $\hat{\tau}^x_\ell$ operators crossed by $L$, and is thus also equal to the product of $\hat{A}_v$ over all vertices $v$ enclosed by $L$.\footnote{This can be viewed as a $\mathbb{Z}_2$ analog of the divergence theorem: $\hat{X}'$ is the net electric flux through $L$, and $\prod_{v\text{ inside }L}\hat{A}_v$ is the volume integral of the divergence of the electric field. By Gauss' law, both give the total $\mathbb{Z}_2$ charge enclosed by $L$.} (Since all links internal to $L$ appear twice in this product, only links on the perimeter remain.) Since $\hat{A}_v=+1$ for all $v$ in the ground state, we have $\hat{X}'\ket{\text{GS}}=\ket{\text{GS}}$, and thus $\ket{\Psi'}=-\ket{\Psi}$ independent of the detailed shape of the contours $C,C',L$, apart from their intersections. Thus, an adiabatic phase of $-1$ is obtained upon braiding an $m$ particle around an $e$ particle, indicating mutual semionic statistics. We conclude that the gapped phase adiabatically connected to the anisotropic coupling limit $J_x,J_y\ll J_z$ is a gapped spin liquid with $\mathbb{Z}_2$ topological order. For recent studies of hyperbolic lattice models displaying $\mathbb{Z}_2$ or other types of topological order (e.g., fracton order), also see Refs.~\onlinecite{Ebisu:2022,Yan:2019,Yan:2022,Yan:2023}.

Finally, a moment's thought reveals that the above analysis can be straightforwardly generalized to hyperbolic $\{2m,3\}$ lattices for any $m\geq 4$, assuming the three-edge coloring discussed in \cref{SM:symmetric-3-coloring}. One first obtains a spin-boson model on the Archimedean $(2m,m,2m,m)$ lattice, which is then projected onto a hyperbolic surface code on the $\{2m,m\}$ lattice.

\labeledsection{Computation of the real-space Chern number}{%
Technical details on the computation of the projector-based real-space Chern number plotted in \cref{main:fig:phase-diagram_K=0.1}, and analysis of convergence.%
}\label{SM:Chern}

\begin{figure}[t]
    \subfloat{\label{fig:chern_region-def}}
    \subfloat{\label{fig:chern_regionsize-dep}}
    \centering
    \includegraphics{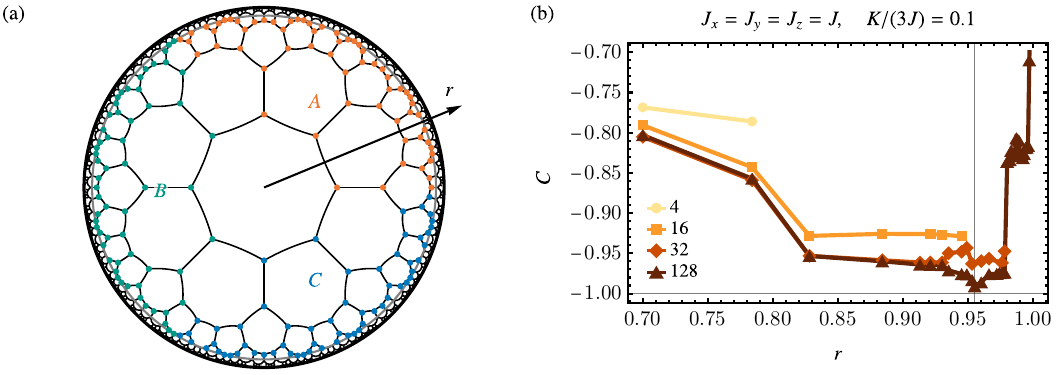}
    \caption{
        Computation of the real-space Chern number.
        (a) $128$-primitive-cell PBC cluster $\Gamma_{\tgquot{33}{1}\cap\tgquot{65}{1}}$ (edges shown in black) and the three regions $A,B,C$ (colored sites) up to a bounding radius of $r=0.955$ (gray circle).
        (b) Real-space Chern number $C$ at the isotropic point for $K\neq 0$ as a function of bounding radius $r$ for different PBC clusters; their size in terms of number of primitive cells is given in the inset legend.
        Note that sites within the bounding circle that have nearest-neighbor bonds that connect to the other side of the projection are discarded.
        The Chern number converges closer to the true value with increasing region size up to a point from which on the result gets poisoned by the periodic boundary conditions.
    }
    \label{fig:chern:regions}
\end{figure}

In this section, we discuss the computation of the Chern number for $K\neq 0$ as performed to obtain the data shown in \cref{main:fig:phase-diagram_K=0.1}.
We use the real-space Chern number~\cite{Kitaev:2006}, which directly applies to any two-dimensional system, including hyperbolic lattices~\cite{Urwyler:2022,Liu:2022,Chen:2023b} and even amorphous media~\cite{Mitchell:2018}.
Given a projector $P$ and three regions $A,B,C$ of sites arranged counterclockwise as depicted in \cref{fig:chern_region-def}, it is defined as
\begin{equation}
    C = 12\pi\i\sum_{j\in A}\sum_{k\in B}\sum_{l\in C}\left(P_{jk}P_{kl}P_{lj}-P_{jl}P_{lk}P_{kj}\right).
    \label{eq:Chern-number}
\end{equation}
While this is often applied to systems with open boundary conditions, we here perform the calculation on PBC clusters.
In the former case, the regions $A,B,C$ have to be chosen sufficiently far away from the boundary, to avoid boundary effects, which would compensate the bulk and result in a trivial Chern number.
On PBC clusters this is similar: we have to constrain the regions such that they are sufficiently separated and do not wrap around the higher-genus surface on which the PBC cluster is embedded.

To select these regions, we work in the Poincaré disk representation and use the fact that any of the supercells described in \cref{SM:SC:sequence} can be interpreted as a finite PBC cluster~\cite{Lenggenhager:PhDThesis} by considering the quotient $\Gamma/\Gamma^{(n)}$ and then following \cref{SM:CF-PBC}.
Starting from the symmetric supercell constructed using the \textsc{HyperCells} package~\cite{HyperCells}, this allows us to immediately obtain a symmetric \emph{projection} of the PBC cluster onto the Poincaré disk.
When defining the regions $A,B,C$, we only include sites lying within a given bounding radius $r$ and explicitly exclude sites that have nearest-neighbor bonds that connect to the other side of the projection.
\Cref{fig:chern_region-def} shows this construction for the $128$-primitive cell PBC cluster defined by $\Gamma_{\tgquot{33}{1}\cap\tgquot{65}{1}}$ given in \cref{eq:supercell-5}.
A priori, it is not clear which $r$ is optimal for approximating the Chern number in the thermodynamic limit, such that we study the Chern number as a function of $r$ on different PBC clusters and choose the maximal $r$ up to which $C(r)$ is monotonic, see \cref{fig:chern_regionsize-dep}.
For the largest PBC cluster, we find an optimal bounding radius of $r=0.955$ (gray vertical line).

To evaluate \cref{eq:Chern-number}, we need the projector onto the occupied subspace, which we obtain by exact diagonalization of the model Hamiltonian $\i A$ on the given PBC cluster.
With eigenstates $\ket{n}$ and eigenenergies $\varepsilon_n\neq 0$ (assuming a spectral gap such that the Chern number is well-defined), the projector is
\begin{equation}
    P = \sum_{n:\varepsilon_n<0}\op{n}{n}.
\end{equation}
Its matrix elements $P_{jk}=\mel*{z_j}{P}{z_k}$ in the real-space basis $\ket*{z_j}$ with $z_j$ the position of site $j$ in the Poincaré disk then enter \cref{eq:Chern-number}.
In practice, we can obtain this projector from the spectrally flattened Hamiltonian
\begin{equation}
    B = -\i\,\mathrm{sgn}(\i A),
\end{equation}
where $\mathrm{sgn}$ only acts on the eigenvalues,
as
\begin{equation}
    P = \frac{1}{2}(1 - \i B).
\end{equation}

\begin{figure}[t]
    \subfloat{\label{fig:chern_convergence}}
    \subfloat{\label{fig:chern_phase-diagram-cut}}
    \centering
    \includegraphics{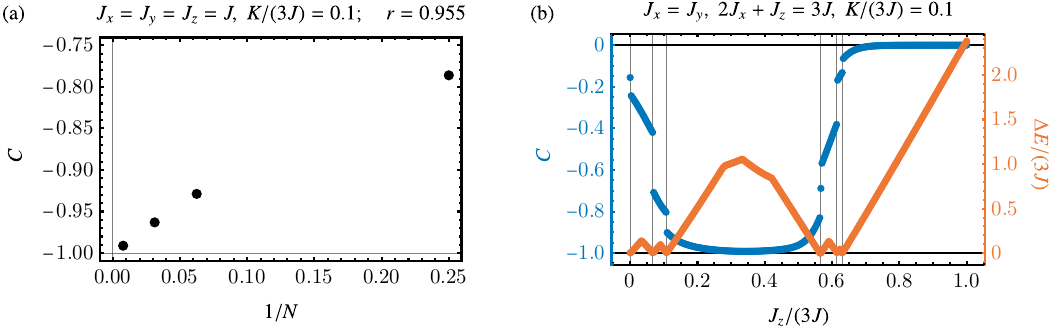}
    \caption{
        Results on the real-space Chern number.
        (a) Convergence with size $N$ (in terms of number of primitive cells) of the PBC cluster at the isotropic point with $K/(3J)=0.1$ and bounding radius $0.955$. The value approaches $-1$.
        (b) Cut through the phase diagram of the Chern number $C$ along $J_x=J_z$ as a function of $J_z$ for $K/(3J)=0.1$ obtained on the $128$-primitive-cell PBC cluster shown in \cref{fig:chern_region-def}.
        In the thermodynamic limit, the Chern number (blue, left axis) is quantized to integers (horizontal black lines at $0,-1$) as long as there is a gap $\Delta E$.
        Due to finite-size effects, there are deviations from the integer values in regions where $\Delta E$ is small.
        More precisely, the closing of the thermodynamic gap is replaced by multiple closings (indicated by vertical gray lines) of the finite-size gap (orange, right axis) resulting in several jumps of $C$ with noninteger values.
    }
    \label{fig:chern:results}
\end{figure}

At the isotropic point $J_x=J_y=J_z=J$, $K/(3J)=0.1$, i.e., the choice of $K$ in \cref{main:fig:phase-diagram_K=0.1} in the Letter, and with the optimal radius $r=0.955$, we find that $C$ converges to $-1$ with increasing supercell size $N$, see \cref{fig:chern_convergence}.
\Cref{fig:chern_phase-diagram-cut} shows a cut through the phase diagram shown in \cref{main:fig:phase-diagram_K=0.1} along $J_x=J_z$.
Besides the Chern number (blue, left axis), we also show the finite-size gap (orange, right axis).
We recognize that the washed-out phase boundary pointed out in the Letter is caused by several jumps of $C$ with intermediate noninteger values.
These jumps occur precisely when the finite-size gap closes (gray vertical lines) and correspond to the transfer of individual states from negative to positive energy or vice versa.

\labeledsection{Extraction of the chiral edge states' dispersion}{%
Technical details on the method for extracting the angular-momentum dispersion and edge weight of the chiral edge states shown in \cref{main:fig:edge-dispersion}, introduced in \cref{App:chiral-edge-states}.%
}

\begin{figure}[t]
    \subfloat{\label{fig:DOS-no-flux}}
    \subfloat{\label{fig:localization-no-flux}}
    \subfloat{\label{fig:DOS-with-flux}}
    \subfloat{\label{fig:localization-with-flux}}
    \subfloat{\label{fig:DOS-trivial-gapped}}
    \subfloat{\label{fig:localization-trivial-gapped}}
    \centering
    \includegraphics[width=1.00\textwidth]{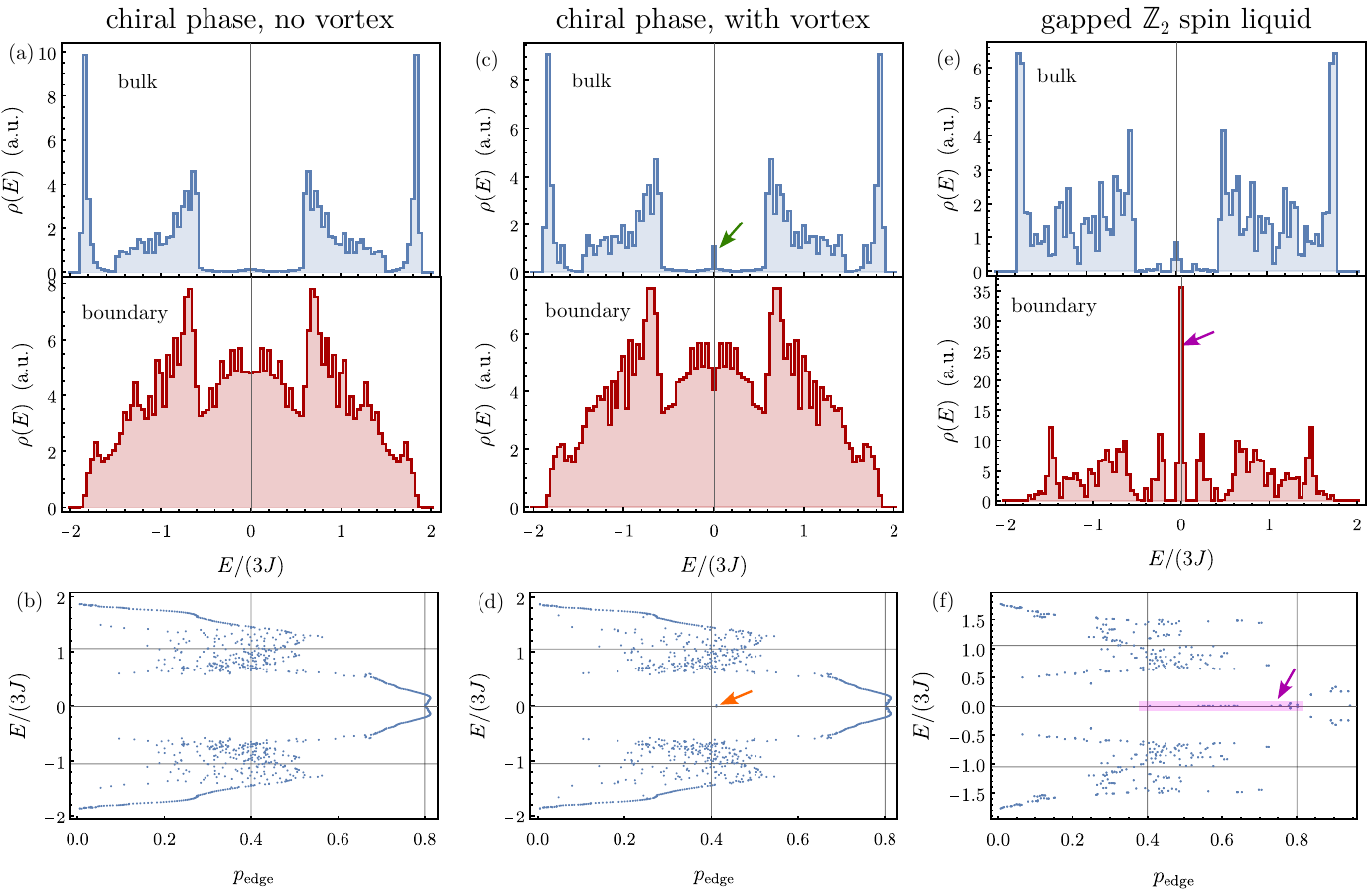}
    \caption{
        (a)~Density of states (DOS) $\rho(E)$ of the chiral gapped phase on a disk with $896$ sites and open boundary conditions, with model parameters set to $J:=J_{x,y,z}=\tfrac{1}{3}$ and $K=\tfrac{1}{10}$, plotted in arbitrary units (a.u.).
        The ``bulk DOS'' (shown in blue) is obtained by summing the local DOS over sites located within the inner $65\,\%$ of the hyperbolic distance to the disk boundary, whereas in ``boundary DOS'' (shown in red) we sum over sites within the outer $10\,\%$ of the hyperbolic distance to the disk boundary. 
        (b) Each eigenstate is represented as a dot with coordinates $(p_\textrm{edge},E)$, where $p_\textrm{edge}$ is a measure of edge localization defined in \cref{eqn:p-edge}.
        Edge states are recognized as the arc of dots on the right side of the plot.
        Vertical grid lines indicate the values at which we saturate the color scheme in \cref{main:fig:edge-dispersion} in the Letter and in \cref{fig:Edge-dispersions}, and the horizontal grid lines indicate the energy range adopted in said figures.
        (c,d) Analogous data in the presence of a $\mathbb{Z}_2$ vortex threaded through the plaquette located at the center of the disk. 
        The vortex binds a zero-energy Majorana mode, resulting in a small peak in the bulk DOS at $E=0$ [indicated with green arrow in panel (c)].
        Due to finite-size effects, the Majorana mode at the vortex hybridizes with a zero-energy Majorana mode at the boundary, thus acquiring an enhanced value of $p_\textrm{edge}$ [indicated with orange arrow in panel (d)] and a small finite energy $E/(3J)\approx \pm 0.0138$.
        (e,f) For comparison, we also show the analogous data for the gapped Abelian $\mathbb{Z}_2$ spin liquid, which carries vanishing Chern number, for parameters $J_x=J_y=1/4$, $J_z=1/2$, and $K=0.$
        We observe that the gapped $\mathbb{Z}_2$ spin liquid phase also exhibits edge states within the bulk energy gap, including a flat band at $E=0$ (indicated with purple arrow). 
        We find that these zero-energy boundary states have a large localization length, making the $E=0$ peak also visible in the numerically computed bulk DOS.
        However, we demonstrate in \cref{fig:trivial-edge-states} that edge states of the trivial gapped phase do not exhibit chiral propagation, in accordance with the vanishing Chern number.
    }
    \label{fig:DOS-bulk-vs-boundary}
\end{figure}

In this section, we clarify the method used to extract the angular dispersion of the edge state, displayed in \cref{main:fig:edge-dispersion} of the main~text. 
When an open boundary is introduced in the chiral gapped ($\chi$) phase, owing to the nonzero Chern number, we expect the model should develop gapless edge states inside the bulk energy gap. 
We illustrate the appearance of such states explicitly in \cref{fig:DOS-no-flux} by comparing the density of states (DOS) in the bulk vs on the boundary of a disk-shaped sample.
To further manifest the boundary character of the in-gap states, we compute for each eigenstate $\ket{\psi_n}$ with energy $E_n$ the quantity:
\begin{equation}
p_{n,\textrm{edge}}=\sum_{j\in\textrm{edge}} \abs{\psi_n(j)}^2,
\label{eqn:p-edge}
\end{equation}
where by ``$j\,{\in}\,\textrm{edge}$'' we mean sites located within the outer $10\,\%$ of the hyperbolic distance to the disk boundary, and $\psi_n(j)=\braket{j}{\psi_n}$. 
Similarly, we define the bulk sites ``$j\,{\in}\, \textrm{bulk}$'' as those located within the inner $65\,\%$ of the hyperbolic distance to the disk boundary.
By representing each state $\ket{\psi_n}$ as a point with coordinates $(p_{\textrm{ed3ge},n},E_n)$ in \cref{fig:localization-no-flux}, we recognize that all states with energy $\abs{E}/(3J)\lesssim 0.6$ exhibit a significantly larger value of $p_{\textrm{edge}}$ than the states at larger values of $\abs{E}$.
All calculations in this section are performed with a system containing all $896$ sites of the $\{8,3\}$ lattice located within $99.14\,\%$ of the radius of the Poincar\'e disk  (the same as used in Ref.~\onlinecite{Urwyler:2022} when extracting the chiral edge mode dispersion of a hyperbolic Haldane model).
To realize the chiral gapped phase, we set the model parameters to $J:=J_{x,y,z}=\tfrac{1}{3}$ and $K=\tfrac{1}{10}$.

Near zero energy, we expect the topological edge mode to exhibit an approximately linear dispersion $E\propto \ell$, where $\ell$ is angular momentum~\cite{Read:2000,stone2004}.
However, owing to the discreteness of the underlying $\{8,3\}$ lattice, angular momentum of the individual eigenstates is well-defined only modulo the order of the rotation symmetry, i.e., mod $8$, implying an impractically narrow edge Brillouin zone. 
To obtain the dispersion over an extended range of angular momenta ($-150\leq \ell \leq 150$ in \cref{main:fig:phase-diagram_K=0.1} in the Letter), we need to devise a physically motivated approximation that treats the discrete lattice as a continuum.

\begin{figure}[t]
    \subfloat{\label{fig:phase-growth-no-flux}}
    \subfloat{\label{fig:phase-growth-with-flux}}
    \centering
    \includegraphics[width=\textwidth]{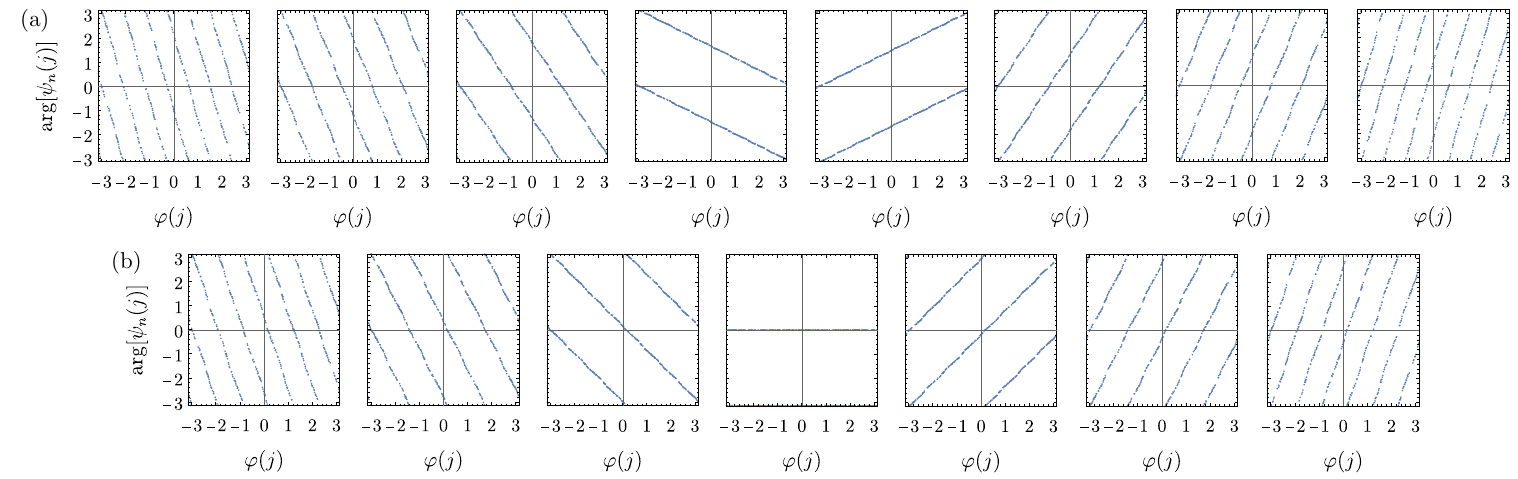}
    \caption{
        (a)~Plots showing the linear growth of the phase $\arg[\psi_n(j)]$ with the angular coordinate $\varphi(j)$ of site $j$ for several eigenstates $\ket{\psi_n}$ with near-zero energy.
        Each plot correspond to a single normalized eigenstate, and the energy of the eigenstates grows from negative (on the left) to positive (on the right). 
        We display data points only for sites $j$ with $\abs{\psi_n(j)}^2>1/(2N)$ where $N=896$ is the number of sites in the disk. 
        The observed linear slopes imply that the chiral edge states carry half-integer angular momentum, $\ell \in \mathbb{Z}+\tfrac{1}{2}$.
        (b)~Analogous data in the presence of a $\mathbb{Z}_2$ vortex threaded through the central plaquette and with a perturbation that shifts the Majorana mode at the vortex to energy $E/(3J)\approx 1$.
        Ins this case, the angular momentum of the chiral edge states is shifted to integer values, $\ell \in \mathbb{Z}$.        
        }
    \label{fig:edge-state-phase-growth}
\end{figure}

Recall that the angular momentum operator is given by $\hat\ell=-\i\partial_\varphi$ where $\varphi$ is the angle in polar coordinates. 
On a disk with continuous $\mathrm{SO}(2)$ rotation symmetry, angular momentum is a well-defined quantum number. 
The complex phase of the eigenstates of $\hat{\ell}$ must grow linearly with $\varphi$, i.e., they are of the form $\ket{\psi_{(\ell)}} \propto \e^{\i \ell \varphi}$, where $\ell$ is the angular momentum eigenvalue.
To verify that this property also holds with good accuracy for the edge states on the discrete graph, we explicitly plot the dependence of the complex phase $\arg[{\psi_n(j)}]$ on the angular coordinate $\varphi(j)$ of sites $j$ for several eigenstates $\ket{\psi_n}$ with energy $E_n$ close to zero. 
The result, shown in \cref{fig:phase-growth-no-flux}, confirms that the complex phase exhibits a well-defined slope corresponding to a half-integer angular momentum, $\ell \in \mathbb{Z}+\tfrac{1}{2}$, as expected for a chiral Majorana edge mode~\cite{Read:2000,stone2004}. The data in \cref{fig:phase-growth-no-flux} also suggests that the slope $\ell$ grows with energy $E$ for the Majorana edge modes near $E=0$.

\begin{figure}[t]
    \subfloat{\label{fig:Edge-dispersion-1}}
    \subfloat{\label{fig:Edge-dispersion-2}}
    \subfloat{\label{fig:Edge-dispersion-3}}
    \centering
    \includegraphics{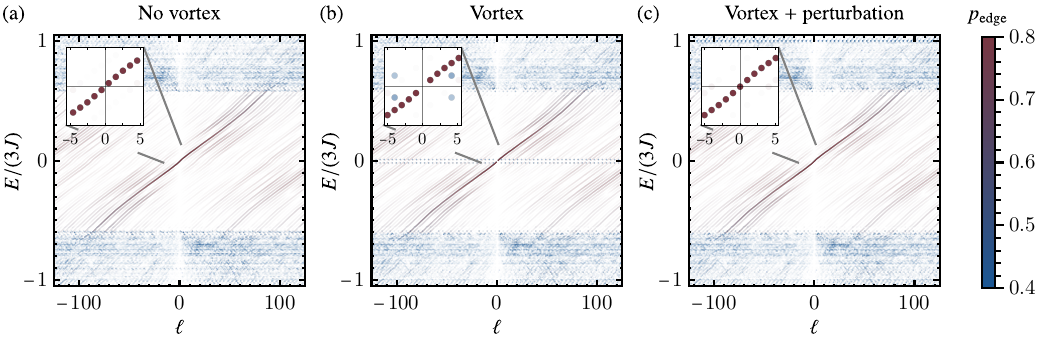}
    \caption{
        Edge state dispersion in the chiral gapped phase.
        For each eigenstate $\ket{\psi_n}$ with energy $E_n$ on the disk-shaped flake and for each angular momentum $\ell\in\tfrac{1}{2}\mathbb{Z}$, we plot at coordinates $(\ell,E_n)$ a data point whose color is set by the edge localization $p_{n,\textrm{edge}}$ [\cref{eqn:p-edge}, see color legend on the right] and whose opacity is given by the coefficient $c_{n,\ell}$ [Eq.~(\ref{eqn:c-n-ell-coeffs})].
        Since most coefficients are close to zero, the vast majority of the data points are fully transparent and thus not visible.
        Only those data points $(\ell,E_n)$ remain visible whose coefficient $c_{n,\ell}\in[0,1]$ is sufficiently larger than zero. 
        (a) In the absence of a vortex, we identify a chiral Majorana mode  localized at the boundary (displayed in red tones) at energies $\abs{E}/(3J)\lesssim 0.6$.
        The inset reveals that the edge states carry half-integer angular momentum $\ell\in\mathbb{Z}+\tfrac{1}{2}$, as expected for chiral Majorana edge modes~\cite{Read:2000,stone2004} and in accordance with the phase growth data in \cref{fig:phase-growth-no-flux}.
        The bulk states generate a featureless signal at $\abs{E}/(3J)\gtrsim 0.6$ (displayed in blue tones).
        (b)~In the presence of a $\mathbb{Z}_2$ vortex threaded through the plaquette at the center of the disk, the angular momentum of the Majorana edge modes is shifted to integer values. 
        In addition, the vortex binds an additional Majorana mode whose coefficients $c_{n,\ell}$ exhibit large values at all $\ell=0 \;(\textrm{mod\,$4$})$. 
        Due to finite-size effects, the Majorana modes expected at zero energy hybridize, resulting in two eigenstates at energy $E/(3J)\approx \pm 0.0138$ and with angular momentum $\ell = 0 \;(\textrm{mod\,$4$})$.
        (c)~To remove the undesired hybridization, we apply the perturbation in Eq.~(\ref{eqn:with-vortex-with-perturbation}) which shifts the Majorana mode at the vortex to energy $E/(3J)~\approx 1$. 
        In this case, we observe an approximately linear dispersion $E(\ell)$ with integer-valued angular momenta $\ell\in\mathbb{Z}$, as expected~\cite{Read:2000,stone2004} and in accordance with~\cref{fig:phase-growth-with-flux}.
        }
    \label{fig:Edge-dispersions}
\end{figure}

To establish the linear dispersion $E(\ell)$ of the edge states over a wider energy range, we devise the following scheme to estimate the angular momentum of the individual eigenstates. 
Assuming that $\ell \in \mathbb{Z}\cup (\mathbb{Z}+\tfrac{1}{2}) \cong \tfrac{1}{2}\mathbb{Z}$, we compute for each normalized eigenstate $\ket{\psi_n}$ and each $\ell\in\tfrac{1}{2}\mathbb{Z}$ the coefficient 
\begin{equation}
c_{n,\ell}=\bigg\lvert\sum_j \e^{-\i 2 \ell\varphi(j)}\psi_n^2(j)\bigg\rvert \in [0,1],\label{eqn:c-n-ell-coeffs}
\end{equation}
which we interpret as the likelihood that the state $\ket{\psi_n}$ carries angular momentum $\ell$. 
This is motivated by the fact that for the exact eigenstate $\ket{\psi_{(\ell')}} \propto \e^{\i \ell' \varphi}$ localized on the boundary of a continuous disk, an analogous integration gives $c_{(\ell'),\ell} = \delta_{\ell',\ell}$.
The result of this analysis, plotted in \cref{fig:Edge-dispersion-1} (also \cref{main:fig:edge-dispersion} in the Letter), reveals that the edge states (displayed in red tones) build up a chiral mode with half-integer angular momentum eigenvalues in the range $-90 \lesssim \ell \lesssim 90$. 
In addition, owing to the discreteness of the lattice, we observe several replicas of the chiral mode whose angular momentum is displaced by integer multiples of $4$ [that the ambiguity is mod~$4$ rather than mod~$8$ follows from the squaring of the wave function amplitude in \cref{eqn:c-n-ell-coeffs}]; nevertheless, the main branch passing through $(E,\ell)=(0,0)$ dominates in intensity over all replicas for all edge states. 
In addition, we clearly recognize that the bulk states (displayed in blue tones) do not carry a well-defined $\ell\in\tfrac{1}{2}\mathbb{Z}$, making the plotted data featureless at energies $\abs{E}/(3J)\gtrsim 0.6$. 

We next investigate the edge states of the chiral gapped phase in the presence of a $\mathbb{Z}_2$ vortex threaded through the plaquette at the center of the disk.
We observe in \cref{fig:DOS-with-flux,fig:localization-with-flux} that this results in a minimal but important change in the bulk and boundary DOS functions.
Namely, the bulk DOS develops a small peak at $E=0$ [indicated by the green arrow in panel~(c)] whereas the boundary DOS develops a correspondingly small dip at the same position.
Furthermore, the dispersion analysis, shown in \cref{fig:Edge-dispersion-2}, reveals that the angular momentum of the Majorana edge states has been shifted to integer values, $\ell \in \mathbb{Z}$, and that the $\ell=0$ Majorana edge state is missing.
These observations are explained by the formation of a Majorana mode bound to the vortex. 
If the vortex were sufficiently far from the boundary, we would expect a vortex-bound Majorana mode $\big\lvert{\psi_{0}^\textrm{vortex}}\big\rangle$ to occur at $E=0$.
Due to finite-size effects (for the adopted system size, the shortest path from the vortex to the boundary consists of only seven edges of the $\{8,3\}$ lattice) we expect the vortex-bound Majorana mode to hybridize with the $E=0,\ell=0$ Majorana state $\big\lvert{\psi_{0}^\textrm{edge}}\big\rangle$ on the boundary. 
The two resulting hybridized states $\big\lvert{\psi_{0}^{\textrm{mixed},1}}\big\rangle$ and $\big\lvert{\psi_{0}^{\textrm{mixed},2}}\big\rangle$ both have a large support on the bulk sites, and we observe them in \cref{fig:localization-with-flux} at $E/(3J)\approx \pm 0.0138$ and $p_\textrm{edge}\approx 0.42$ (indicated by the orange arrow). 
This shared bulk character also explains the dip observed in the boundary DOS in \cref{fig:DOS-with-flux}.
In addition, since the vortex-bound Majorana mode $\big\lvert{\psi_{0}^\textrm{vortex}}\big\rangle$ is strongly localized on a single octagonal plaquette at the center, its coefficients $c_{n,\ell}$ computed according to \cref{eqn:c-n-ell-coeffs} will exhibit strong peaks at all $\ell = 0 \;(\textrm{mod\,$4$})$.
This ill-defined value of the angular momentum is shared by the hybridized Majorana states $\big\lvert{\psi_{0}^{\textrm{mixed},1}}\big\rangle$ and $\big\lvert{\psi_{0}^{\textrm{mixed},2}}\big\rangle$ found in the numerics, as visible in the data near $E=0$ in \cref{fig:Edge-dispersion-2}.

\begin{figure}[t]
    \subfloat{\label{fig:bound-Majorana}}
    \subfloat{\label{fig:edge-state-evolution}}
    \subfloat{\label{fig:trivial-edge-states}}
    \centering
    \includegraphics[width=1.00\textwidth]{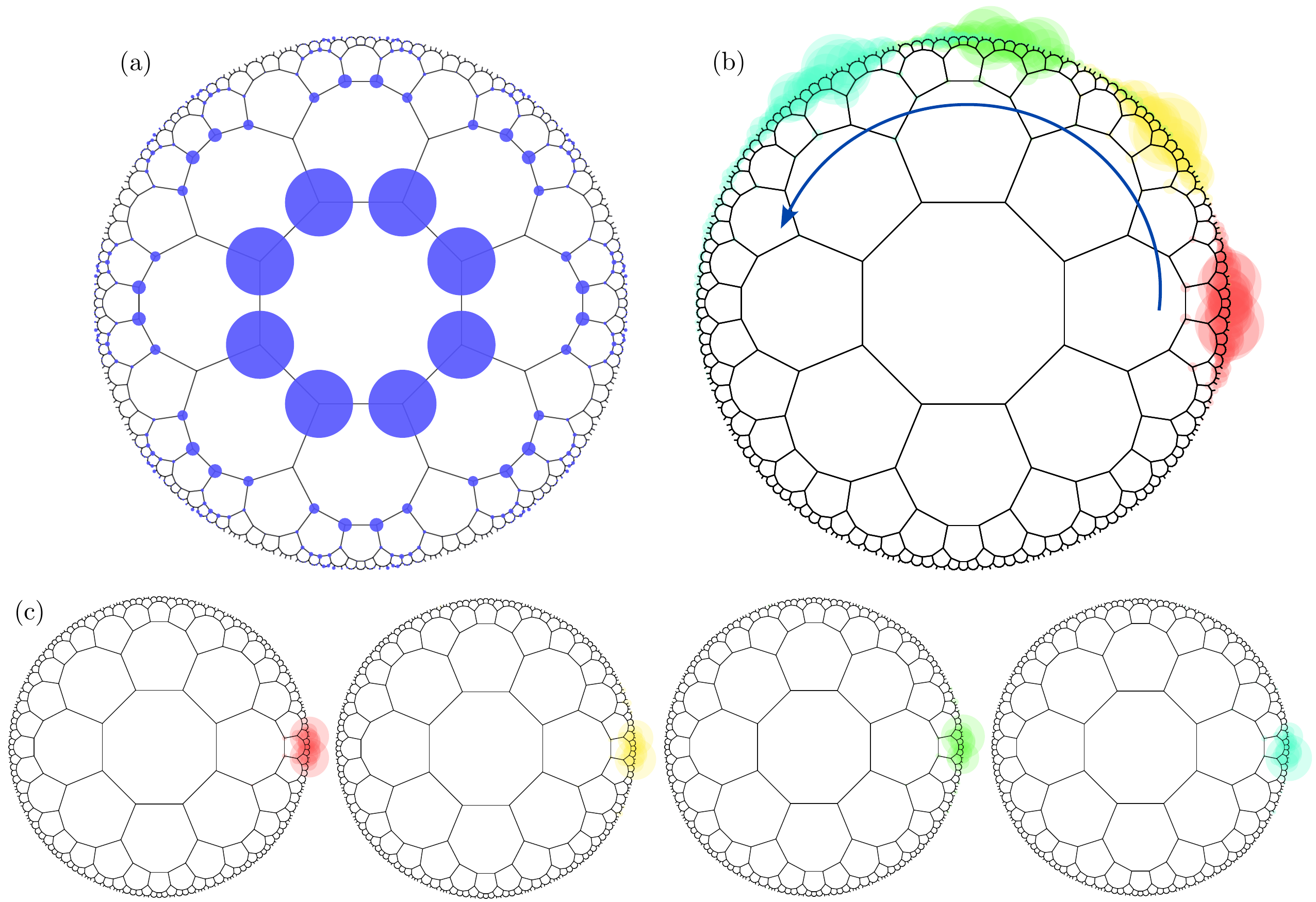}
    \caption{
        (a) Plot of the Majorana mode bound to the $\mathbb{Z}_2$ vortex threaded through the central plaquette, computed using the procedure described around \cref{eqn:mixing-E=0-Majoranas}.
        The area of the blue disk at site $j$ is proportional to the probability density $|\psi_0^{\textrm{unmixed},2}(j)|^2$.
        (b) Propagation of the Majorana wave packet with $\mu_E=0$ and $\sigma_E=0.06$ in the chiral gapped phase. 
        The red data show the initial wave packet as defined in \cref{eqn:wave-packet}, and the subsequent yellow/green/cyan data show snapshots of the wave packet at equally spaced later times $3JT\in\{100,200,300\}$, respectively. 
        The blue arrow indicates the counterclockwise propagation of the wave packet, with the direction fixed by $\partial E/\partial \ell > 0$.
        (c)~Propagation of the wave packet with $\mu_E=0$ and $\sigma_E=0.12$ in the gapped Abelian $\mathbb{Z}_2$ spin liquid phase with parameters $J_x = J_y = 1/4$, $J_z=1/2$ and $K=0$.
        The red data show the initial wave packet, and the subsequent yellow/green/cyan data display snapshots of the wave packet at equally spaced later times $3JT\in\{100,200,300\}$, respectively. 
        The wave packet evolution in this case exhibits no noticeable propagation nor broadening.
        }
    \label{fig:states-profiles}
\end{figure}

To remove these finite-size effects from the numerics, we apply a perturbation to the Hamiltonian as follows.
First, we numerically find linear combinations 
\begin{equation}
\label{eqn:mixing-E=0-Majoranas}
\left(\begin{array}{c}
\big\lvert{\psi_0^{\textrm{unmixed},1}}\big\rangle \\   
\big\lvert{\psi_0^{\textrm{unmixed},2}}\big\rangle
\end{array}\right) = M \left(\begin{array}{c}
\big\lvert{\psi_0^{\textrm{mixed},1}} \big\rangle\\   
\big\lvert{\psi_0^{\textrm{mixed},2}}\big\rangle
\end{array}\right)
\end{equation}
with $M\in\mathrm{U}(2)$ such that $\big\lvert{\psi_0^{\textrm{unmixed},1}}\big\rangle$ exhibits the maximal possible value of localization to the boundary $p_\textrm{edge}$ [\cref{eqn:p-edge}].
We interpret the constructed $\big\lvert{\psi_0^{\textrm{unmixed},1}}\big\rangle$ as $\big\lvert{\psi_0^\textrm{edge}}\big\rangle$ and the constructed orthogonal $\big\lvert{\psi_0^{\textrm{unmixed},2}}\big\rangle$ as $\big\lvert{\psi_0^\textrm{vortex}}\big\rangle$.
We plot the computed state $\big\lvert{\psi_0^{\textrm{unmixed},2}}\big\rangle\approx \big\lvert{\psi_0^\textrm{vortex}} \big\rangle$ in \cref{fig:bound-Majorana}.
Then we perturb the Hamiltonian by adding a term proportional to the projector onto $\big\lvert{\psi_0^\textrm{vortex}}\big\rangle$; specifically:
\begin{equation}
H'=H+H_\textrm{pert.}\qquad \textrm{where} \quad H_\textrm{pert.}=3J\big\lvert{\psi_0^{\textrm{unmixed},2}}\big\rangle\big\langle{\psi_0^{\textrm{unmixed},2}}\big\rvert.\label{eqn:with-vortex-with-perturbation}
\end{equation}
We anticipate such a perturbation to shift the vortex-bound Majorana mode to finite energy $E/(3J)\approx 1$, leaving behind a well-defined Majorana edge state with $E=0,\ell=0$. 
This expectation is confirmed by the edge dispersion data shown in \cref{fig:Edge-dispersion-3}. 
Correspondingly, the complex phase $\arg[\psi_n(j)]$, plotted as a function of the angular coordinate $\varphi(j)$ of sites $j$ in \cref{fig:phase-growth-with-flux}, exhibits for eigenstates $\ket{\psi_n}$ with energy near $E=0$ the expected linear growth with integer slopes, in accordance with $\ell\in\mathbb{Z}$.

The chiral character of the Majorana edge modes in the chiral gapped phase can also be illustrated through the propagation of a wave packet localized at the boundary. 
To that end, we return to the simpler case without a vortex, and we construct the Gaussian projector operator~\cite{Urwyler:2022}
\begin{equation}
\label{eqn:Gaussian-projector}
P_{\mu_E,\sigma_E}=\sum_n \exp\left[-\frac{(E_n-\mu_E)^2}{2\sigma_E^2}\right]\ket{\psi_n}\bra{\psi_n},
\end{equation}
where we sum over all eigenstate labels $n$.
Starting with a state $\ket{\psi_\textrm{loc.}}$ localized on a single site at the boundary of the disk (we specifically select the site with the largest horizontal coordinate $\Re(z)$ and $\Im(z)>0$), the application of the Gaussian projector operator generates a state
\begin{equation}
\label{eqn:wave-packet}
\big\lvert{\psi^\textrm{w.p.}_{\mu_E,\sigma_E}}\big\rangle = P_{\mu_E,\sigma_E}\ket{\psi_\textrm{loc.}}.
\end{equation}
Provided that the range $\abs{\mu_E \pm \sigma_E}$ is sufficiently smaller than the bulk energy gap, the operator $P_{\mu_E,\sigma_E}$ effectively projects onto the edge states. 
Because of the approximately linear dispersion $E(\ell)$ of the edge states in the chiral gapped phase, the Gaussian function of the eigenstate energy $E_n$ in \cref{eqn:Gaussian-projector} is simultaneously a Gaussian function of the angular momentum $\ell$, i.e., the state $\big\lvert{\psi^\textrm{w.p.}_{\mu_E,\sigma_E}}\big\rangle$ constructed in \cref{eqn:wave-packet} is a wave packet.
As such, we expect the wave packet to propagate along the circumference of the disk with angular velocity $\omega = \partial E / \partial \ell$, which due to the chiral dispersion has a definite positive or negative sign. 
To test this prediction numerically, we choose $\mu_E=0$ and $\sigma_E=0.06$, which is one order of magnitude smaller than the bulk energy gap.
The initial wave packet $\big\lvert{\psi^\textrm{w.p.}_{\mu_E=0,\sigma_E=0.06}}\big\rangle$ is plotted in red in \cref{fig:edge-state-evolution}. 
The evolution of this wave packet under unitary time evolution $e^{-\i H T}$ at times $3JT\in\{100,200,300\}$ is depicted in yellow/green/cyan in \cref{fig:edge-state-evolution}.
We observe the propagation of the Majorana wave packet in the counterclockwise (positive) direction, consistent with $\partial E/\partial\ell >0$.

Let us finally contrast the Majorana edge states of the chiral gapped phase to edge states realized in the gapped Abelian $\mathbb{Z}_2$ spin liquid phase. 
For concreteness, we consider the trivial gapped phase at parameters $J_x = J_y = 1/4$, $J_z=1/2$ and $K=0$, which also corresponds to the data in the bottom of \cref{main:fig:DOS_K=0} in the Letter.
The bulk and boundary DOS for this phase are displayed in \cref{fig:DOS-trivial-gapped}.
We observe the formation of a zero-energy flat band on the boundary.
Due to the somewhat large localization length of the corresponding eigenstates (indicated with purple arrow in \cref{fig:localization-trivial-gapped}), the flat band remains visible as a small peak at $E=0$ also in the numerically computed bulk DOS.

To illustrate the absence of chiral edge states in this phase, we consider again the wave packet construction in Eq.~(\ref{eqn:wave-packet}), in this case taking as the seed the localized state $\ket{\psi_\textrm{loc.}}$ with equal-amplitude support on the two sites with the largest value of $\Re(z)$ (one site has $\Im(z)>0$ while the other one has $\Im(z)<0$).
Choosing $\mu_E=0$ and $\sigma_E=0.12$, we obtain the initial state plotted in red in \cref{fig:trivial-edge-states}.
Unitary time evolution with $e^{-\i H T}$ at times $3JT=\{100,200,300\}$, shown in yellow/green/cyan in \cref{fig:trivial-edge-states}, indicates that (\emph{i})~the state does not propagate in any definite direction around the disk and that (\emph{ii})~the state does not noticeably broaden on the chosen time scale.
The absence of clockwise/counter-clockwise propagation confirms the nonchiral character of the $\mathbb{Z}_2$ spin liquid phase, whereas the absence of broadening can be seen as a consequence of the flat-band (i.e., dispersionless) character of the edge states near $E=0$.

\end{document}